\documentclass[preprint,preprintnumbers,aps,nofootinbib,tightenlines,floatfix,groupedaddress]{revtex4-1}

\usepackage{hyperref}
\usepackage{amsmath,amssymb}
\usepackage{cleveref}
\usepackage{graphicx}
\usepackage{xcolor}
\usepackage[utf8]{inputenc}
\usepackage{bm}
\usepackage{comment}
\usepackage[percent]{overpic}
\usepackage{qcircuit}
\usepackage{csquotes}
\usepackage[section]{placeins}
\usepackage{rotating}

\allowdisplaybreaks

\newcount\hour \newcount\hourminute \newcount\minute
\hour=\time \divide \hour by 60
\hourminute=\hour \multiply \hourminute by 60
\minute=\time \advance \minute by -\hourminute
\newcommand{\mydate}{\ \today \ - \number\hour :\number\minute}

\begin{document}

%%%%%%%%%
\title{Minimally-Entangled State Preparation of Localized Wavefunctions on Quantum Computers }
%%%%%%%%%

\author{Natalie Klco}
\email{klcon@uw.edu}
\affiliation{Institute for Nuclear Theory, University of Washington, Seattle, WA 98195-1550, USA}
\author{Martin J.~Savage}
\email{mjs5@uw.edu}
\affiliation{Institute for Nuclear Theory, University of Washington, Seattle, WA 98195-1550, USA}

\date{\mydate}

\preprint{INT-PUB-19-013}

\begin{figure}[!t]
 \vspace{-1.5cm} \leftline{
 	\includegraphics[width=0.2\textwidth]{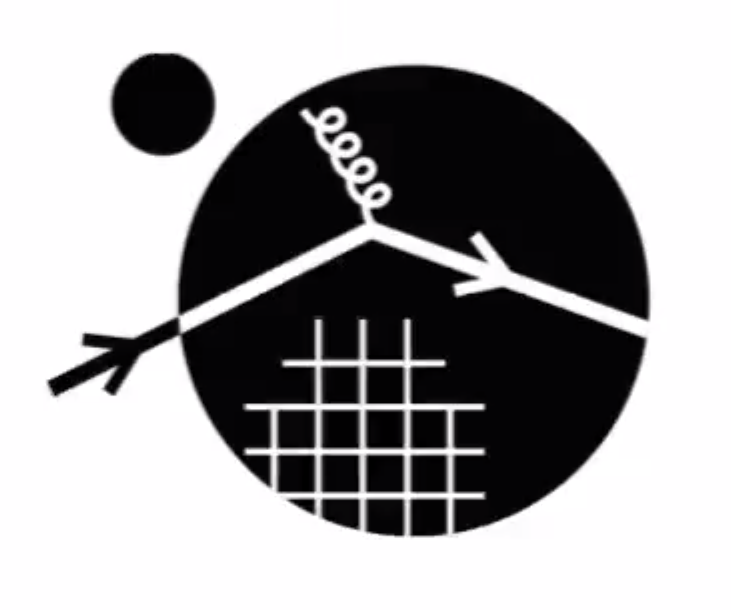}}
\end{figure}

\begin{abstract}
Initializing a single site of a lattice scalar field theory into
an arbitrary state with support throughout the quantum register requires ${\cal O}(2^{n_Q})$ entangling gates
on a quantum computer with $n_Q$ qubits per site.
It is conceivable that, instead, initializing to functions that are good approximations to states may have
utility in reducing the number of required entangling gates.
In the case of a single site of a non-interacting scalar field theory,
initializing to a symmetric exponential wavefunction requires $n_Q-1$ entangling gates, compared
with the $2^{n_Q-1} + n_Q-3 + \delta_{n_Q,1}$ required for a symmetric Gaussian wavefunction.
In this work,
we explore the initialization of 1-site ($n_Q=4$), 2-site ($n_Q=3$) and 3-site ($n_Q=3$) non-interacting scalar field theories
with symmetric exponential wavefunctions using IBM's quantum simulators and quantum devices
({\tt Poughkeepsie} and {\tt Tokyo}).
With the digitizations attainable with $n_Q = 3,4$,
these tensor-product wavefunctions are
found to have large overlap with a Gaussian wavefunction, and provide a suitable low-noise initialization for improvement and  \emph{Somma Inflation}.
In performing these simulations, we have employed a workflow that interleaves calibrations to
mitigate  systematic errors in production.
The calibrations allow tolerance cuts on gate performance
including the fidelity of the symmetrizing Hadamard gate, both in vacuum ($|{\bf 0}\rangle^{\otimes n_Q}$)
and in medium ($n_Q-1$ qubits initialized to an exponential function).
The results obtained in this work are relevant to systems beyond scalar field theories,
such as the deuteron radial wavefunction,  2- and 3-dimensional cartesian-space wavefunctions,
and non-relativistic multi-nucleon systems built on a localized eigenbasis.
\end{abstract}
\maketitle

\tableofcontents
\vfill\eject

%%%%%%%%%%%%%%%%
\section{Introduction}

Quantum simulations of quantum field theories in condensed matter, high-energy and nuclear physics hold the potential to address Grand Challenge problems in each of these science domains that cannot be addressed using classical computation.
Typical situations where a quantum advantage is expected to be achieved include the real-time dynamics of non-equilibrium multi-particle systems and high-density systems of fermions.
Classical calculations of quantities in such systems typically encounter a sign problem in sampling over trajectories in a path integral.  Quantum computing offers the potential to work with Hilbert space dimensionalities that are exponentially larger than those accessible with classical computers,
and the real-time evolution of quantum wavefunctions in both non-relativistic systems and in quantum field theories.
During the last two decades, impressive work has been accomplished in designing algorithms to
simulate quantum field theories in the Hamiltonian framework, including scalar field theories, Abelian gauge theories and non-Abelian gauge theories~\cite{Lloyd1996,SommaPhysRevA2002,Byrnes:2005qx,Jordan:2011ci,Jordan:2011ne,Zohar:2012xf,Zohar:2012ay,Banerjee:2012pg,Banerjee:2012xg,Wiese:2013uua,Marcos:2014lda,Wiese:2014rla,Jordan:2014tma,Bazavov:2015kka,Zohar:2016iic,Somma:2016:QSO:3179430.3179434,Jordan:2017lea,Bermudez:2017yrq,Banuls:2017ena,PhysRevLett.121.110504,Klco:2018zqz,Kaplan:2018vnj,Stryker:2018efp,Kokail:2018eiw,Raychowdhury:2018osk,Zhang:2018ufj,Hackett:2018cel,Gustafson:2019mpk,Lamm:2019bik,Alexandru:2019ozf}, as well as quantum many-body systems~\cite{doi:10.1063/1.3115177,OrtizPhysRevA2001,OvrumHjorthJensen,doi:10.1063/1.3236685,Jiang:2017pyp,Roggero:2018hrn,Bauer:2019qxa}.
These efforts represent the first critical steps toward quantum simulation of such theories, but significant further efforts are required to realize a quantum simulation of quantum field theories that can be compared with experiment.
The rapidly-improving control of entanglement over macroscopic volumes of spacetime has enabled first calculations of simple spin-systems, field theories and non-relativistic effective field theories (EFT) to be performed on small trapped-ion systems, cold-atom systems, annealing devices and superconducting quantum
devices~\cite{Martinez2016,OMalleyPhysRevX,Kandala2017,PhysRevLett.120.210501,PhysRevA.98.032331,Hempel2018,PhysRevA.99.032306}.

The ability to efficiently prepare physically-relevant wavefunctions on quantum devices is
imperative for the success of quantum simulation.
Good progress is being made in understanding how to accomplish effective state preparation in a number of systems~\cite{2002quant.ph..8112G,PhysRevA.71.052330,2008arXiv0801.0342K,doi:10.1063/1.3115177,PhysRevA.83.032302,Somma:2016:QSO:3179430.3179434,2018arXiv180710781A,2018arXiv180302466N}.
An intriguing proposal is to evolve a narrow Gaussian wavefunction to a broad Gaussian wavefunction through application of polynomially-efficient, tuned time evolution operators~\cite{Somma:2016:QSO:3179430.3179434}.
This exploits the fact that the two-qubit gate requirements for arbitrary wavefunction preparation scales inefficiently only with the dimension of the Hilbert space where the wavefunction has support.
We will refer to this as \emph{Somma Inflation}~\cite{Somma:2016:QSO:3179430.3179434}.
For the Gaussian wavefunction, it has been shown that implementation of Somma Inflation on a quantum device requires complexity scaling polynomially with $n_Q$ and $\log \epsilon$ in the large-$n_Q$ limit, consistent with the $n_Q^2$ cost of the Trotterized time evolution operator~\cite{Klco:2018zqz}.
A Gaussian wavefunction prepared on a small subset of the intended qubits may be expanded to the full digitized space of large qubit number with polynomial cost.
It is the preparation on this small subset in the NISQ era that is the focus of this work.

In the case of  scalar quantum field theory,
as for many problems of interest,
the wavefunctions of the low-lying states of the non-interacting
theory can be constructed from the eigenstates of the harmonic oscillator,
the ground state of which is a Gaussian distribution.
In working with a scalar field theory discretized on a lattice,
described by a spatial grid of points with a real scalar field residing at each grid point,
the field at each grid point is represented by a wavefunction distributed across a number of qubits, $n_Q$.
One basis choice is defined by the eigenstates of the field operator at each spatial site and
is efficient for Hamiltonian evolution~\cite{Jordan:2011ci}.
In this basis, the ground state of the field theory is an entangled state of position-space field wavefunctions.
In the study of the ground state or low-lying states of the scalar field theory, the system would ideally be initialized
into one of, or a superposition of, these states.
The tensor-product state with each spatial site in its ground state, is a (simple)
superposition of states that has a significant overlap with the ground state of the field theory and low-lying excitations.

Previously,
we built upon the seminal work of Jordan, Lee and Preskill, and subsequent works related to digitizing scalar field theories~\cite{Somma:2016:QSO:3179430.3179434,PhysRevLett.121.110504},
to understand what might be achievable with NISQ-era quantum devices in the context of
the number of qubits required to suppress digitization artifacts below expected quantum noise levels and in the impact of gate fidelity~\cite{Klco:2018zqz}.
The digitizations of  scalar fields with a small number of qubits per spatial site
indicate that the theoretical systematic errors introduced through digitization will not constrain the precision and accuracy of such computations, which instead are expected to be limited by the capabilities of the quantum hardware.
We also detailed the qubit interactions and gate operations required to accomplish time-evolution of
1-site and 2-site $\lambda\phi^4$ theory in symmetric localized and delocalized  states~\cite{Klco:2018zqz}.
The gate operations for time evolving larger lattices can be easily built from these basic elements.
In this work, we turn to the issue of state preparation for scalar field theory.
At each site of a non-interacting massive scalar field, the field space ($\phi$-space) is spanned by the eigenstates of
a harmonic oscillator,
and the ground state of the $d$-dimensional quantum field theory is an entangled state of these $\phi$-space eigenstates induced by the  $({\bm \nabla}\phi)^2$ operator.
Considering a 1-site system defined with $n_Q$ qubits,
initializing an arbitrary state requires $2^{n_Q}-2$ entangling gates, while
a wavefunction with definite reflection symmetry, such as the ground state,
requires  $2^{n_Q-1} + n_Q - 3 + \delta_{n_Q,1}$
entangling gates.
As a symmetric exponential wavefunction requires only $n_Q-1$ entangling gates to initialize,
and can be tuned to have substantial overlap with the desired digitized ground state wavefunction,
we suggest that such wavefunctions may provide an efficient means of state preparation for  scalar field theory.
It is this thesis that we examine more closely within the IBM~Q~Experience quantum ecosystem~\cite{IBMdevices,IBMQExpe34:online}.

%%%%%%%%%%%%%%%%%%%%%%%%%%%%%%%%%%%%%%
\section{Preparation of a Symmetric Exponential Wavefunction}
\label{sec:ExpPrep}

\subsection{Quantum Circuits}
\label{sec:ExpCircuits}

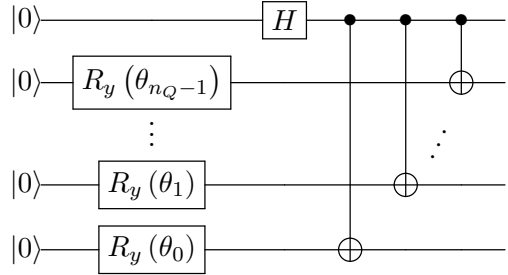
\begin{figure}
\[
  \begin{gathered}
  \Qcircuit @C=1.1em @R=.6em {
  |0 \rangle \quad & \qw & \gate{H} & \ctrl{5} & \ctrl{4} & \ctrl{1} & \qw \\
  |0 \rangle \quad & \gate{R_y \left( \theta_{n_Q-1} \right)} & \qw & \qw & \qw & \targ & \qw \\
  &\vdots &\hspace{3.7cm} \text{\begin{rotate}{-80}$\ddots$\end{rotate}} \\ \\
  |0 \rangle \quad & \gate{R_y \left( \theta_1 \right)} & \qw & \qw & \targ & \qw & \qw \\
  |0 \rangle \quad & \gate{R_y\left( \theta_0 \right)} & \qw & \targ &  \qw & \qw & \qw \\
  }
  \end{gathered}
\]
  \caption{Quantum circuit for implementation of the symmetrized exponential wavefunction.  The single qubit rotations, with angles defined by
  Eq.~\eqref{eq:expangles}, prepare an exponential wavefunction on $n_Q-1$ qubits while the Hadamard and CNOTs are responsible for duplication and subsequent symmetrization.}
  \label{fig:symmetrizedExpcircuit}
\end{figure}

Due to the translational invariance of the ratio of function evaluations between binary-interpreted neighbors
that are spaced equally on a one-dimensional grid, a wavefunction with positive and real amplitudes distributed as an exponential of the form
\begin{equation}
  |\psi\rangle =
  \frac{1}{\sqrt{\sum\limits_{y = 0}^{2^{n_Q}-1} \psi^2(y) }}
  \sum\limits_{x = 0}^{2^{n_Q}-1} \psi (x) |x\rangle \qquad \psi(x) = e^{\alpha x}
\end{equation}
can be prepared on $n_Q$ qubits without entanglement.
This can be implemented through a particular choice of angle for each of $n_Q$ qubit rotations about the $y$-axis,
\begin{equation}
  R_y(\theta) = e^{-i \theta \sigma_y} = \begin{pmatrix}
    \cos \theta & -\sin \theta \\ \sin \theta & \cos \theta
  \end{pmatrix}
  \ \ \ .
\end{equation}
The first qubit, representing the nearest-neighbor connectivity on the grid and thus the least significant bit in the binary representation, is rotated by an angle $ \theta_0 = \arctan\left( e^{\alpha}\right)$.
Subsequent qubits can be added to copy this ratio structure into the volume-doubled Hilbert space with a rotation on the newest qubit that determines the ratio at the boundary between the two half spaces.
Fixing this ratio across the binary-interpreted boundary (between states $|2^{n_Q-1}-1\rangle$ and $|2^{n_Q-1}\rangle$),
determines the angles necessary to prepare an exponential function across the amplitudes of a digitally-controllable quantum state,
\begin{equation}
  \theta_\ell =  \arctan \exp\left(2^\ell \alpha\right)
  \ \ \ .
  \label{eq:expangles}
\end{equation}
If an exact copy (rescaled for normalization) of this exponential is desired, this can be achieved through
the addition of a qubit in the $|+\rangle = H|0\rangle$ state ($H$ is a Hadamard gate)
as the most significant bit.  Once this copy is created, the second half of the Hilbert space can be reversed in its binary interpretation through the implementation of 2-qubit CNOT gates.
This procedure creates a wavefunction localized in the center of the binary Hilbert space with exponentially-decaying tails,
and is summarized in the circuit diagram of Fig.~\ref{fig:symmetrizedExpcircuit}.  Implementing this spatial parity is a simple case of the general reflection operators discussed in Ref.~\cite{2018arXiv180302466N}.
To antisymmetrize the wavefunction rather than symmetrize, the last introduced qubit should be in the state $|-\rangle$,
rather than $|+\rangle$, which can be achieved through the implementation of an $\hat X$ Pauli operator
before the Hadamard, or a $\hat Z$ Pauli operator following the Hadamard, as shown in Fig.~\ref{fig:symmetrizedExpcircuit}.

As will be emphasized in the following section on Gaussian initialization, the number of CNOT gates required to prepare the localized exponential wavefunction is significantly reduced (growing as $n_Q-1$) with respect to that required to prepare arbitrarily-structured localized wavefunctions, which scales as $\mathcal{O}(2^{n_Q})$.
This feature makes the preparation of the symmetrized exponential wavefunction amenable to NISQ-era digital quantum devices.
In addition to its low CNOT cost, the connectivity required for this implementation is significantly reduced.
The qubit associated with the most significant qubit (the last qubit introduced for symmetrization) appears at the control bit for all CNOT gates.
This requires one qubit to have connectivity with all other qubits but does not require connectivity amongst the $n_Q-1$ previous qubits.
Then by connectivity, the digitization achievable on the preparation of a localized symmetric exponential on a quantum device is set by the connectivity of the most-connected qubit.
For example, on IBM's {\tt Poughkeepsie}({\tt Tokyo}) device, the most-connected qubits have 6(3) nearest neighbors allowing a localized exponential to be prepared with $n_Q = 7(4)$.
If one doubles the number of CNOT gates required for the symmetrization, the CNOTs can be organized to appear only between nearest-neighbor qubits (see Appendix~\ref{app:Connectivity} for details).
With this choice, the connectivity of {\tt Poughkeepsie}({\tt Tokyo}) allows symmetric exponentials digitized with up to $n_Q = 18(17)$.

\begin{figure}
  \centering
  \includegraphics[width=0.3\textwidth]{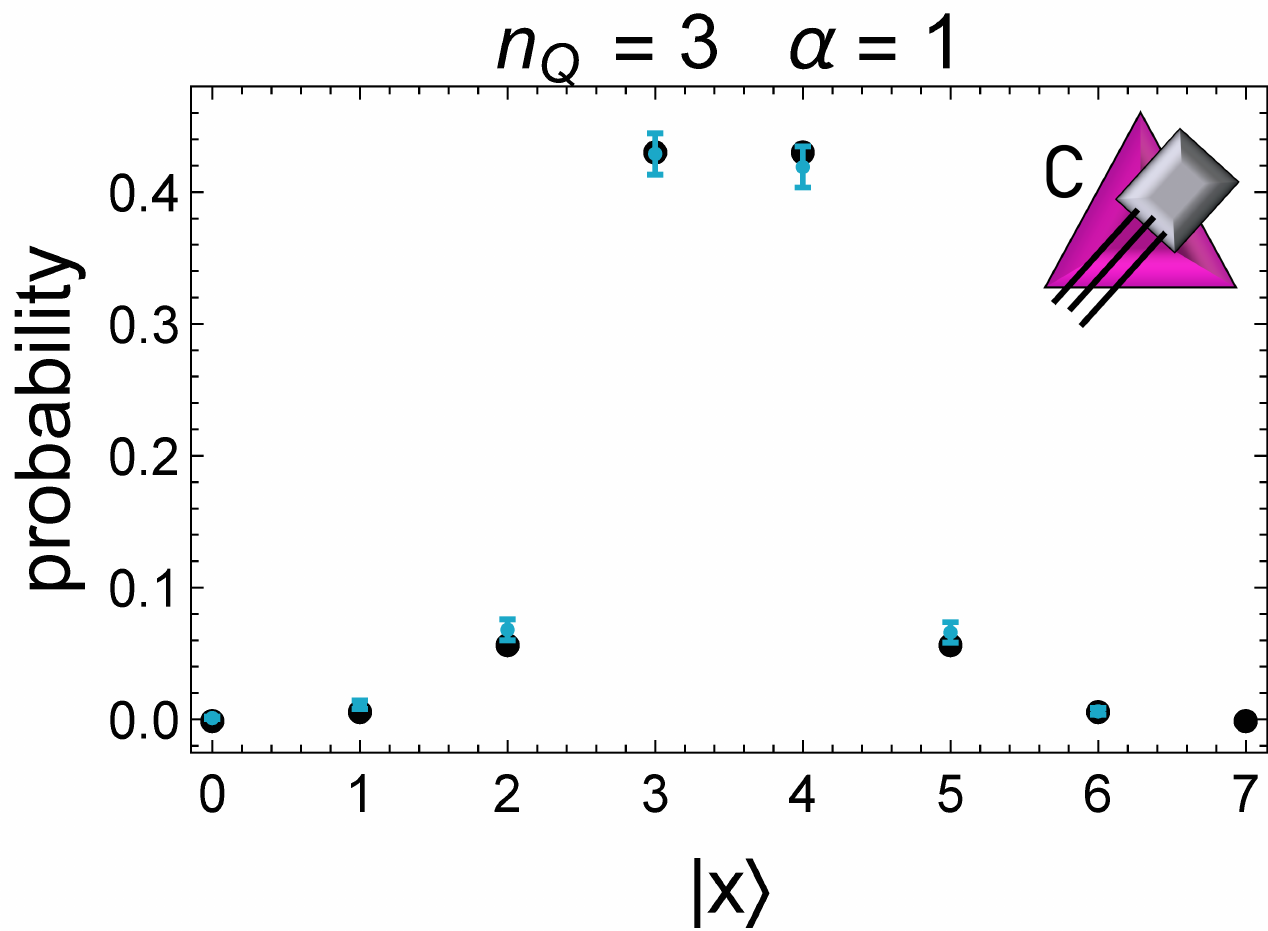}
  \includegraphics[width=0.3\textwidth]{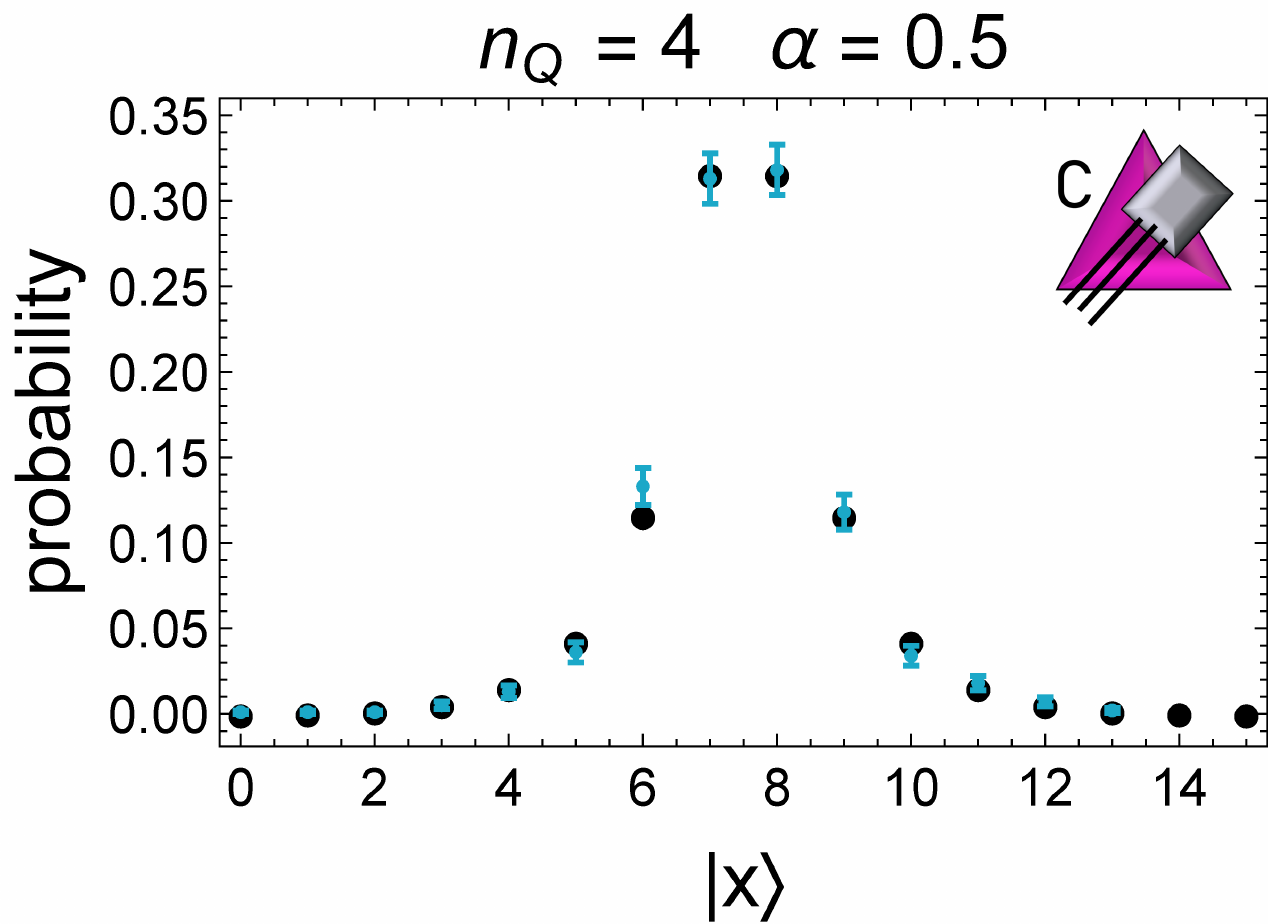}
  \includegraphics[width=0.3\textwidth]{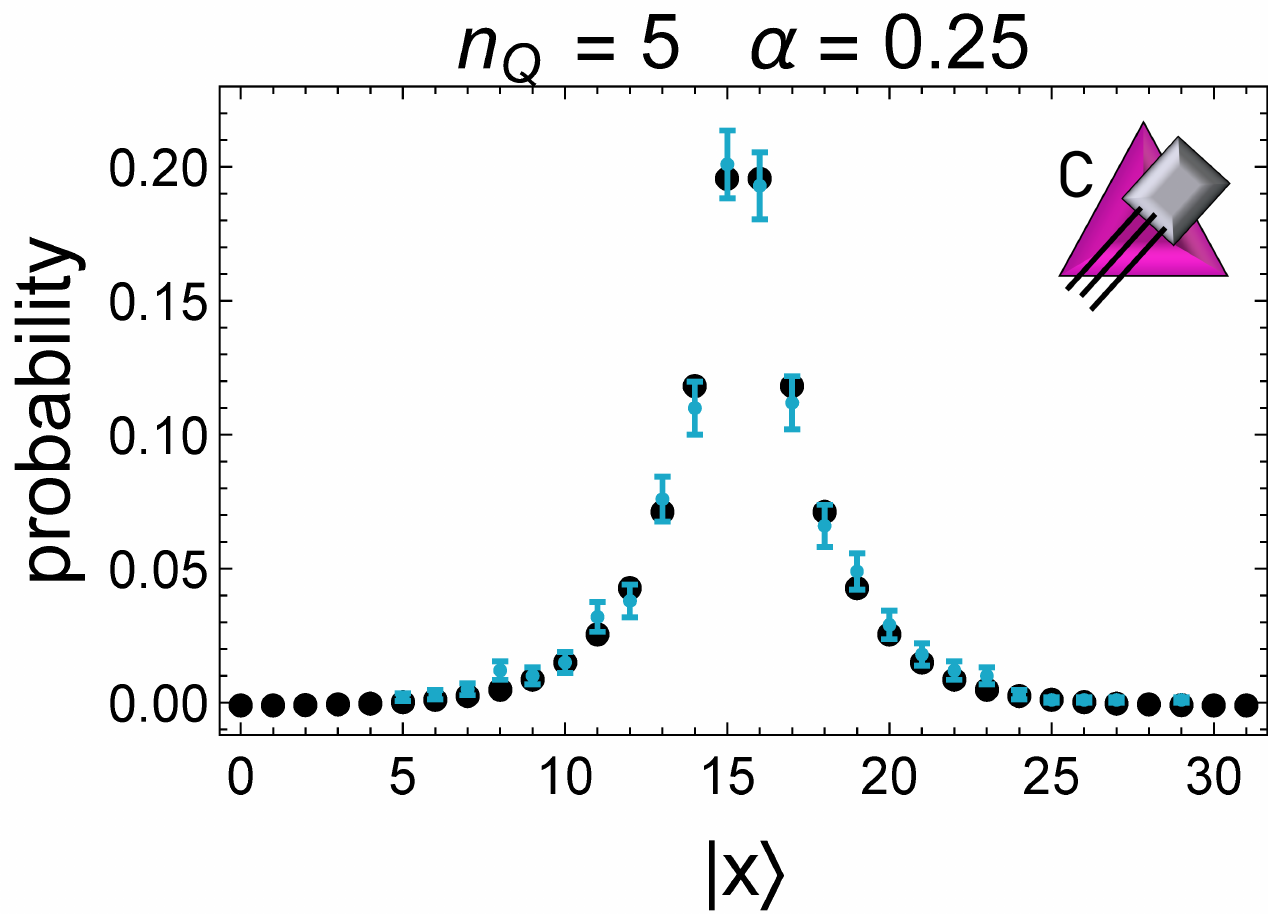}\\
  \includegraphics[width=0.3\textwidth]{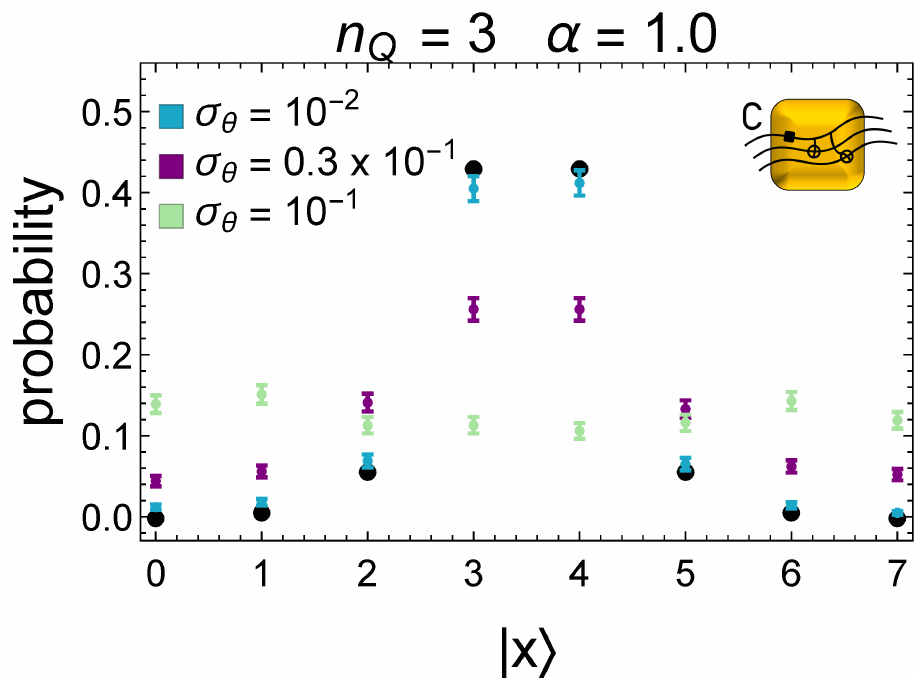}
  \includegraphics[width=0.3\textwidth]{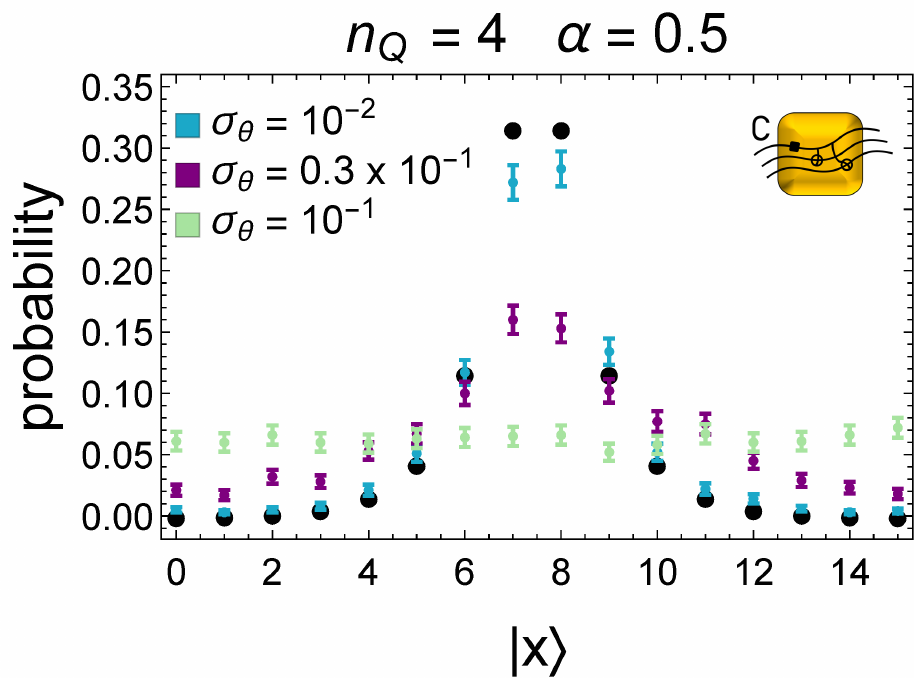}
  \includegraphics[width=0.3\textwidth]{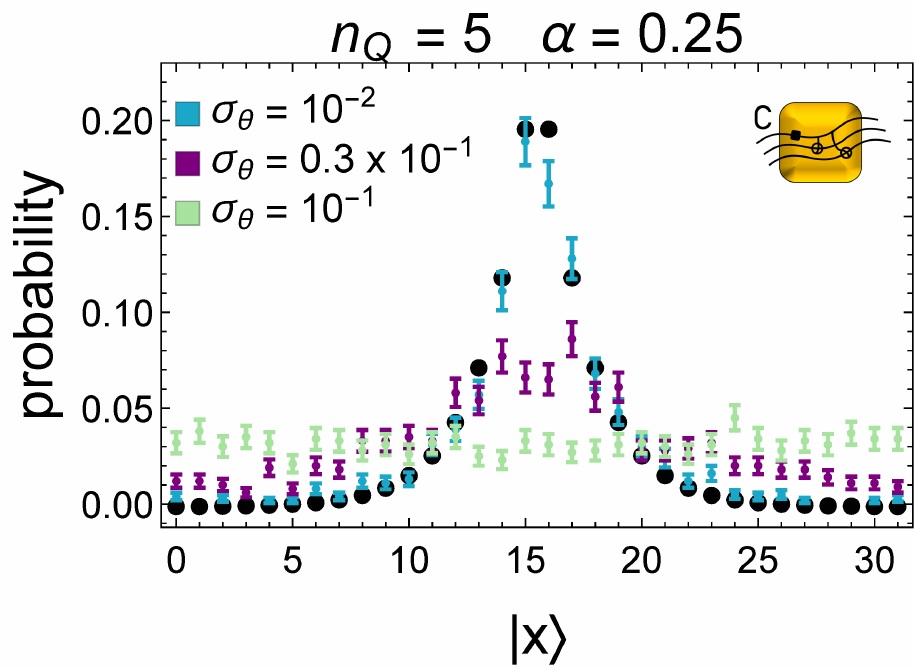}
  \caption{
  Simulations preparing a localized exponential on an ideal quantum device with 1000 samples.  The top row demonstrates the functionality of the method on an ideal quantum simulator.
  The bottom row explores the effects of noisy quantum gates with Gaussian-distributed random errors with
  standard deviations of  $\sigma_\theta$ for single-qubit rotations and  $\sigma_{CNOT} = 10 \sigma_\theta$ for  2-qubit entangling gates. Details of the resource icons used in figures throughout this work can be found in Appendix~\ref{app:ResourceIcons}.
  }
  \label{fig:ExpSim}
\end{figure}
Because the scheme of Fig.~\ref{fig:symmetrizedExpcircuit} introduces additional samples of the wavefunction at the end of the binary-interpreted register rather than increasing the density between existing samples, the value of $\alpha_\ell$ is $n_Q$ dependent if increased digitization is desired (i.e., the scale of $\alpha$ is measured against the digitization of the Hilbert space).
To increase the digitization of an exponential distribution constant in $n_Q$,
$\alpha$ is scaled by a factor of 1/2 with each added qubit.
This is shown in Fig.~\ref{fig:ExpSim} for a progression from 3 to 5 qubits.
In this figure, an ideal quantum device~(\includegraphics[width=5mm]{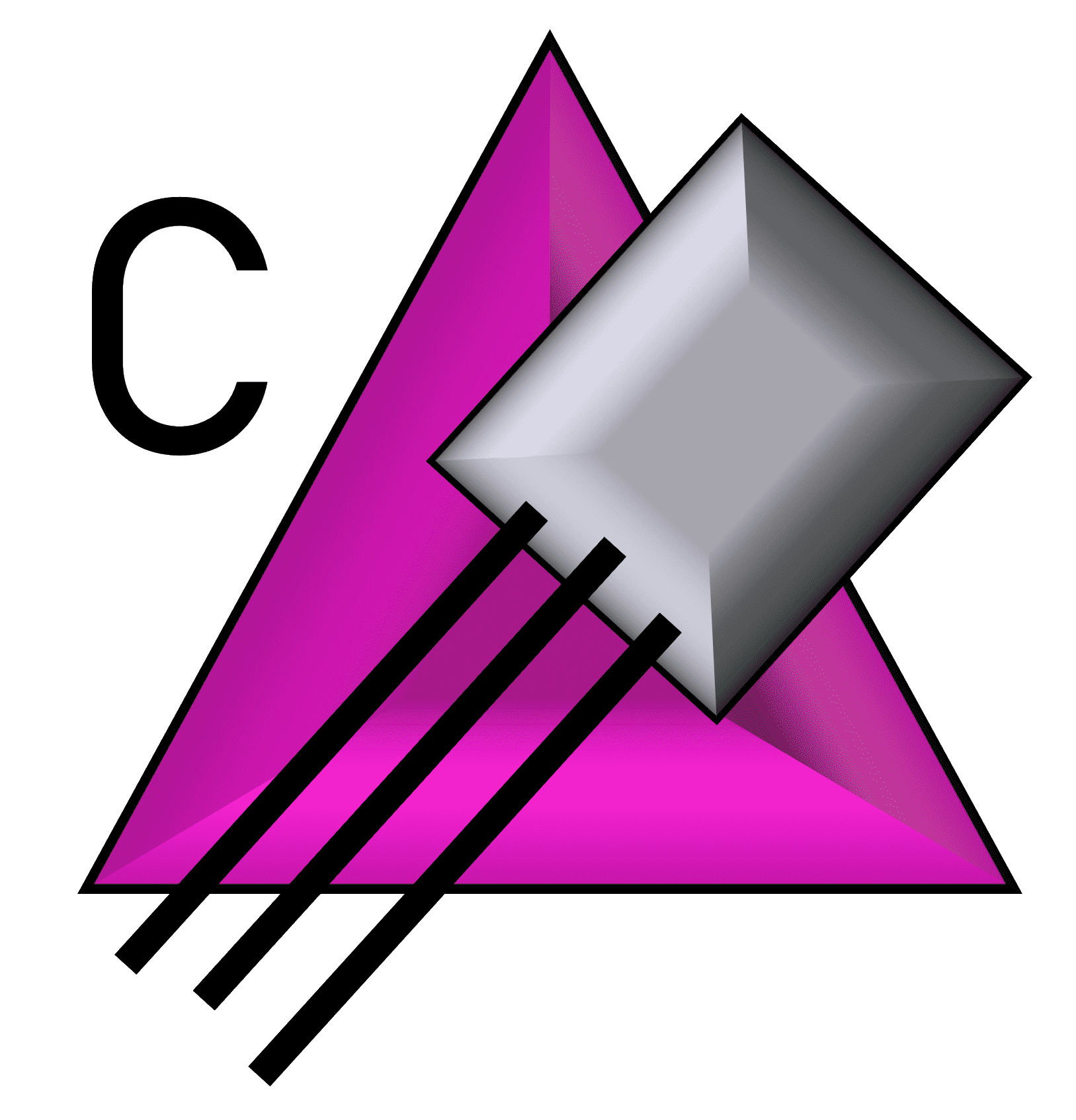} see Appendix~\ref{app:ResourceIcons} for details of the resource icons used throughout this work.) is simulated with 1000 samples in the top row.
Notice that this choice of rescaling $\alpha$ to increase the digitization of the localized exponential results in an increased number of basis states within the wavefunction's region of support and thus a reduction in the signal-to-noise (StN) ratio.
The bottom row of Fig.~\ref{fig:ExpSim} shows the same progression with three levels of Gaussian noise on each gate~(\includegraphics[width=5mm]{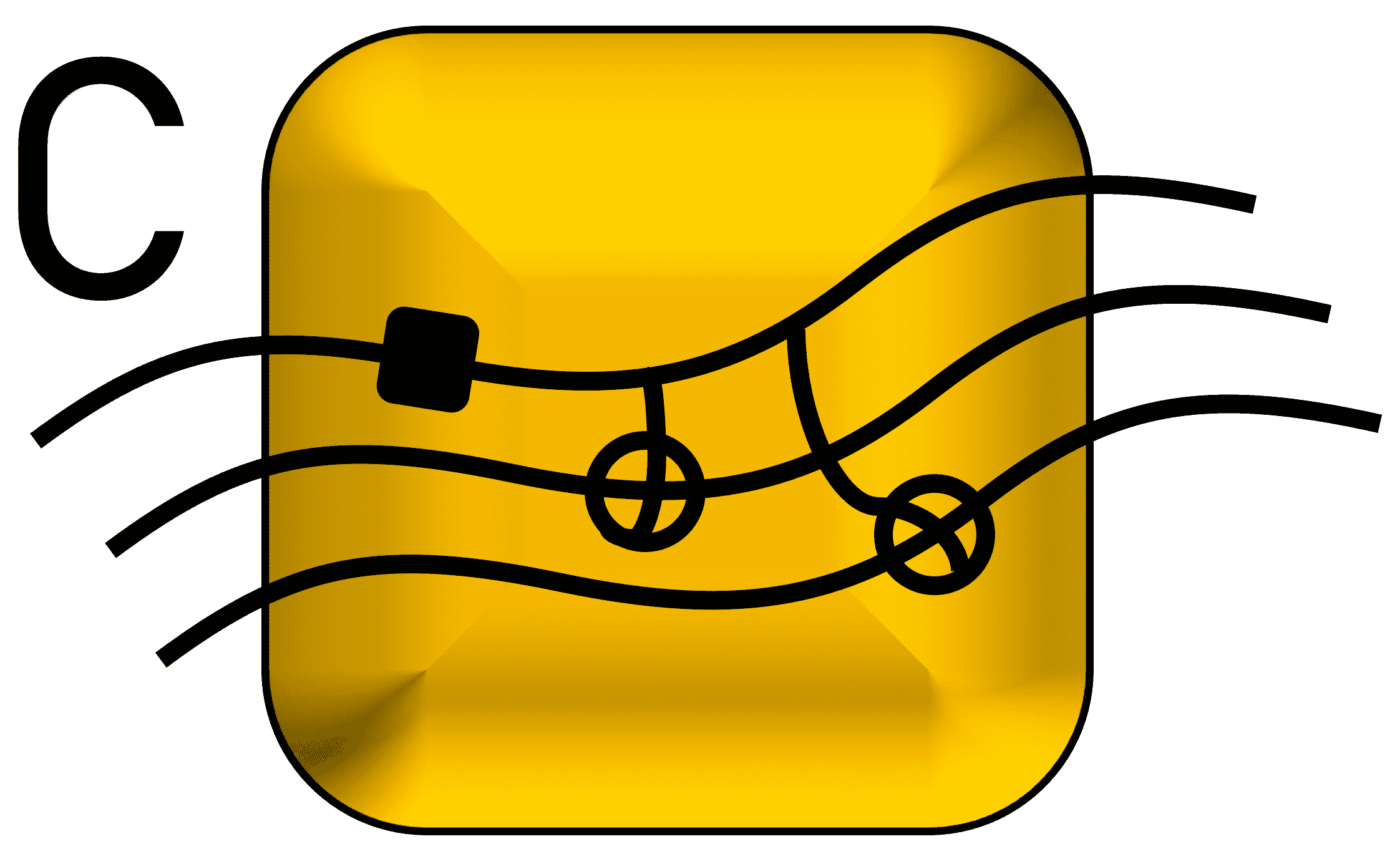}).
As the CNOT errors tend to dominate current error rates, we set $\sigma_{CNOT} = 10 \sigma_\theta$.
Each single-qubit rotation is surrounded by SU(2) operators with angles sampled from a normal distribution of mean zero and standard deviation $\sigma_\theta$.
For CNOT gates, the control and target are surrounded by SU(2) $\otimes$ SU(2) operators with angles sampled from a normal distribution of mean zero and standard deviation $\sigma_{CNOT}$.
While formulating a fairly primitive quantum noise model, this shows the sensitivity of the state preparation to the precision of angles achievable in quantum gate implementation.

\subsection{Quantum Methods}

Our state preparation calculations were performed on
IBM's {\tt Tokyo} and {\tt Poughkeepsie} 20-qubit superconducting quantum devices~\cite{IBMdevices}
(\includegraphics[width=5mm]{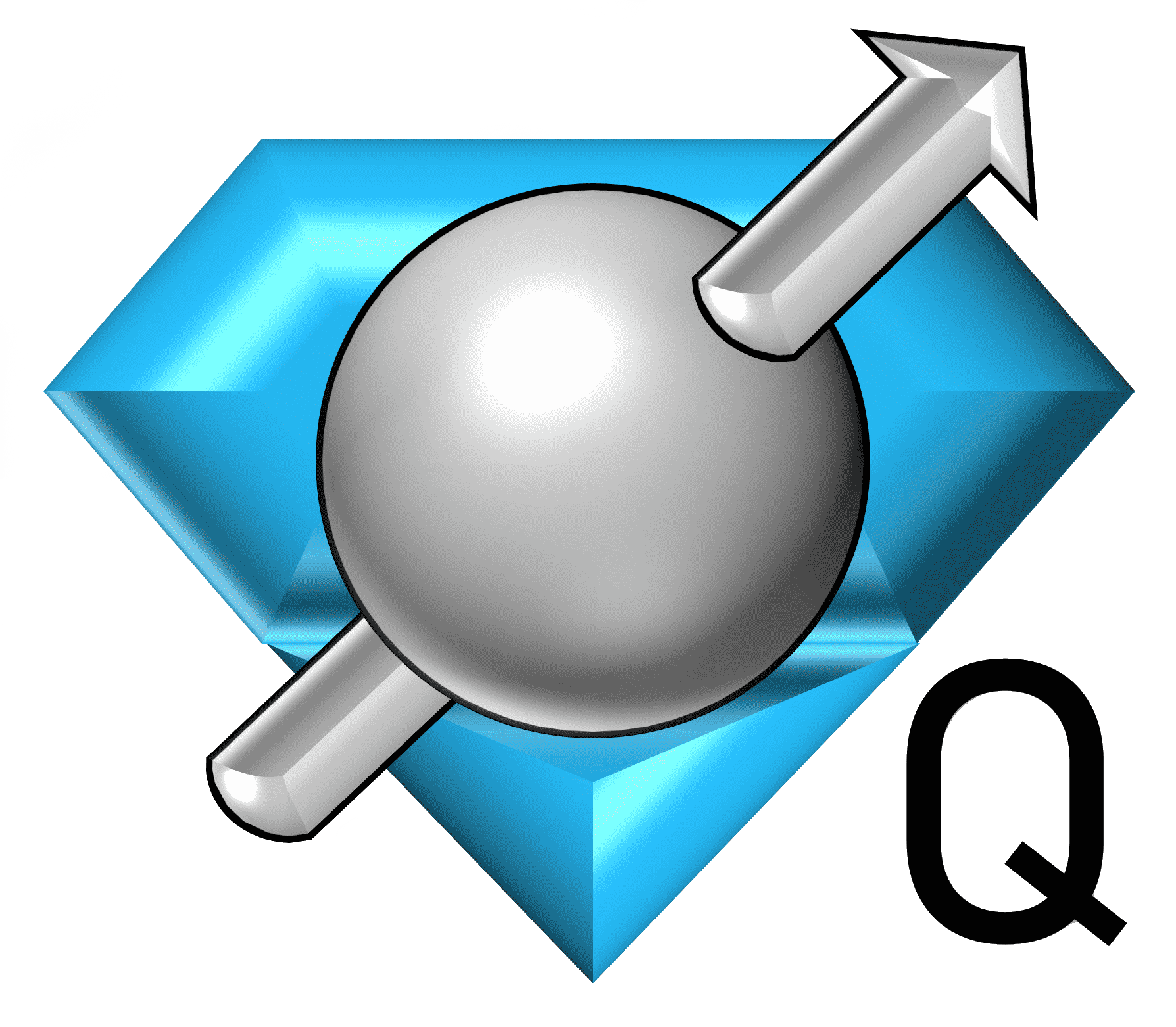}),
on IBM's default {\tt qiskit} noisy quantum simulator~\cite{Qiskit} (\includegraphics[width=5mm]{iconC2N_bf.png})
with the noise models from {\tt Tokyo} and {\tt Poughkeepsie},
on {\tt Qiskit}'s default ideal quantum simulator (\includegraphics[width=5mm]{iconC1_bf.png}),
and on classical computers (laptops)(\includegraphics[width=5mm]{iconC1_bf.png}).
Results obtained using {\tt Poughkeepsie} are presented in this paper only.
After developing the quantum logic and circuits using {\tt Mathematica}~\cite{Mathematica},
quantum circuits were compiled within {\tt Qiskit}~\cite{Qiskit} using python notebooks, and executed through
cloud-based access to IBM simulators and quantum devices through the IBM Q Hub at Oak Ridge National Laboratory.
For each computation on a quantum device or simulator, 20 qubits were allocated to the process,
while only 4, 6, or 9 ``active'' qubits and classical registers
were acted on with gates or measured, depending on the particulars of the computation.
For a given problem size, the selection of active qubits was determined by qubit connectivity and the
results of the most recent IBM device calibration~\cite{IBMdevices}.
\begin{table}[!ht]
  \begin{tabular}{c|c|c|c|c}
  \hline
  \hline
   &  spatial sites  & $n_Q$ qubits per site  &   active qubit partition & date\\
  \hline
  \hline
 \includegraphics[width=6mm]{iconQ1_bf.png}&1 &  4 & [5,11,15,{\bf 10}] & March 31, 2019\\
 \includegraphics[width=6mm]{iconC2N_bf.png}&2 &  3 & [11,15,{\bf 10}]  [8,14,{\bf 9}] & April 2, 2019\\
 \includegraphics[width=6mm]{iconQ1_bf.png}&2 &  3 & [15, 17, {\bf 16}] [14, 18, {\bf 19}] & April 9, 2019\\
 \includegraphics[width=6mm]{iconQ1_bf.png}&3 &  3 & [15, 17, {\bf 16}] [14, 18, {\bf 19}] [1, 3, {\bf 2}] & April 9, 2019 \\
   \hline
  \hline
  \end{tabular}
  \caption{(Color online) The map of active qubits of the {\tt Poughkeepsie} chip employed for the
  1-site, 2-site, and 3-site calculations.
  The bold face qubits denote those with appropriate connectivity to the other qubits in the bracket for the application of symmetrizing entangling gates.  }
  \label{tab:qubitmaps}
\end{table}

Two distinct workflows, {\it Calibration Production} (W1) and {\it State Preparation Production}  (W2),
were used to establish a symmetric exponential wavefunction centered on the mid-point of the Hilbert space.
The calibration workflow of W1 involved:
\begin{enumerate}
\item
{\bf Multi-qubit measurement-error correction}:
This was
accomplished through an ensemble of measurements of the $2^{N_Q}$
state probabilities resulting from the quantum register being initialized into each of the possible computational basis states.
\item
{\bf Vacuum Hadamard calibration}: A sweep over a limited range of SU(2) y-angles in the neighborhood of $\pi/2$ corresponding to a Hadamard gate on the last qubit with the first $n_Q-1$ qubits remaining in the $|{\bf 0}\rangle^{\otimes {n_Q-1}}$ state.
\item
{\bf In-medium Hadamard calibration}: A sweep over a limited range of SU(2) y-angles in the neighborhood of $\pi/2$ corresponding to a Hadamard gate on the last qubit with the first $n_Q-1$ qubits transformed to establish an exponential profile on the first $2^{n_Q-1}$ states in the space.
\end{enumerate}
Analysis of the results obtained with W1,
including the shape of the two exponentials created by the Hadamard gate on the last qubit,
revealed an optimal, shape-preserving, SU(2) y-angle used in the subsequent W2 workflow.
The state preparation workflow of W2 involved:
\begin{enumerate}
\item
{\bf Duplicate exponentials}:
An exponential in the first half of the register was duplicated in the second half using the tuned SU(2) y-rotation
(to recover a Hadamard gate) on the last qubit (as in the Vacuum and In-medium calibration steps in the previous workflow).
\item
{\bf State Preparation}:
A ``reflection symmetric'' exponential distribution, centered on the mid-point of the state-space,
is produced by acting with $n_Q-1$ CNOT gates controlled by the last qubit.
\item
{\bf CNOT error mitigation circuits}: the state preparation circuits are executed again,
 except with each CNOT gate replaced by an odd number, $r$, of such gates.
 Extrapolations in distributions obtained from these calculations allow for mitigation of the
 CNOT  gate noise.
\item
{\bf Reproducibility runs}: The state preparation circuits are repeated
multiple  times to explore variability in the production,
providing an estimate of the systematic error associated with device drift.
\end{enumerate}
Each of the circuits in W1 and W2 are measured with $8000$~shots for the single-site implementations.

For each observable in each production run, the number of counts accumulated in each state in the Hilbert space were recorded within the {\tt Qiskit} suite.
Standard statistical tools were used to convert the accumulated counts into probability distributions.
The multi-qubit measurement-error correction results obtained in W1 are combined to form a
$2^{n_Q}\times 2^{n_Q}$ matrix used to mitigate systematic errors inherent to the quantum device,
and approximately simulated in the noisy simulator, that modify all other calibration and production measurements in W1 and W2,
as described in Appendix~\ref{app:MeasurementCorrection}.

%%%%%%%%%%%%%%%%%%%%%%%%%%%%%%%%%%%%%%%%%%%%%%%%%
\FloatBarrier
\subsection{Hadamard Tuning}
\label{sec:HadamardTuning}
\FloatBarrier

\begin{figure}[!ht]
  \includegraphics[width = 0.7\textwidth]{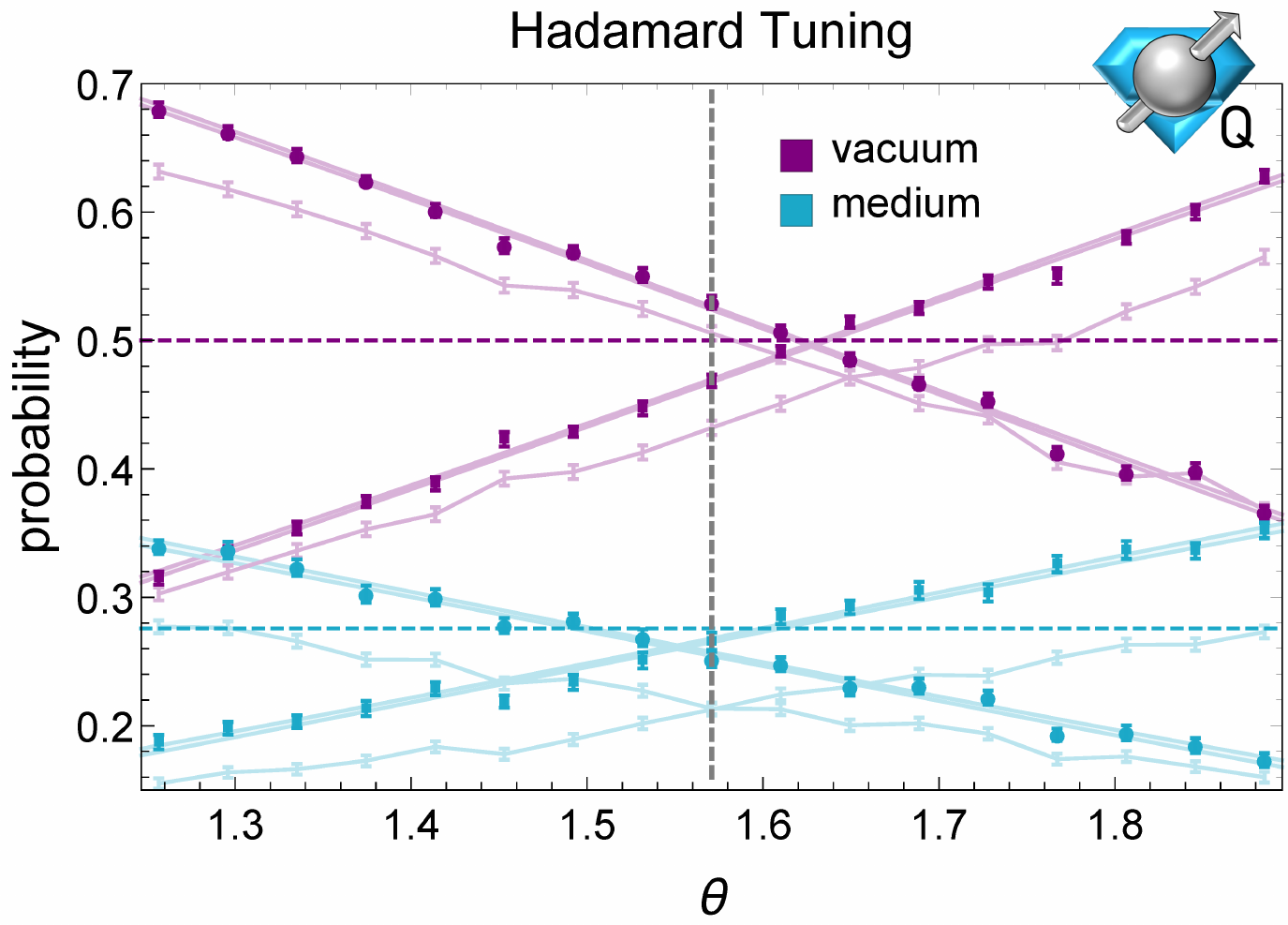}
  \caption{(Color online)
  For vacuum tuning of the Hadamard, probabilities in states $|0\rangle$ and $|8\rangle$ are measured
  after implementation of IBM's $U_3( \theta, \phi, \lambda) = U_3\left(\theta,0,\pi\right)$ gate.
  The active qubits were: 5, 11, 15, and 10 of IBM chip {\tt Poughkeepsie}.
  For in-medium tuning, probabilities in states $|7\rangle$ and $|15\rangle$ are shown after
  preparation of the half-space exponential and $U_3\left(\theta,0,\pi\right)$ on qubit 10.
  Horizontal dashed lines indicate theoretical probabilities at which these states cross for the ideal Hadamard.
  When $ \theta = \frac{\pi}{2}$, the gate theoretically implements the Hadamard gate, $U_3(\pi/2, 0, \pi) = H$.
  The light/dark points correspond to before/after measurement-error correction.
  Theoretically, this observable is insensitive to tuning $\phi$ and $\lambda$.}
  \label{fig:HadamardTuning}
\end{figure}

The symmetrizing circuit of Fig.~\ref{fig:symmetrizedExpcircuit} relies upon the quality of the Hadamard operation
and the stability of the control in CNOT operations.
This section explores the former of these sensitivities by scanning the first angle, $\theta$, of the IBM
$U_3(\theta,\phi,\lambda)$ operator with $\phi=0$ and $\lambda=\pi$ (their values for Hadamard implementation).
The angle, $\theta$ is responsible for modifications to the amplitude beyond phases and theoretically produces the Hadamard operation when $\theta = \pi/2$.
For $n_Q=4$,
the action of the Hadamard in vacuum (i.e., implemented directly from the initial state $|0\rangle$) gives
the probabilities of measuring states $|0\rangle$ and $|8\rangle$ that are shown as the
purple points of Fig.~\ref{fig:HadamardTuning}.
The lightly-colored, joined (purple) points are the raw data from the device
before correcting for measurement errors.
It can be seen that the intersection of probabilities for these two states occurs at a probability $\sim 3\%$
lower than expected for the Hadamard operation.
This is due to excitations of the other 14 states in the active Hilbert space (or even the other $2^{20} - 2$ states
accessible by the {\tt Poughkeepsie} chip) and is found to be well corrected by the measurement-error
correction procedure proposed by IBM for their device (see Appendix~\ref{app:MeasurementCorrection}).
The measured data with statistical uncertainties from 8000 shots at each angle and 8000 shots for each of the 16 calibration circuits used in the measurement-error  correction are shown in the dark (purple) points of Fig.~\ref{fig:HadamardTuning}.
The measurement-error corrected rotation of the $|0\rangle$ state has been fit to its theoretical sinusoidal form,
with the a determination of $\cos^2\left(0.503(7) \theta -0.03(1) \right)$, in good agreement
with the expected $\theta$-dependence though conclusively shifted by fractions of a radian.
A compensating shift to the definition of the digital Hadamard gate can be implemented to restore the expected symmetric $z$-basis superposition property of the gate.

The in-medium Hadamard tuning shown in Fig.~\ref{fig:HadamardTuning} has the same interpretation as the
vacuum measurements.
The lightly-colored, joined points are the data before measurement-error correction,
 while the dark points are the measurement-error corrected values with statistical uncertainties
 resulting  from the 8000 samples of each Hadamard as well as each measurement-error  correction calibration circuit.
The description "in-medium" indicates that the corresponding asymmetry of the Hadamard gate (now observed from states $|7\rangle$ and $|15\rangle$) is probed in a background of the exponential wavefunction preparation---the other three qubits (5, 11, and 15) having been acted upon by the single-qubit rotation gates necessary to prepare the exponential wavefunction (see the circuit in Fig.~\ref{fig:symmetrizedExpcircuit}).
If the device was an ideal quantum simulator, the tuning procedure for the in-medium Hadamard would be equivalent to that in vacuum.
However, due to residual many-body interactions between the qubits, the in-medium measurement requires a different angular shift to instantiate the copied exponential.
Constrained by the theoretically-calculated in-medium amplitude of 0.55, the input-angle-dependence for the evolution of the $|7\rangle$-state probability with the in-medium Hadamard was determined to be
$0.55\ \cos^2 \left(0.49(2) \theta +  0.05(2) \right)$.
Again, a  shift in the definition of the Hadamard angle is found,  but of opposite sign to that found in vacuum.

The $Z$-basis probability measurements implemented here are theoretically only sensitive to modifications of the angle
$\theta$ and insensitive to the phase angles $\phi$ and $\lambda$, while the quantum noise occurring in the device is capable of producing phases and rotations in any direction.
Generically, the correction procedure we have used would be insufficient to correct the Hadamard gate; to do so would require more substantial calibrations e.g., gate set tomography \cite{PhysRevA.87.062119,2013arXiv1310.4492B,2015arXiv150902921G,2017NatCo...8R....B}.
For our purposes in minimal-entanglement state preparation, however, where the current goal is preparation of localized probability distributions regardless of their phase content, it is possible to improve the implementation of the gate as observed in the $Z$-basis with significantly-reduced quantum computational requirements.

With the observations above, it becomes relevant to define a smearing function intended to capture the distribution of quantum states produced when implementing an operator in hardware.  We define the effective gate application (EGA) function, $f$, as the distribution of unitaries effectively implemented in a noisy quantum device
\begin{equation}
  U_{\boldsymbol{\theta}} |\psi\rangle = \int \text{d} \boldsymbol{\theta}' \ f\left(\boldsymbol{\theta},\boldsymbol{\theta}'\right) U_{\boldsymbol{\theta}'} |\psi\rangle \ \ \ .
\end{equation}
For the Hadamard implementation, a 1-parameter family of gate operations is currently considered, but the EGA may be defined to be a higher-dimension distribution in which case $\boldsymbol\theta$ becomes a vector of parameters.
The EGA may also be state- or medium-dependent as we have seen above.
For an ideal quantum device, the EGA is the Dirac delta function.
For the purposes of characterizing the $U_3$ gate in a way relevant to our goals, we calculate the EGA for both the vacuum and in-medium rotations.
The data for this calculation can be seen in Fig.~\ref{fig:measurementcorrection}. This figure shows the deviation between the angle implied by the measured ratio of the two quantum states of dominant probability and the input angle to the $U_3$ gate as a function of the input angle.
For the vacuum measurement, the two dominant quantum states are $|0\rangle $ and $|8\rangle$; for the in-medium measurement, the two dominant quantum states are $|7\rangle $ and $|15\rangle$.
Before measurement-error correction, the device appears to under rotate by an amount scaling with the input angle.
After measurement-error  correction, this angular dependence is removed in both in vacuum and in medium.
From the 17 results in vacuum and 56 results in medium, the Gaussian approximations to the EGA functions are
found to have  the following parameters:
\begin{align}
  \mu_{\rm vac} &= 0.0280(9) \qquad  &\sigma_{\rm vac} &= 0.0051(7) \nonumber \\
  \mu_{\rm med} &= -0.012(1) \qquad  & \sigma_{\rm med} &= 0.045(1) \ \ \ .
  \label{eq:EGAparameters}
\end{align}
Notice that the measurement in vacuum was found to be shifted from its expected value,
but remained fairly stable,
while the EGA in medium has only a small shift with a larger distribution over the implemented angles.
In the next section, the in-medium Hadamard gate is corrected  to prepare a symmetrized exponential.
On-line confirmation of this tuning is performed by interspersing calibration measurements in the production workflow
to allow for post-processing cuts to be applied.

%%%%%%%%%%%%%%%%%%%%%%%%%%%%%%%%%%%%%%%%
\FloatBarrier
\subsection{Event Structure and Cuts}
\FloatBarrier
In this section, preparations of the symmetrized exponential are explored with two distinct agendas.
The first agenda seeks to interpret the prepared quantum state to the highest possible accuracy.
This will involve extrapolation over multiple preparation parameters to constrain our knowledge of the structure that the quantum state was intended to have.  This is the type of procedure that would be desired if the tomography of the final wavefunction contained a desired computational result.
However, it is often the case in state preparation that it is not the extrapolated wavefunction that is of interest,
but rather the quality of the wavefunction existing in hardware in a single instantiation of the system.
This is required for time evolving the created instance to explore the system dynamics.
This second agenda does not allow for error extrapolation and instead, we propose to monitor and control its errors
dynamically by collecting temporally-correlated diagnostic information about a quantum device,
parallel to the procedures utilized for the analysis of high-energy particle or nuclear scattering data.
\begin{figure}[!ht]
\centering
  \includegraphics[width=0.32\textwidth]{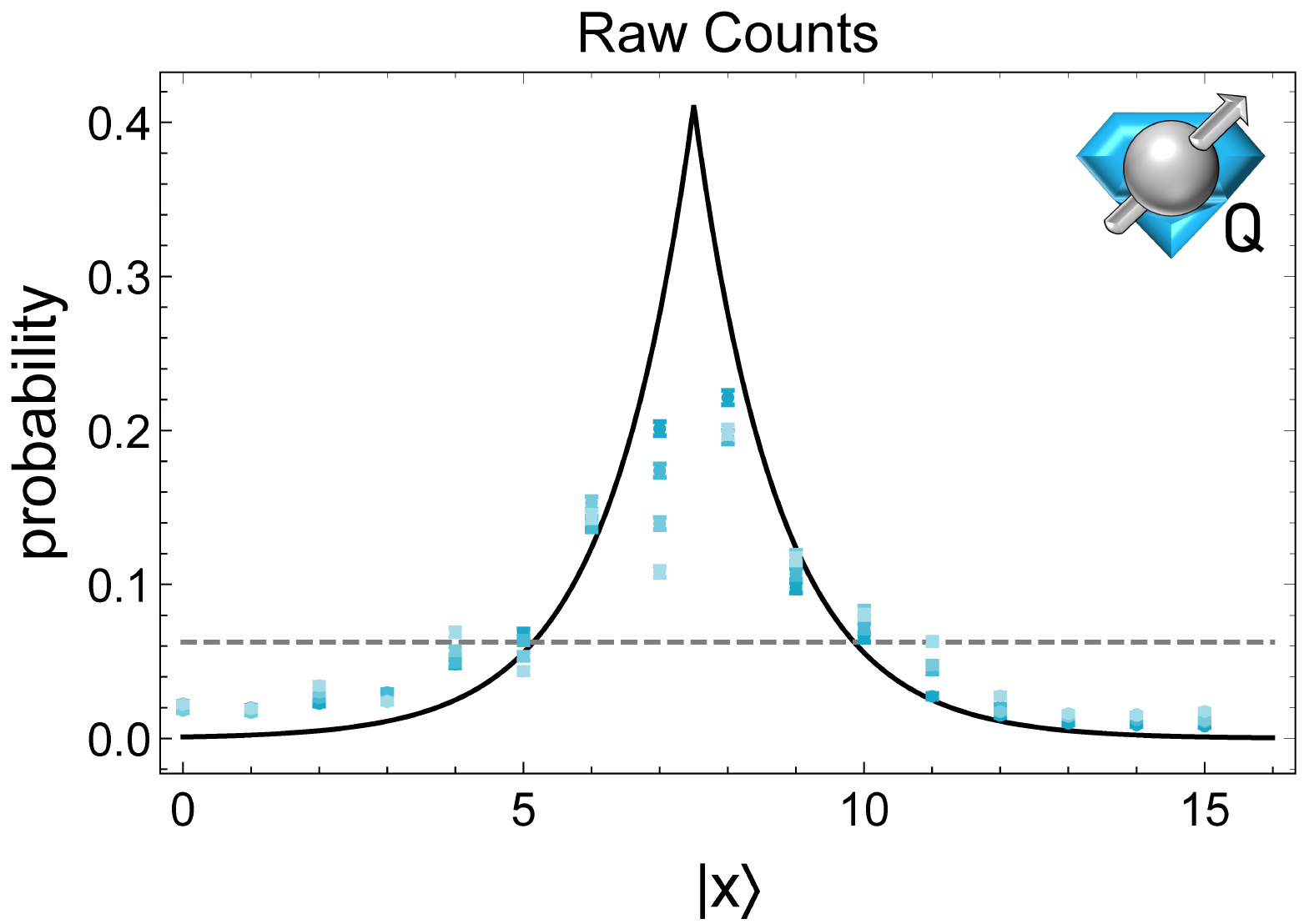}
  \includegraphics[width=0.32\textwidth]{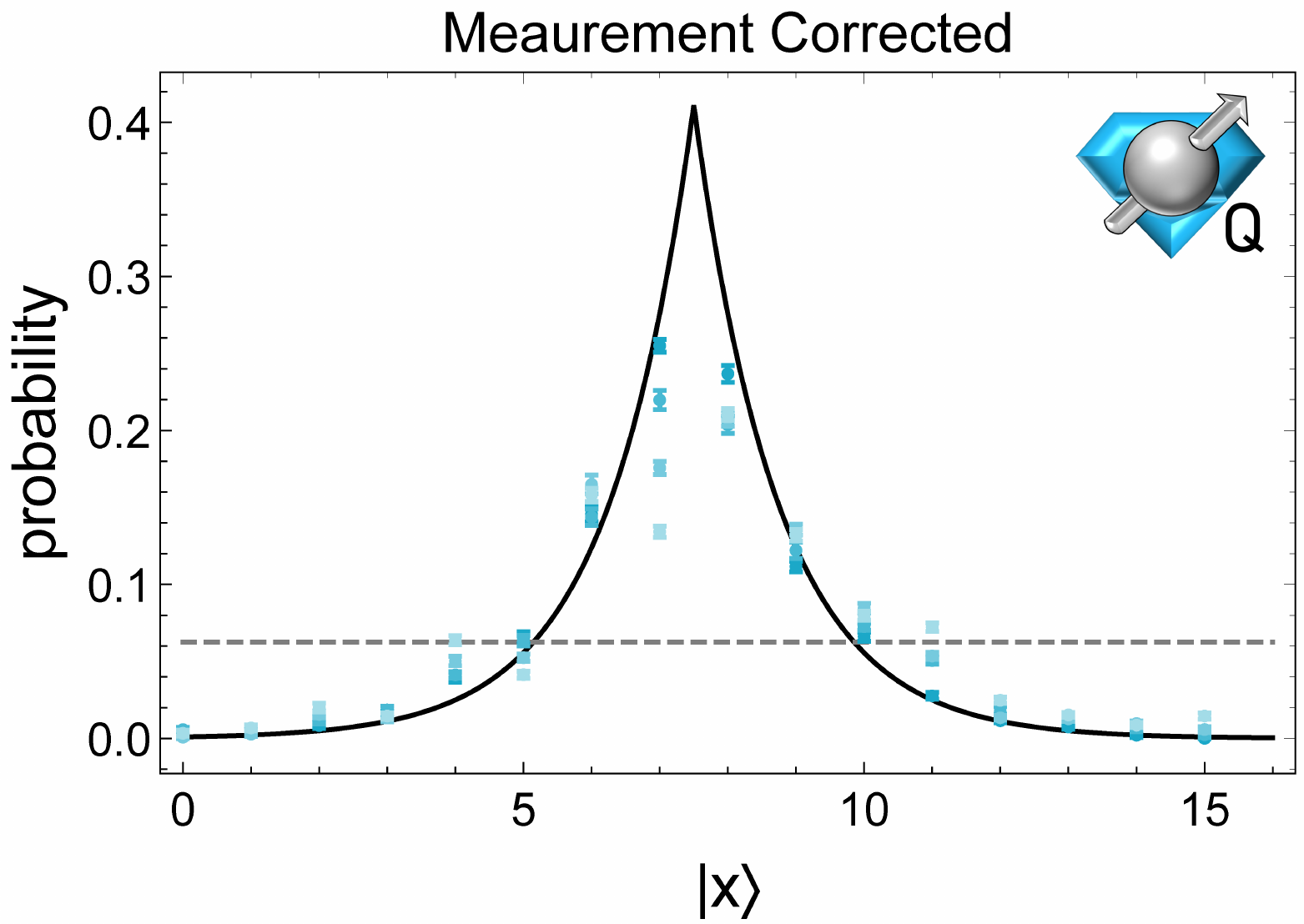}
  \includegraphics[width=0.32\textwidth]{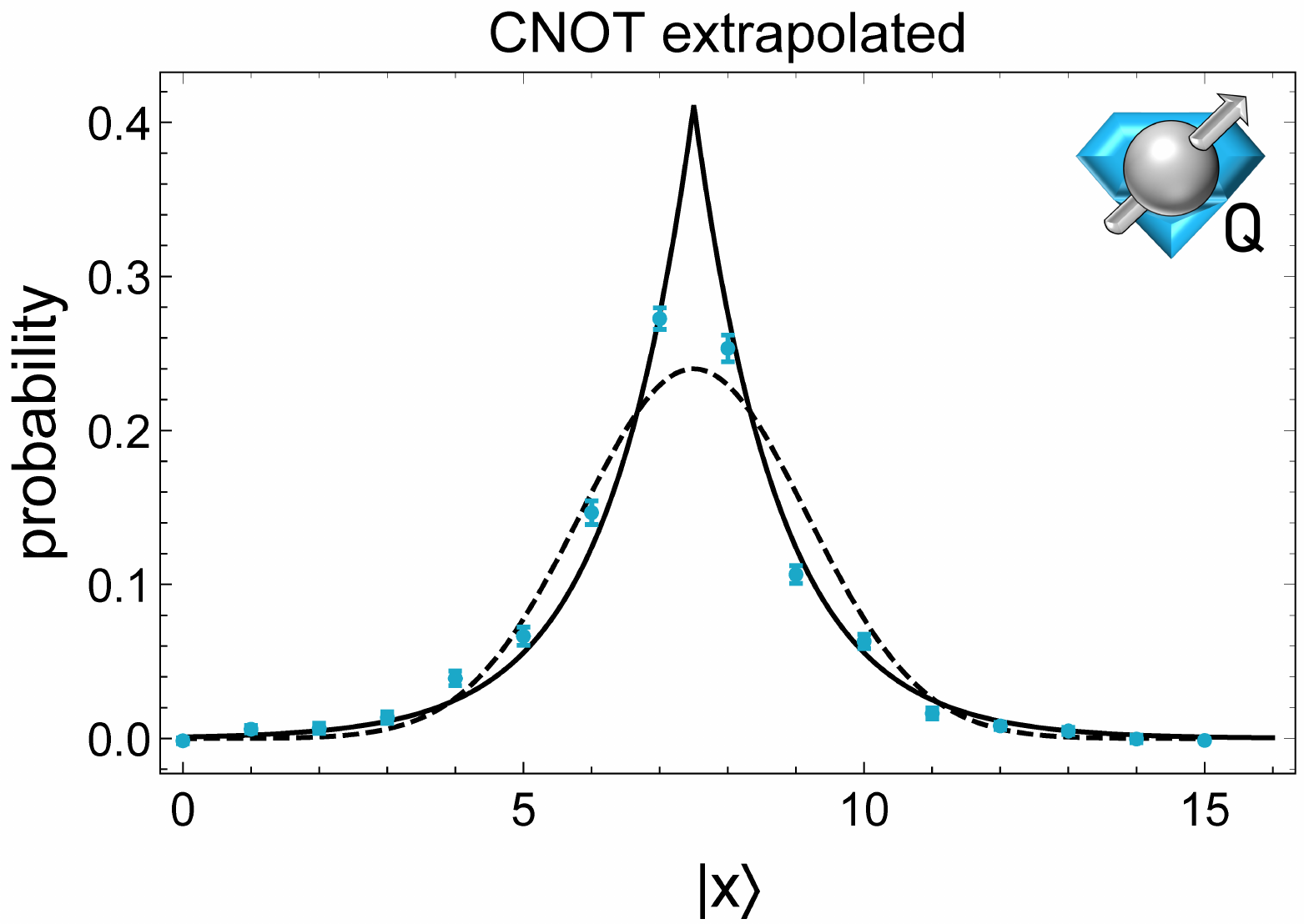}
  \caption{
  Progression of results for the symmetrized exponential wavefunction.
  The  probabilities measured directly from the quantum device (left),
  the measurement-error  corrected probabilities (middle)
  and the linearly-extrapolated probabilities (right) implemented with a Hadamard angle of $\theta = 1.52$ in the $U_3(\theta,0,\pi)$ gate.
  In the left two panels, the data points become lighter in color as $r = 1, 3, 5, 7$ with $r = 1$ the darkest of the four measurements.  The horizontal dashed line shown at $2^{-4}$ is the classical saturation level.  In addition to the theory curve for the implemented exponential at right, the dashed curve is the theoretically-closest Gaussian state when digitized onto four qubits.}
  \label{fig:CNOTextrapolation}
\end{figure}
\begin{figure}[!ht]
  \centering
  \includegraphics[height = 6cm]{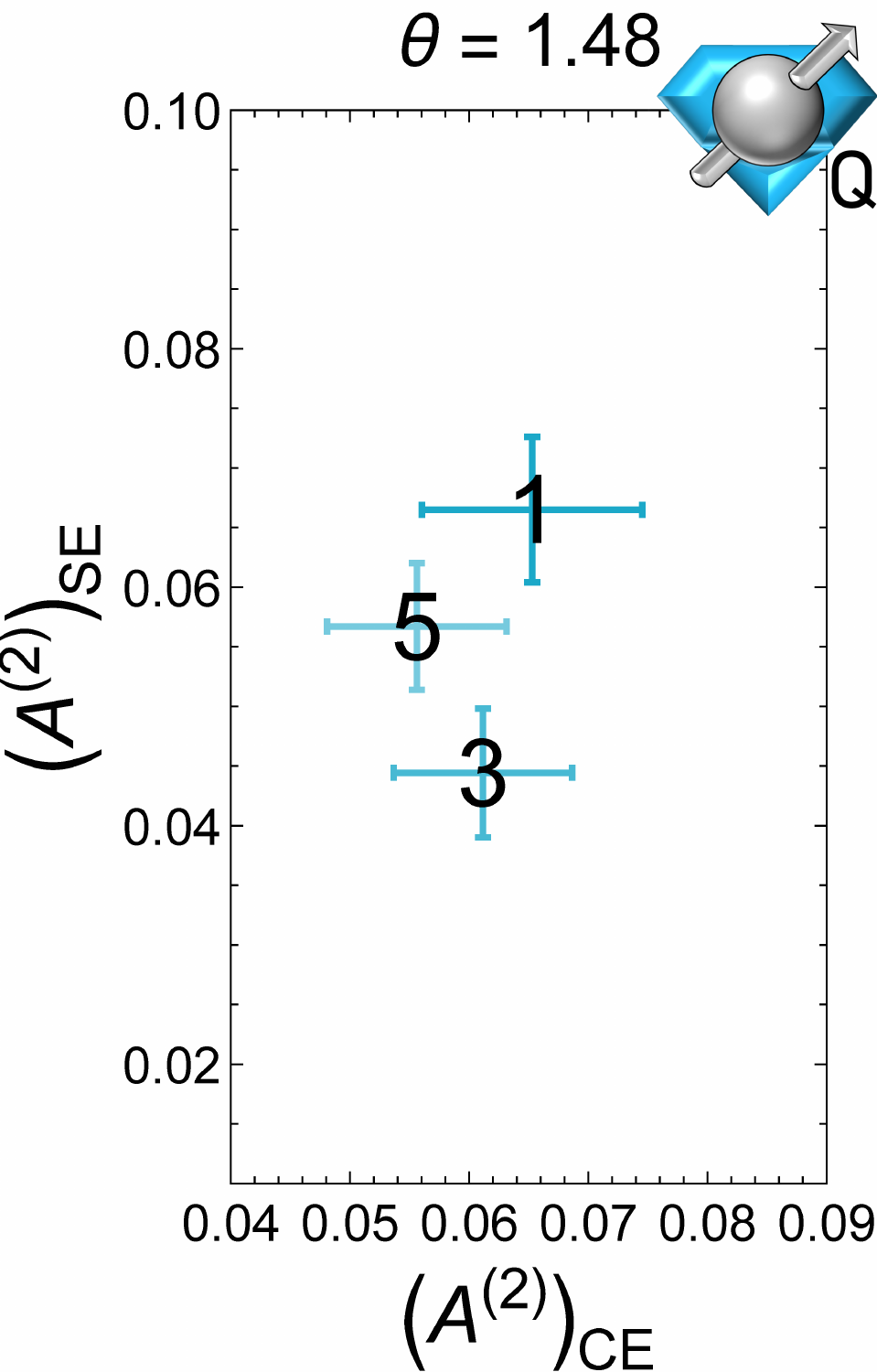}
  \includegraphics[height = 6cm]{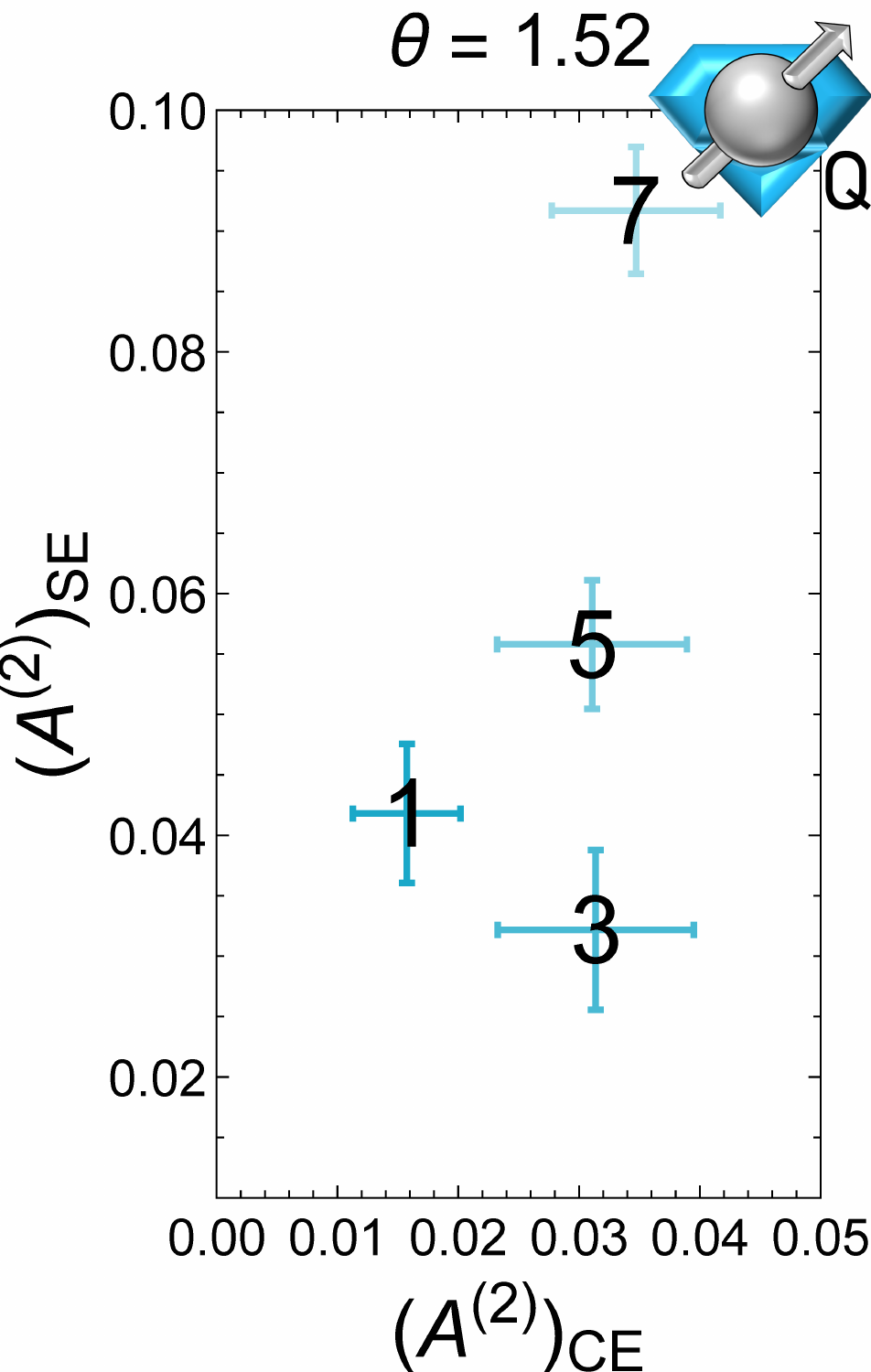}
  \includegraphics[height = 6cm]{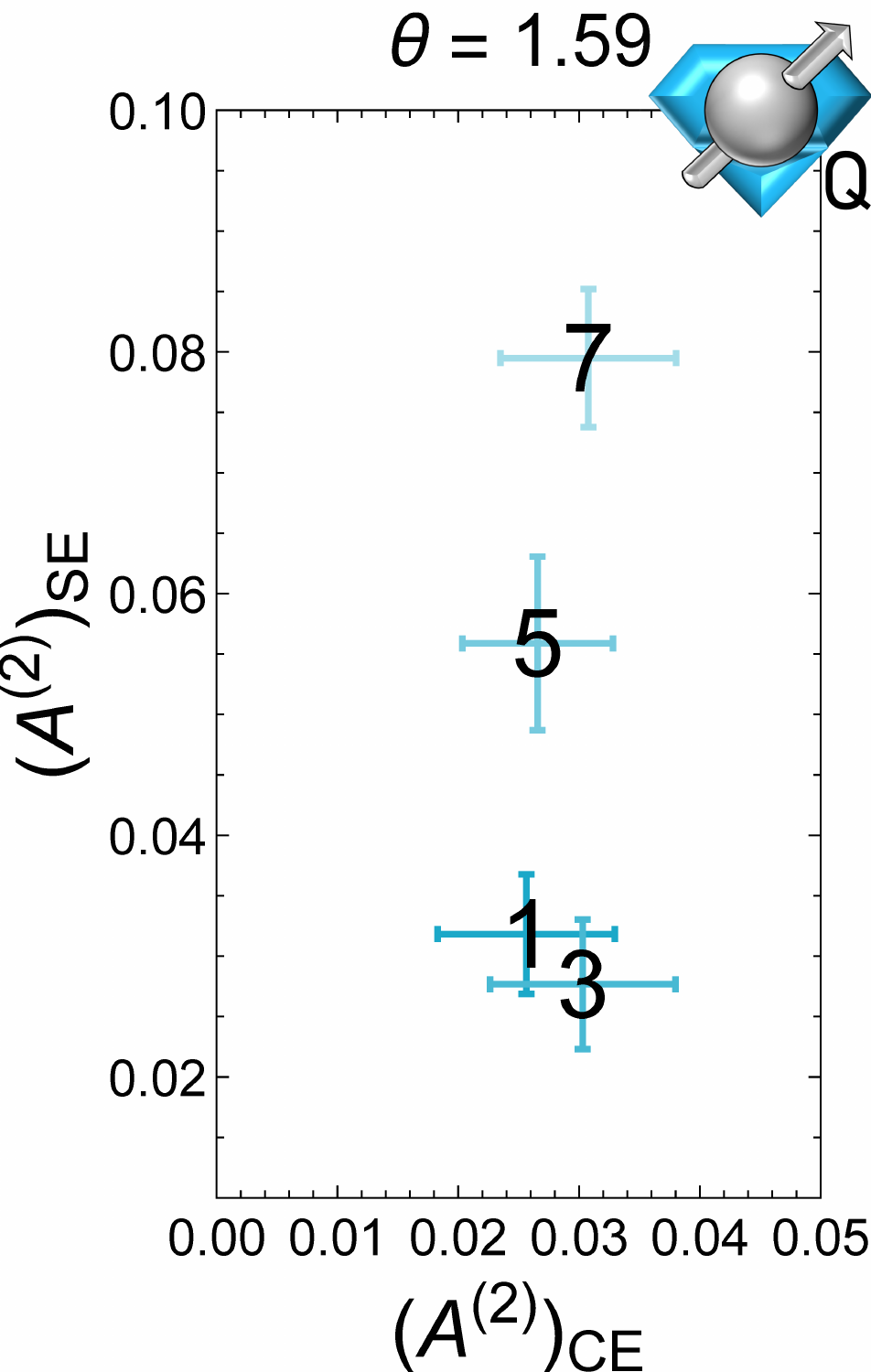}
  \caption{
  Two-norm asymmetry of the localized, symmetrized exponential wavefunction, $A^{(2)}_{SE}$,
  as a function of the time-correlated calibration asymmetry of the copied exponential, $A^{(2)}_{CE}$.
  These results are determined from CNOT-extrapolations of calculations at Hadamard $\theta$ values in the
  $U_3(\theta,0,\pi)$ gate of $\theta = 1.48, 1.52, $ and $1.59$.
  The $r$-value (number of CNOTs per CNOT) is indicated on each point.
  }
  \label{fig:CNOTAsymmetry}
\end{figure}

When performing calculations on quantum devices, the 2-qubit gates are often found to be a dominant source of systematic error
and mitigation has been explored through extrapolation of results with a range of exacerbated 2-qubit errors \cite{Kandala2017,2017PhRvL.119r0509T,PhysRevLett.120.210501,PhysRevA.98.032331,Kandala2019}.
This approach is demonstrated in Fig.~\ref{fig:CNOTextrapolation} where the $\alpha = 0.4$ exponential is prepared in hardware with a
Hadamard angle of $\theta = 1.52$.
The left panel shows, against the theoretical curve, the results as it is measured off the quantum device with error bars representing only statistical uncertainties.
Increasing values of $r$, the number of CNOTs per CNOT within the circuit of Fig.~\ref{fig:symmetrizedExpcircuit}, with $r = 1, 3, 5, 7$
become increasingly lighter in color.
The introduction of additional CNOT pairs reduces the coherence of the quantum state and tends to move the probabilities
towards their classical values depicted by the horizontal dashed line at a probability of $2^{-4}$.
Recall that the structure of this circuit places all CNOTs in the same direction with the control acting on the most significant qubit in the binary interpretation.
If the control of the CNOT operation was truly left invariant with the operation and only consulted for its value, the left half of this curve would be completely stable in the CNOT extrapolation.
As seen in the $r$-dependence of the probabilities in state $|7\rangle$, it is not currently appropriate to assume this stability---the quantum state experiences non-negligible effects in the 0-controlled portion of the Hilbert space.
The visual isolation of this feature in the symmetrized exponential state preparation makes this system a useful tool for future hardware diagnostics pertaining to the implementation of controlled operations.

In addition to the non-trivial interaction of the control, the range of CNOT results in Fig.~\ref{fig:CNOTextrapolation}
indicates that the 2-qubit gates, whose role in this state preparation is to simply reflect the binary interpretation of the
second half of the Hilbert space, also twists the wavefunction.
This twisting is seen naturally in the peak of  Fig.~\ref{fig:CNOTextrapolation} where the values in states $|7\rangle$ and $|8\rangle$ trade prominence with increasing $r$ and is quantified in Fig.~\ref{fig:CNOTAsymmetry}.
The two-norm asymmetry of the copied exponential, $A^{(2)}_{CE}$ on the $x$-axis of Fig.~\ref{fig:CNOTAsymmetry}, that is time-correlated with each of the $r$-value data sets of Fig.~\ref{fig:CNOTextrapolation} shows stability of the in-medium Hadamard.
This is equivalent to stability in the implementation of the circuit of Fig.~\ref{fig:symmetrizedExpcircuit} with the CNOTs removed.
As $r$ is increased, the two-norm asymmetry of the symmetrized exponential, $A^{(2)}_{SE}$, is seen to systematically increase.
Thus, the reduction in the symmetry at the peak between states $|7\rangle $ and $|8\rangle$ can be isolated as a result of the increased number of CNOTs and not temporal fluctuations of the in-medium Hadamard.
Had any of the calibration implementations of the copied exponential (in-medium Hadamard) yielded $A^{(2)}_{CE}$ values outside the determined regime of acceptable asymmetry, the temporally-correlated symmetrized exponential would be cut and only implemented once calibration stability is re-established.
The pair of calibration and subsequent implementation of a symmetrized exponential is termed an \emph{event},
and events are accepted or rejected based on the quality of their calibration.
In this way, qualities of a quantum state can be monitored and maintained through temporal fluctuations of the device specifications without direct measurement of the calculation wavefunction.
After the Hadamard tuning of section~\ref{sec:HadamardTuning}, no events were removed due to this threshold in the 1-site implementation.

In the second panel of Fig.~\ref{fig:CNOTextrapolation},
the measurement-error correction determined using the technique recommended by IBM has been implemented (see Appendix~\ref{app:MeasurementCorrection} for details).
The uncertainties are larger as they now include the statistical uncertainties in the calibration matrix used in the measurement-error correction inversion, informed by the implementation of $2^4$ low-depth circuits.
The measurement-error correction can modify the symmetry determined for the localized wavefunction, implying the same for the imperfect measurement.
Notice that an important role of the measurement-error correction is to reassign the background of \enquote{randomly-recorded} measurements that would otherwise obscure the observation of the strongly-suppressed tails of a localized wavefunction.
In the third panel of Fig.~\ref{fig:CNOTextrapolation}, a linear extrapolation is performed with the first two $r$ values (1,3) and uncertainties are extrapolated to the probability at $r = 0$.
In addition to the intended continuous exponential wavefunction, a Gaussian wavefunction with $\sigma = 0.29$ is shown.
This is the Gaussian that, when digitized onto four qubits, has the maximal overlap, $|\langle \psi^{g}|\psi^{e}\rangle |^2 = 0.98$, with the Gaussian at $\alpha = 0.4$.
The motivation for assessing the state's relation to the Gaussian, as mentioned previously, is in the relevance of this form for the ground state of the scalar field.
\begin{table}[!ht]
  \begin{tabular}{c|cccc}
  \hline
  \hline
  \includegraphics[width=6mm]{iconQ1_bf.png} &$A^{(1)}_{SE}$ & $A^{(2)}_{SE}$  &  $ ||\psi^2_{\rm hw} - \psi^2_{\rm exp}||_1$ & $|| \psi_{\rm hw}^2 - \psi^2_{\rm Gauss} ||_1 $\\
  \hline
  \hline
  CNOT extrapolation $\theta = 1.48$ & 0.14(2) & 0.08(1)  &  0.14(2) & 0.21(2)\\
  CNOT extrapolation $\theta = 1.52$ & 0.10(2) & 0.053(9)  & 0.12(2) & 0.21(2)\\
  CNOT extrapolation $\theta = 1.59$ & 0.08(2) & 0.040(7)  & 0.16(2) & 0.18(2) \\
  Calibration Window & 0.10(3) & 0.06(2)  & 0.14(3) & 0.16(3) \\
  \hline
  \hline
  \end{tabular}
  \caption{
  Properties of the symmetric exponential wavefunction with $\alpha = 0.4$ prepared on IBM's {\tt Poughkeepsie}.
  The first and second column quantify the asymmetry of the wavefunction while the third and fourth quantify its overlap (with assumptions on the wavefunction's distribution of phases) with the intended exponential and the associated theoretically-nearest
  Gaussian state when digitized onto four qubits. }
  \label{tab:expstats}
\end{table}
As can be seen visually, the state prepared in the quantum device is an approximation to the Gaussian wavefunction.
The third and fourth columns of Table~\ref{tab:expstats} calculate the one-norm deviation of the measured probability distributions with those associated with the exponential and Gaussian wavefunctons.
If one makes the non-trivial assumption that the hardware wavefunction was composed of real, positive amplitudes as intended with the circuit of Fig.~\ref{fig:symmetrizedExpcircuit},
the norm-squared overlap of the states prepared in hardware to the exponential and Gaussian wavefunctions would be calculated to be $\gtrsim 0.97$ for each system represented in Table~\ref{tab:expstats}.
Of course, simply extending the observations of the EGA function defined in Sec.~\ref{sec:HadamardTuning}
to the second and third angles of the $U_3$ gate would indicate that this assumption is invalid at the level of fractions of radians per gate.
These two quantifications of the $Z$-basis distribution of probability in the localized wavefunction are expected to correlate with the projection onto eigenstates of this form in an implementation of quantum phase estimation.  For four qubits, it is seen that a noisy implementation of a localized, symmetrized exponential is a viable option for approximating a Gaussian wavefunction with reduced quantum resources.

\begin{figure}
  \includegraphics[height = 5cm]{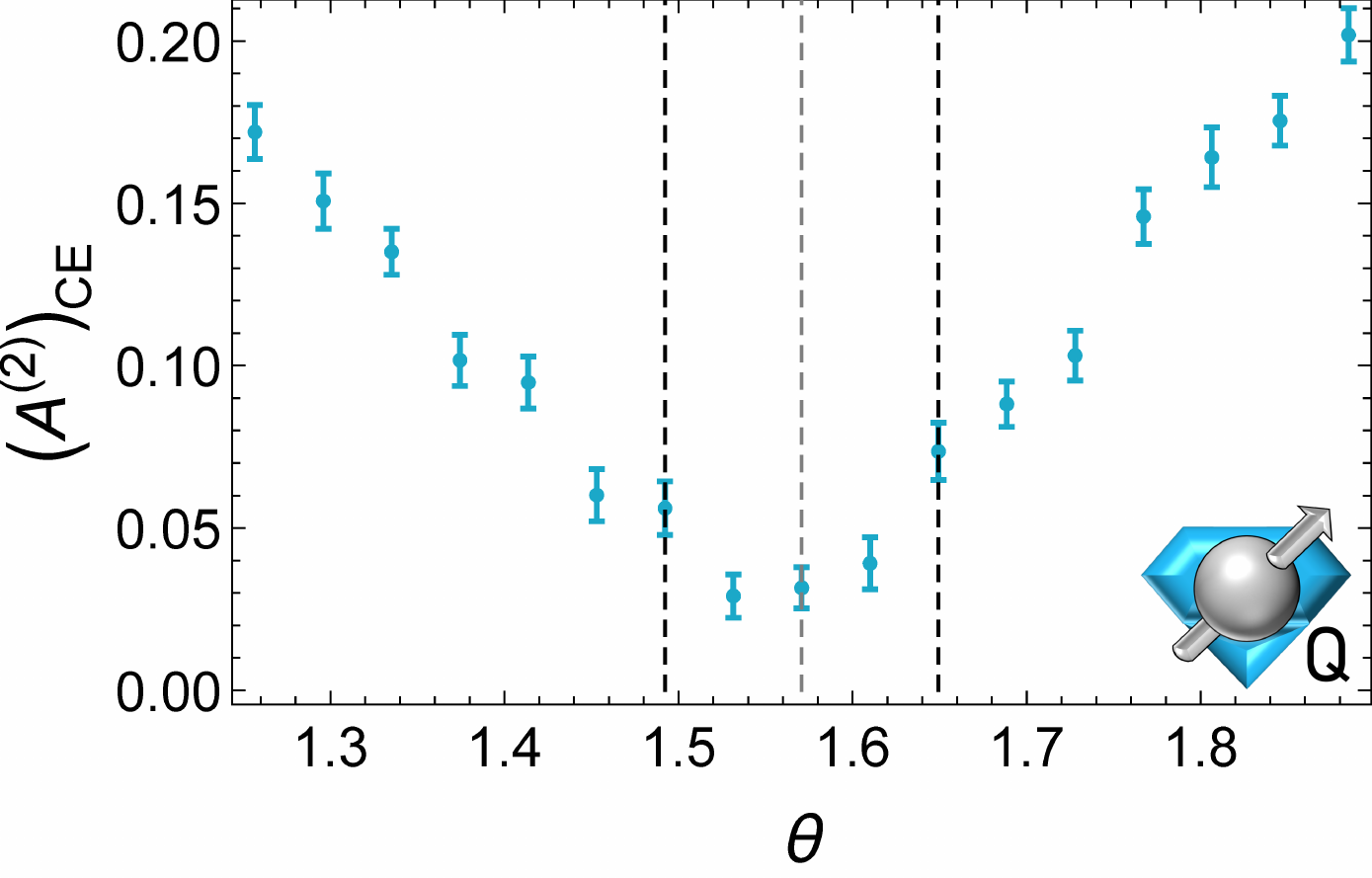}
  \includegraphics[height = 5.1cm]{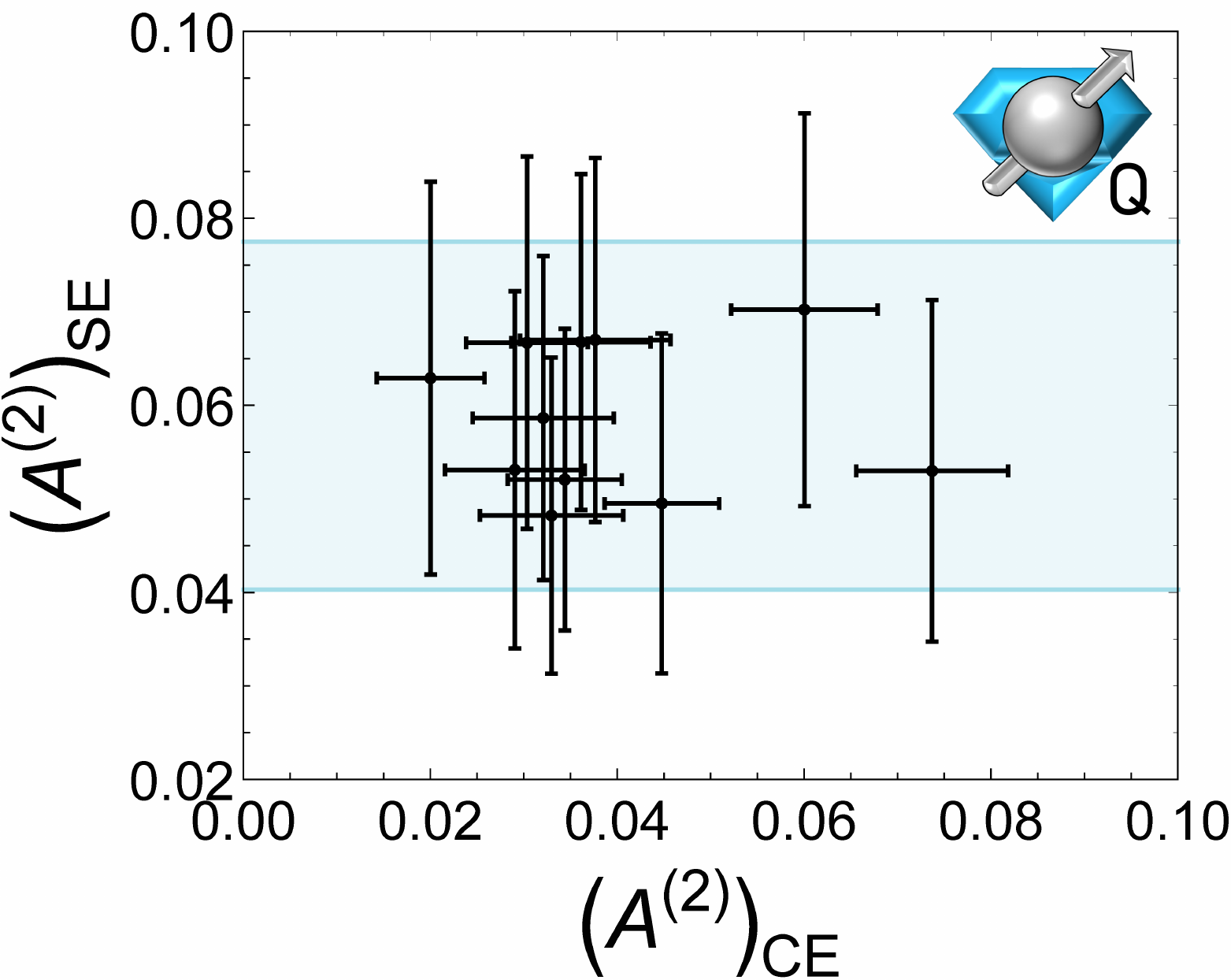}
  \includegraphics[height = 6cm]{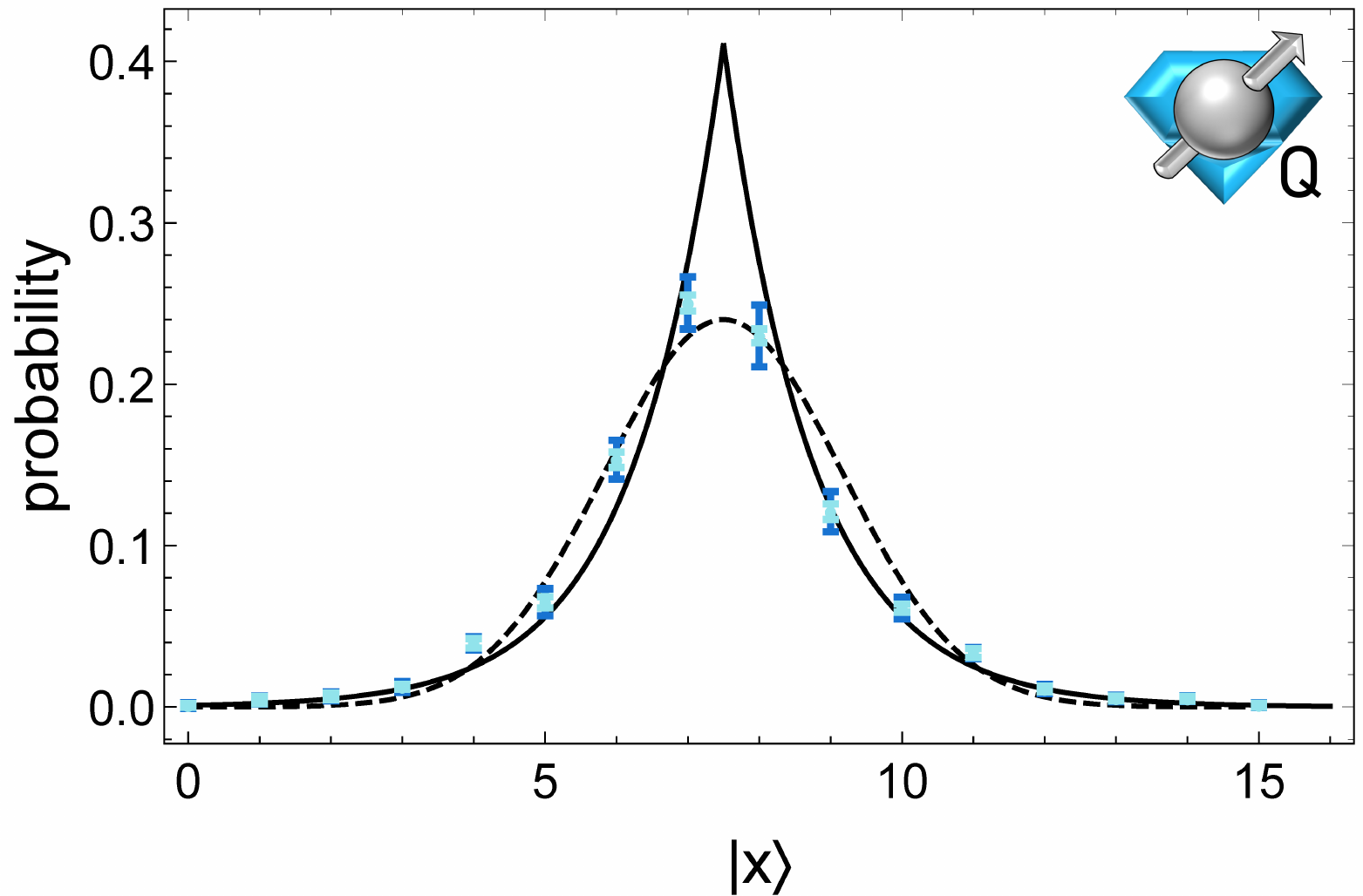}
  \caption{(Color Online) Top left: Asymmetry calculated for the in-medium Hadamard copying the exponential distribution to the second half of the Hilbert space.  Top right: Asymmetry of the localized, symmetrized exponential distribution, $A_{SE}$,  as a function of the time-correlated calibration asymmetry for $\theta$ between the vertical dashed lines at left.  Bottom: Representative probability distribution after preparation of an exponential wavefunction.  Light blue error bars represent statistical uncertainties including that from the measurement-error
  correction.  Dark blue error bars represent the full uncertainty including systematics of the in-medium Hadamard.  Black curves are theoretical probability distributions for an exponential(solid) with $\alpha = 0.4$ and Gaussian(dashed) with $\sigma = 0.29$.
  }
  \label{fig:HadamardWindowExp}
\end{figure}
To explore the properties of the wavefunction that can be prepared in the device rather than the quality of the extrapolation that can be learned over a range of wavefunctions, we place a cut on the Hadamard angle used to copy the exponential into the second half of the Hilbert space as seen in the circuit of Fig.~\ref{fig:symmetrizedExpcircuit}.  This cut is directly related to a cut on the asymmetry allowed in the in-medium Hadamard as shown in the top left panel of Fig.~\ref{fig:HadamardWindowExp}
where the two-norm asymmetry, $A^{(2)}_{CE}$, is shown as a function of the input angle.
The vertical lines indicate the region of $\theta$ allowed for accepted \emph{events}.  The two-norm is monitored between each set of 8000 samples and shown in the $x$-axis of the top right panel of Fig.~\ref{fig:HadamardWindowExp} for the 11 events measured in this work.  Depending on the desired level of stringency placed on the calibration's asymmetry measurement, a few events with $A^{(2)}_{CE}$ 1-$\sigma$ higher than the rest could be removed for concern about fluctuations in the device properties.  Due to the previous observation that asymmetries at the level $~0.05$ can be extrapolated to a good approximation of the exponential distribution, we choose a cut window on the calibration asymmetry of $A^{(2)}_{CE} = 0.1$, thus retaining all 11 \emph{events} recorded.  This retention is a testament to the stability of the four chosen qubits during the time period of this production ($\sim 3$h) and is not expected to be maintained in the use of different qubits or as calculation times grow with circuit depth and qubit number.

Using \emph{events} produced with Hadamard angles in the window shown in the top left panel of Fig.~\ref{fig:HadamardWindowExp} with on-line monitoring of the in-medium Hadamard's asymmetry, the probability distribution in the lower panel of Fig.~\ref{fig:HadamardWindowExp} is one representative distribution determined to have been prepared within the quantum device.
The light-blue error bars indicate only statistical errors from the collection of 8000 samples in both the production and calibrations including propagation through the measurement-error correction procedure.
The dark blue error bars include also systematic uncertainties on the probabilities in each state informed by variation in the calibration measurements of the copied exponential produced within the Hadamard angle window and surviving the asymmetry cut.
The systematics include time fluctuations in the device as well as the EGA distribution of the in-medium Hadamard as discussed in Section~\ref{sec:HadamardTuning}.
Looking to Table~\ref{tab:expstats}, it can be seen that this single-circuit calibration window works quite well with comparable deviations of the probability distribution from the exponential and Gaussian state to those of the CNOT-extrapolated distributions.
This success is largely due to the quality of 1- and 2-qubit gates on the set of qubits chosen for the collection of this data, which can be seen visually in the middle panel of Fig.~\ref{fig:CNOTextrapolation} where the measurement-error corrected, $r = 1$ probability distribution is already quite representative of the intended exponential.
It would appear that the preparation of high-quality states within calibration windows will be a vital capability
before evolving wavefunctions to study quantum
dynamics~\footnote{
While the fidelities of {\tt Poughkeepsie} quantum gates allow for the initialization of an approximate
wavefunction for the ground state of the scalar field, the connectivity of the device (see Appendix~\ref{app:Connectivity})
does not allow for subsequent time evolution.
As shown in Ref.~\cite{Klco:2018zqz}, all-to-all
connectivity within the four qubits digitizing the field space of the single site would be required to
implement the symmetric QuFoTr as well as the phases in position and momentum space.
}.

Before going further, it is interesting to realize why this procedure of preparing an exponential wavefunction rather than the desired Gaussian works so well in this application and where the advantage ceases.
The momentum-space distribution of the exponential wavefunction is significantly larger than that of the Gaussian.
Specifically, the cusp in the center of the space provides a large derivative contribution.
The Nyquist-Shannon (NS) sampling theorem (applied in this context in Refs.~\cite{Klco:2018zqz,PhysRevLett.121.110504}) indicates that to capture more faithfully the behaviour of the cusp, the wavefunction needs to be sampled at a higher rate in field space---specifically at twice the rate of the highest frequency in fourier space.
If the wavefunction is sampled at a lower rate than dictated by the NS saturation point, the continuous reconstruction implied by the digitized samples is not unique.
We leverage this lack of uniqueness by acknowledging the existence of a family of functions distinct in their ultraviolet (UV) completions, but equivalent in their IR digitizations up to the accuracy capable of the noisy quantum device.
By shifting the samples away from the peak of the exponential, the UV cutoff on the exponential is lowered.
When digitizing with more qubits, the UV cutoff will necessarily increase as the exponential is sampled more densely.
While the overlap with the Gaussian remains high---e.g., digitization with $n_Q = 12$ yields $|\langle \psi^e | \psi^g\rangle|^2 \gtrsim 0.97$ with small $n_Q$-dependent tuning of the width parameter---the exponential wavefunction's precision on the ground state energy of the free scalar field is found to be saturated at $\sim 1\%$  achieved at the low digitization of $n_Q = 3$.
If a higher precision is desired, the $n_Q$-stable, substantial overlap of the symmetrized exponential with the ground state of the free scalar field makes it a convenient starting point for more advanced preparation techniques~\cite{NielsenChuang,1995quant.ph.11026K,PhysRevLett.83.5162,2000quant.ph..1106F,2016PhRvL.117a0503W,Kaplan:2017ccd,2019arXiv190107653M}
in which the necessary resources often scale with the initial state's overlap with the target wavefunction.
To put the value of this achievable precision into perspective, the explorations of Ref.~\cite{Klco:2018zqz} determined that field-space wavefunctions digitized with $n_Q = 3$ were capable of achieving $~0.01\%$ precision on the ground state energy when utilizing the digitized Hamiltonian's exact ground state wavefunction.
By acknowledging that the wavefunction does not need to be implemented to higher accuracy than the noisy quantum device is capable of preparing, it is possible to hide controlled theoretical systematics beneath the hardware noise floor---preparing a wavefunction with equivalent accuracy in hardware but requiring significantly fewer quantum resources, which scale only linearly with the number of qubits for the exponential wavefunction.
The localized exponential-Gaussian pair is one example of a successful entanglement-minimized, digital approximation.

%%%%%%%%%%%%%%%%%%%%%%%%%%%%%%%%%%%%%%%%%%%%%
\FloatBarrier
\subsection{Multi-Site Preparation}

\begin{figure}[!ht]
  \includegraphics[height = 0.5 \textwidth]{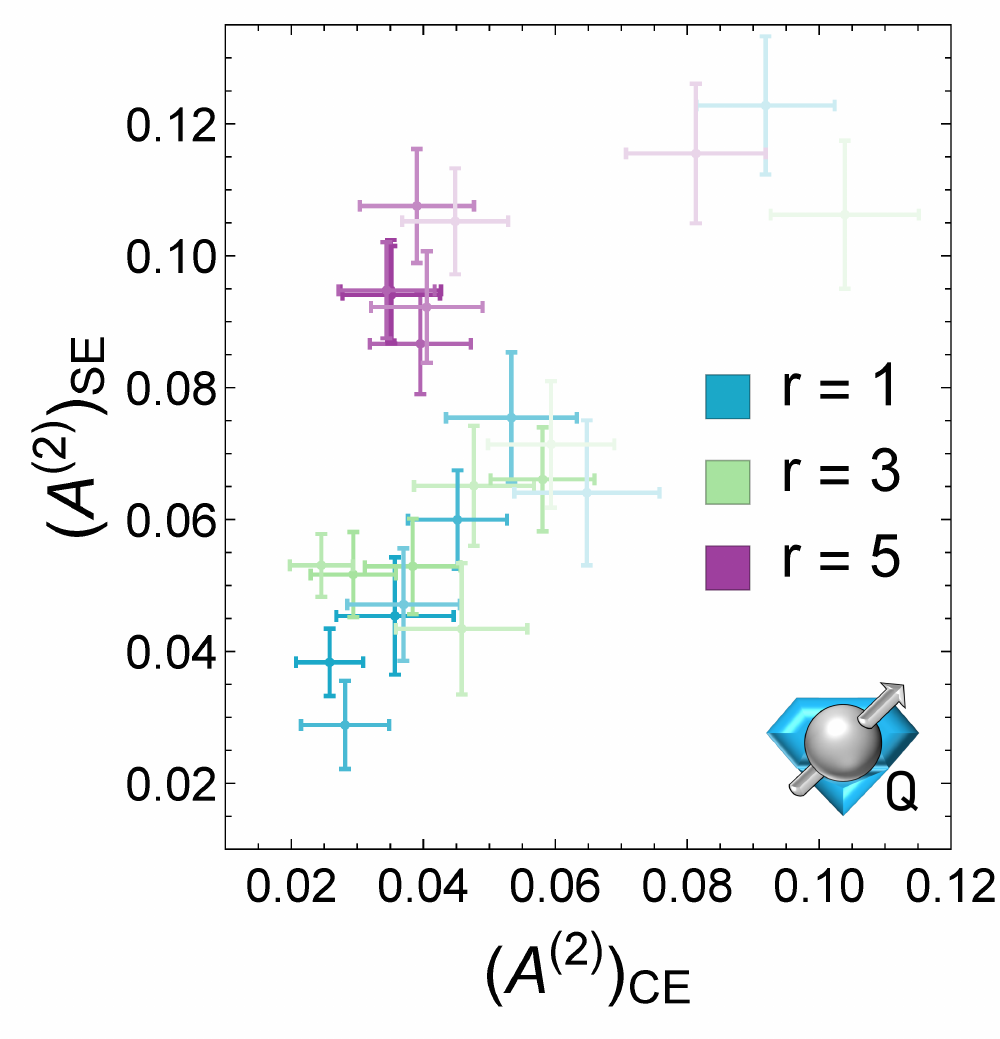}
  \hspace{1cm}
  \begin{overpic}[height = 0.5\textwidth]{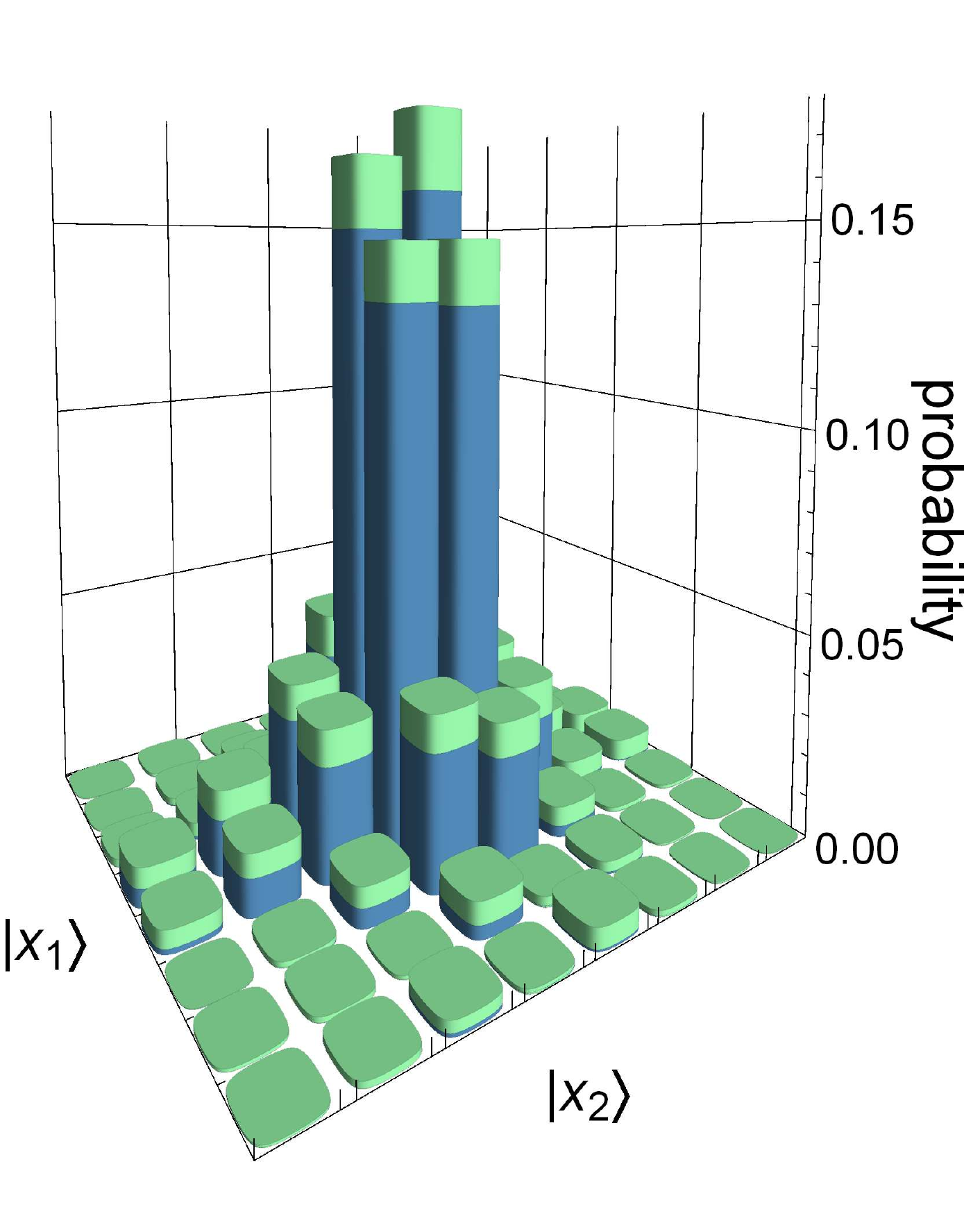}
     \put(5,82){\includegraphics[height=0.07\textwidth]{iconQ1_bf.png}}
  \end{overpic}
  \\
  \includegraphics[width = 0.8\textwidth]{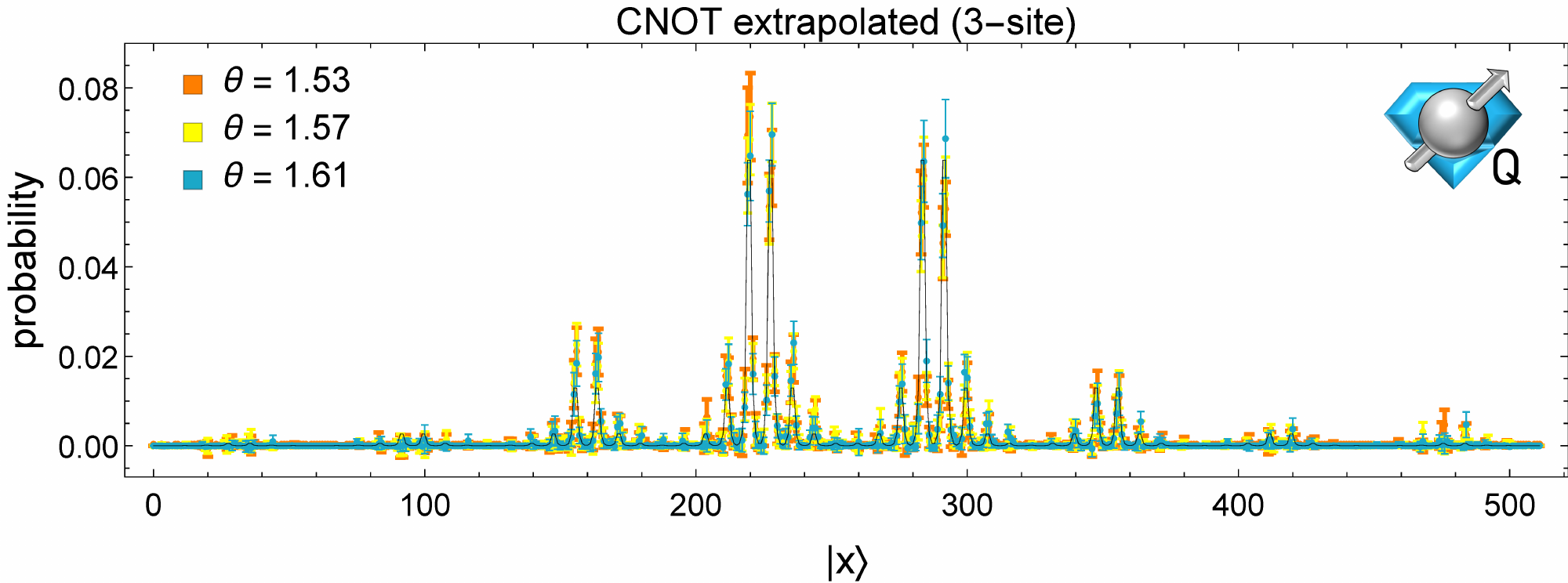}
  \caption{(Color Online)
  The 2-site and 3-site probability distributions of a symmetric exponential wavefunction initialized on each spatial site.
  The top left panel shows the two-norm asymmetry of the 2-site symmetrized exponential, $A^{(2)}_{SE}$, as a function of the angle-correlated calibration two-norm asymmetry of the copied exponential, $A^{(2)}_{CE}$.
  The colors indicate the CNOT extrapolation level while their saturation decreases with increasing distance of the $U_3( \theta, 0, \pi)$ input angle from $\pi/2$ for the symmetrizing Hadamard.
  The top right panel shows the CNOT-extrapolated probability distribution for $\theta = 1.57$ across the full 64-dimensional Hilbert space of the $n_Q = 3$ symmetrized exponential prepared over two sites and arranged into a grid.  The light green regions span the 1-$\sigma$ confidence intervals on the probability at each grid location.
  In the lower panel, three sites of non-interacting symmetrized exponential are prepared and extrapolated with data at $r = 1, 3, 5$ at three different values of the $U_3(\theta,0,\pi)$ symmetrization angle within the tolerance cuts.
  }
  \label{fig:2site}
\end{figure}
As a first step toward initializing a lattice scalar field theory into its ground state, we extend the work in the previous sections to
initialize a 2- and 3-site system $(n_s = 2, 3)$ into a tensor product state of symmetric exponentials,
$\langle x_1 x_2|\Psi\rangle = \psi_{\rm exp}(x_1)\ \psi_{\rm exp}(x_2) $ and
$\langle x_1 x_2 x_3|\Psi\rangle = \psi_{\rm exp}(x_1)\ \psi_{\rm exp}(x_2)\ \psi_{\rm exp}(x_3) $.
This state is neither the true ground state for each site, nor the true ground state of the 2- or 3-site systems
due to the lack of entanglement.
We have explored the preparations of these systems in a number of different ways: with Gaussian gate noise, with the {\tt Qiskit} noisy quantum simulator using the {\tt Poughkeepsie} noise model parameters, and directly with the {\tt Poughkeepsie} quantum device.
Both the 2- and 3-site systems were implemented with $n_Q = 3$ qubits per spatial site, corresponding to 64- and 512-dimensional Hilbert spaces.
The reduced probability distributions for each site are expected to correspond to symmetric exponentials,
while the distribution across the entire Hilbert space corresponds to the convolution of two or three
symmetric exponentials.
As the initialized wavefunctions are the tensor product of symmetric exponentials, without entanglement, the circuits for preparing these states correspond to parallel implementations of the gates necessary to prepare the state on a single site.

Figure~\ref{fig:2site} shows the results of preparations implemented on the quantum device {\tt Poughkeepsie}, corrected for measurement errors.
The top left panel of this figure shows the two-norm asymmetry in the 2-site symmetrized exponential, $A^{(2)}_{SE}$, correlated with the same quantity calculated for the copied exponential calibration, $A^{(2)}_{CE}$.  The Hadamard $\theta$-angles represented in this panel were scanned over a range of 1.41 to 1.73, spanning the theoretical value of $\pi/2$.
For the blue points at $r = 1$, the two asymmetries are seen to be highly correlated with an asymmetry that generically increases in both the calibration and production as the Hadamard angle in the $U_3(\theta,0,\pi)$ operator deviates from $\pi/2$.
This angular distance from $\pi/2$ is represented in the saturation of the data points such that $\theta = 1.41$ is lightly shaded appearing in the top right corner while the $r = 1$ point for $\theta = 1.57$ is darkly shaded and appears in the lower left.
This direct correlation does not persist for the purple points at $r = 5$.
Instead, a vertical clustering is observed located at a large value of $A_{SE}^{(2)}$ indicating, as seen in the 1-site data,
that the additional CNOTs produce an asymmetry that overshadows that from the  Hadamard gate.
In this case, all angles paired with $r = 5$ were unable to achieve symmetrized exponentials with two-norm asymmetries below $A^{(2)}_{SE} = 0.08$, in spite of the stability of their angle-correlated calibration asymmetries, $A{(2)}_{CE}$.

In the top right panel of Fig.~\ref{fig:2site}, the probability distribution for the 2-site exponential preparation is shown after
measurement-error correction and CNOT extrapolation with $r = 1, 3$ for a Hadamard implemented with $U_3\left(\pi/2, 0, \pi\right)$.
The presentation of probabilities obtained from 2-site systems has been discussed previously,
e.g. Ref.~\cite{Klco:2018zqz}, where it is seen to be useful to decompose the system
into the probabilities at each site and represent the data in a 2D grid.
When viewed in this way, the four-fold discrete rotational symmetry is apparent in the data, though imperfections are discernable outside the light-green bands representing the 1-$\sigma$ confidence interval on the probability at each state in the Hilbert space.

Finally, the bottom panel of Fig.~\ref{fig:2site} shows the linearized Hilbert space representation of the 3-site symmetrized exponential.
For three different Hadamard angles within a window defined about $\theta = \pi/2$, the probability distributions are shown after constrained-inversion measurement-error correction and CNOT extrapolation using values of $r = 1, 3, 5$.
The measurement calibration matrix created after implementation of the 512 circuits initializing computational basis states
(with 400 shots each)  can be seen in Fig.~\ref{fig:PouCalibrationMatrices} of Appendix~\ref{app:MeasurementCorrection}.
With 8000 shots per angle and  a signal maximally rising to a probability of 0.08, it is encouraging that these preparations are capable of conclusively identifying the central and neighboring peak structures of the 3-site symmetrized exponential wavefunction.

By adding additional sites each with $n_Q$ qubits, increasing the dimensionality of the system from 8 for 1 site to 64 for 2 sites to 512 for 3 sites, the statistical precision of the measurements is reduced for a fixed number of shots.
For the chosen width of the exponential, where most of the measurements are localized within a few states,
the precision is reduced by approximately a factor of two, as opposed to the ratio of Hilbert-space dimensionalities.
More generally, the precision is determined by the number of states where the wavefunction has support, rather than the dimensionality of the full Hilbert space.
Relying on the localization of the wavefunction to arrive at statistically significant results
may be important when working with larger lattices of scalar fields.

\begin{table}[!ht]
  \begin{tabular}{c|cccc}
  \hline
  \hline
   two site& $A^{(1)}_{SE}$ & $A^{(2)}_{SE}$ &  $ ||\psi^2_{\rm hw} - \psi^2_{\rm exp}||_1$ & $|| \psi_{\rm hw}^2 - \psi^2_{\rm Gauss} ||_1 $ \\
  \hline
  \hline
  Gaussian Noise $\sigma_\theta = 0.3 \times 10^{-1}$ \ \includegraphics[width=6mm]{iconC2N_bf.png} & 0.059(8) & 0.014(2) &  0.78(1)  &  0.79(1)  \\
   Gaussian Noise $\sigma_\theta = 10^{-2}$ \ \includegraphics[width=6mm]{iconC2N_bf.png} & 0.041(7) &  0.013(3)   &  0.14(1)  &  0.18(1)  \\
    {\tt Poughkeepsie} Noise Model \ \includegraphics[width=6mm]{iconC2N_bf.png} & 0.07(1) &  0.019(3) & 0.10(1)  &  0.16(1)   \\
    Calibration Window $A^{(2)}_{CE}\leq 0.4$\ \includegraphics[width=6mm]{iconQ1_bf.png} & 0.13(3) &  0.04(1)   &  0.21(4)  &  0.18(4)   \\
    CNOT extrapolation $\theta = 1.57$\ \includegraphics[width=6mm]{iconQ1_bf.png} & 0.15(2) &  0.05(1)   &  0.18(3)  &  0.21(3)   \\
  \hline
  \hline
  three site &  & &  & \\
  \hline
  \hline
  CNOT extrapolation $\theta = 1.53$ \ \includegraphics[width=6mm]{iconQ1_bf.png} & 0.29(2) & 0.053(6) & 0.30(4) & 0.32(4) \\
  CNOT extrapolation $\theta = 1.57$ \ \includegraphics[width=6mm]{iconQ1_bf.png} & 0.30(2) & 0.053(7)& 0.31(4) & 0.34(4) \\
  CNOT extrapolation $\theta = 1.61 $ \ \includegraphics[width=6mm]{iconQ1_bf.png} & 0.29(2) & 0.049(7) & 0.30(4) & 0.31(4) \\
  \hline
  \hline
  \end{tabular}
  \caption{
  (Color online) Properties of localized exponentials on two and three sites prepared with Gaussian noise, a quantum noise simulator with {\tt Poughkeepsie}'s noise parameters, and with the quantum device {\tt Poughkeepsie}. The first and second column quantify the asymmetry of the wavefunction while the third and fourth quantify the measured probability distribution's distance from the intended exponential and the associated theoretically-nearest Gaussian state. }
\label{tab:twositeExpGauss}
\end{table}
As in the 1-site ($n_s = 1$) system, it is interesting to quantify the asymmetry of the measured probability distributions as well as their distance from both the intended symmetric exponential probability distribution and that of the nearest tensor product of Gaussian wavefunctions.
These quantities calculated  \ 1.) with Gaussian gate noise, \ 2.) on the quantum noise simulator with {\tt Poughkeepsie}'s noise parameters extrapolated to the limit of vanishing CNOT noise,  and \ 3.) on the {\tt Poughkeepsie} hardware are shown in Table~\ref{tab:twositeExpGauss}.
Focusing on the 2-site implementation first, it is interesting to observe that none of the three classical noisy simulations (\includegraphics[width=5mm]{iconC2N_bf.png}) effectively capture the level of asymmetry measured on the quantum device.
Increasing the Gaussian gate noise,
as implemented in Ref.~\cite{Klco:2018zqz},
to $\sigma_\theta = 0.03$ dramatically increases the deviation of the probability distribution from the localized exponential and
Gaussian,
as seen in the third and fourth columns, but produces a wavefunction that retains a level of symmetry greater than that of the hardware implementations.
The same is true for the {\tt Qiskit} noise simulator defined by {\tt Poughkeepsie} noise parameters.
Once again, the preparation of the exponential wavefunction allows an isolated environment for exploring the capabilities of the quantum
device, this time suggesting
that the classical simulations do not include all of the sources of asymmetry present in the quantum hardware.
As noted previously noted, the $n_s(n_Q-1)$ CNOT gates are  identified as a source of additional asymmetry.
Generic improvements to classical (\includegraphics[width=5mm]{iconC2N_bf.png}) and quantum (\includegraphics[width=5mm]{iconQ1_bf.png}) simulations will benefit from understanding, incorporating, and mitigating these identified effects,
which are seen to be exacerbated with an increasing number of spatial sites prepared in parallel.

The second two columns of Table~\ref{tab:twositeExpGauss} indicate the distance of the measured probability
distributions from the theoretical symmetrized exponential and Gaussian distributions.
The chosen metric is the one norm of the state-by-state deviations.
It can be seen from the 3-site CNOT extrapolations that the probability distribution's deviation from the intended exponential is fairly stable over a range of $\theta$ angles in the $U_3(\theta, 0, \pi)$ gate used to symmetrize the exponential.
The presence of this stability is not surprising in that a noisy quantum environment will be somewhat insensitive
to \emph{small} modifications in the digital circuitry.
This exploration has quantified \emph{small} in this context for currently-available superconducting devices giving effectively a resolution scale for implemented angles as in Eq.~\eqref{eq:EGAparameters}.

While this gate resolution scale limits the precision of prepared wavefunctions, it also provides the logic of a calibration window used in this work to monitor system performance in real time. Preparations passing the cuts are considered identical instances of a noisy preparation.  This allows error mitigation to effectively be applied to the wavefunction prepared in the quantum device without post-process extrapolation.
This, of course, will be necessary if time evolution of the prepared state is subsequently desired.
As performed for the 2-site system, the window was set to allow asymmetries of the 2-site copied exponential $A_{CE}^{(2)} \leq 0.04$ for time-correlated calibrations.
As seen for the $r = 1$ results in the top left panel of Fig.~\ref{fig:2site}, this cut retains four prepared distributions at angles (1.57, 1.57, 1.61, 1.64), a set located near, but not centered upon the theoretically exact angle of $\theta = 1.57$ (implementations with angles smaller than 1.57 are also present in Fig.~\ref{fig:2site} but do not pass the calibration asymmetry cut).
The resulting wavefunctions may be combined to represent the average wavefunction the device is capable of preparing in hardware.
The properties of the resulting probability distribution are comparable to those extrapolated from multiple preparations at increasing $r$ values as shown in Table~\ref{tab:twositeExpGauss}.  This on-line method of calibration monitoring is expected specifically to aid in simulations of time evolution in which control of the state in hardware, rather than the extrapolated state, is vital.

The last column of Table~\ref{tab:twositeExpGauss} calculates the deviation of the various probabilty distributions from the theoretically-nearest tensor-product Gaussian distributions ($\sigma = 0.31$).
This state is also not the true ground state for the multi-site scalar lattice due to the absence of site-wise entanglement produced by interactions of the gradient operator, but better represent the ground state for each non-interacting site.
For the 2-site and 3-site systems in Table~\ref{tab:twositeExpGauss}, the one-norm probability distribution deviations are calculated to be only marginally larger than their intended-exponential counterparts.
The precision of the device is insufficient to conclusively distinguish between these two wavefunction structures by deviations of their probability distributions and thus, the exponential is proposed as an appropriate low-noise, low-entanglement approximation for the Gaussian wavefunction.

In general, on a quantum device with all-to-all connectivity, once the state is prepared it can then be time evolved with e.g., the
Trotterized time evolution operator.
However, the connectivity of {\tt Poughkeepsie} is not conducive to implementation of the time evolution operator as discussed in Ref.~\cite{Klco:2018zqz}.
Not only is sufficient connectivity between sites absent, but the all-to-all connectivity needed for the evolution of a single site is not present for $n_Q\ge 3$.
This work has shown how the available connectivity of {\tt Poughkeepsie} could be used to approximate a wavefunction technically requiring much higher degree of 2-qubit entanglement.
Unless a similar approximation can be made to the required gates present in the time evolution operator, additional qubit connectivity will be required to time evolve instances of a scalar field.

%%%%%%%%%%%%%%%%%%%%%%%%%%%%%%%%%%%%%%%%%%%%%%%%%%
\FloatBarrier
\section{Real Wavefunctions}
\FloatBarrier

\subsection{Circuits}
\label{sec:GaussianCircuits}

\begin{figure}[!ht]
\[
    |\psi\rangle = \quad \begin{gathered}
    \scalebox{0.9}{
   \Qcircuit @C=0.3em @R=.5em {
   |0\rangle \quad & \gate{R\left(\theta_{0,0}\right)} & \ctrlo{2} & \ctrl{2} & \ctrlo{2} & \ctrlo{2} & \ctrl{2} & \ctrl{2} & \ctrlco{2} & \qw  \\
   &&&&&&& \hspace{2.5cm}\cdots \\
   |0\rangle \quad & \qw & \gate{R\left(\theta_{1,0}\right)} & \gate{R\left(\theta_{1,1} \right)} & \ctrlo{2} & \ctrl{2} & \ctrlo{2} & \ctrl{2} & \ctrlco{2} & \qw \\
   &&&&&&& \hspace{2.5cm}\cdots \\
   |0\rangle \quad & \qw & \qw & \qw & \gate{R\left(\theta_{2,0}\right)} & \gate{R\left(\theta_{2,1}\right)} & \gate{R\left(\theta_{2,2}\right)} & \gate{R\left(\theta_{2,3}\right)} & \ctrlco{2}  & \qw \\
   & \vdots &\vdots&\vdots&&&& \hspace{2cm} \stackrel{\text{\begin{rotate}{0}$\ddots$\end{rotate}}}{} \\
   |0\rangle \quad & \qw & \qw & \qw & \qw & \qw & \qw & \qw & \gate{R\left(\vec{\theta}_{n_Q-1}\right)} & \qw \\
   }}
   \end{gathered}
   \]
   \caption{
   A circuit that can prepare an arbitrary real wavefunction by acting individually in increasing bipartitions of the Hilbert space.
   }
\label{fig:circNctrlArb}
\end{figure}
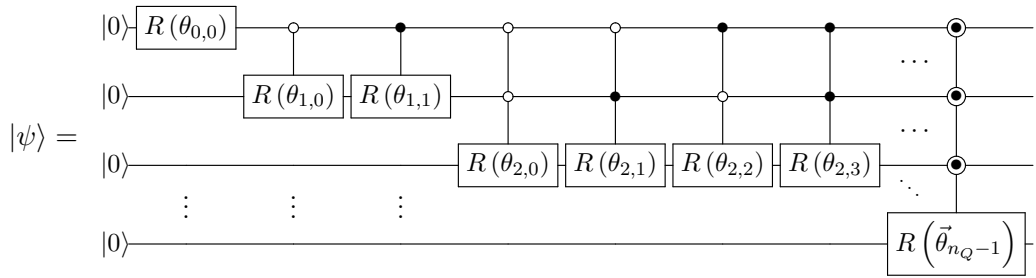

The preparation of an arbitrary real wavefunction $|\psi\rangle$ on $n_Q$ qubits decomposed as
$|\psi\rangle = \sum\limits_{x=0}^{2^{n_Q}-1} \psi(x)|x\rangle $
can be implemented by allowing a rotational degree of freedom for each introduced subspace of
Hilbert space, similar to the procedure described in Ref.~\cite{2002quant.ph..8112G}, as demonstrated in the circuit of Fig.~\ref{fig:circNctrlArb}.
In this design, new qubits are introduced at the bottom of the circuit and become the least significant bit in the binary interpretation of the linearized Hilbert space~\footnote{Note that this is the opposite scheme to that implemented in Sec.~\ref{sec:ExpCircuits} above for the exponential, where qubits were introduced at the top of the qubit staff becoming the most significant qubit.  This choice results from the decision to build the exponential from left to right in the binary-interpreted Hilbert space while here additional qubits are used to increase the digitization rather than expand the space.}.
Rotation operators are $y$-axis rotations, as above, and the extension of the circuit is shown pictorially with the last circuit element representing $2^{n_Q-1}$ $y$-axis rotations, each with ($n_Q-1$) controls.
The angles are notated as $\theta_{\ell,k}$
where $\ell$ indicates the level that runs from 0 to $n_Q-1$ and $k$ indexes the angles within the level running from 0 to $2^{\ell}-1$.
For example, the angle $\theta_{2,1}$ is the second angle in level 2 and is thus located in the doubly-controlled rotation gate with controls active on the state $|k\rangle = |01\rangle$.
Thus, the associated gate may be read directly from the angle label where the number of controls is $\ell$ and the binary for which those controls are active is the binary representation of $k$.
The collection of gates at each level are compacted graphically into a single gate with partially-filled controls and a vector of angles as shown at the right side of Fig.~\ref{fig:circNctrlArb}.  This circuit element is equivalent to the uniformly-controlled gates defined in Ref.~\cite{PhysRevA.71.052330}.
Note that the gates in a particular level commute (explicitly acting on different sectors of the Hilbert space) so may be implemented in any order while levels must be implemented in increasing sequential order.
Solving for the angles $\theta_{\ell,k}$ may be done through ratios of various bipartitions of the wavefunction samples,
\begin{eqnarray}
& &  \hspace{2.1cm} \theta_{\ell, k}  \ = \ \arctan
  \sqrt{
  \frac{
  \sum\limits_{x = x_{\rm min}}^{x = x_{\rm max}}
  \psi(x)^2}{\sum\limits_{y = y_{\rm min}}^{y = y_{\rm max}} \psi(y)^2}}
  \nonumber\\
&&   \begin{array}{ll}
  x_{\rm min}  =    2^{{n_Q}-\ell-1} (2k+1) & \ \ ,\ \   x_{\rm max} =    2^{{n_Q}-\ell } (1+k) -1 \\
  y_{\rm min}  =  k 2^{{n_Q}-\ell} & \ \ ,\ \   y_{\rm max} =    2^{{n_Q}-\ell-1} (2k+1)-1
  \end{array}
  \ \ \ .
  \label{eq:thetas}
\end{eqnarray}

On occasion, it may be convenient to solve for the angles corresponding to the wavefunction interpreted in reverse binary order.
On three qubits, for example, this distorts the wavefunction through the following interchanges of the quantum states
\begin{equation}
  \psi(0)\leftrightarrow\psi(0) \ \ \  \psi(1) \leftrightarrow \psi(4) \ \ \ \psi(2) \leftrightarrow \psi(2) \ \ \  \psi(3) \leftrightarrow \psi(6) \ \ \  \psi(5) \leftrightarrow \psi(5) \ \ \ \psi(7) \leftrightarrow \psi(7) \ \ \ .
\end{equation}
If this choice is made for the angle calculation, a swap network is needed at the end of the circuit to reorient the qubits.
If the state is simply being measured after preparation (rather than, for example,  time evolved), this reversal may be handled by simply reading the computational basis states backwards when classically recording measurements.

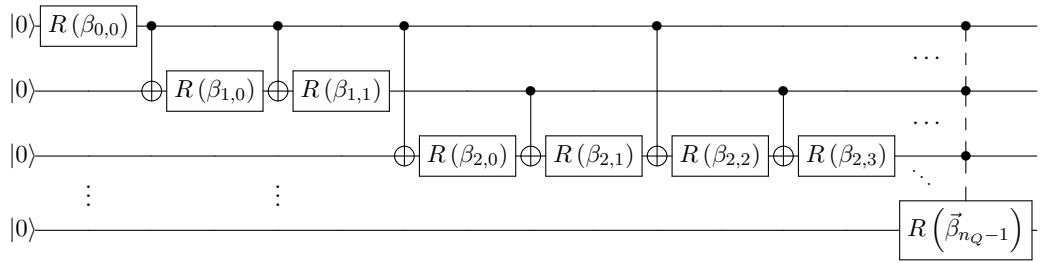
\begin{figure}[!ht]
\[
  \begin{gathered}
  \scalebox{0.85}{
   \Qcircuit @C=0.2em @R=.5em {
   |0\rangle \quad & \gate{R\left(\beta_{0,0}\right)} & \ctrl{2} & \qw & \ctrl{2} &\qw & \ctrl{4} & \qw & \qw & \qw & \ctrl{4} & \qw & \qw & \qw & \ctrl{0}  & \qw \\
   &&&&&&&&&&&&& \hspace{2.5cm} \cdots
   \\
   |0\rangle \quad & \qw & \targ & \gate{R\left(\beta_{1,0}\right)} & \targ & \gate{R\left(\beta_{1,1} \right)}  & \qw & \qw & \ctrl{2} & \qw & \qw & \qw & \ctrl{2} & \qw & \ctrl{0} \qwxdt & \qw \\
   &&&&&&&&&&&&& \hspace{2.5cm} \cdots
   \\
   |0\rangle \quad & \qw & \qw & \qw &\qw & \qw & \targ & \gate{R\left(\beta_{2,0}\right)} & \targ & \gate{R\left(\beta_{2,1}\right)} & \targ & \gate{R\left(\beta_{2,2}\right)} & \targ & \gate{R\left(\beta_{2,3}\right)} & \ctrl{0} \qwxdt & \qw \\
   & \vdots &&&\vdots&&& &&&&& & \hspace{2cm} \stackrel{\text{\begin{rotate}{0}$\ddots$\end{rotate}}}{}
   \\
   |0\rangle \quad & \qw & \qw & \qw & \qw & \qw & \qw & \qw & \qw & \qw & \qw & \qw & \qw & \qw & \gate{R\left( \vec{\beta}_{n_Q-1} \right) } \qwxdt & \qw \\
   }}
   \end{gathered}
   \]
   \caption{
   A circuit  with ${\cal O}(2^{n_Q})$ CNOTs capable
   of preparing an arbitrary real wavefunction by implementing single-qubit gates defined by linear combinations of the rotation angles used in Fig.~\ref{fig:circNctrlArb} interleaved with CNOTs of varying control and target placements.
    }
   \label{fig:circTreeArb}
\end{figure}

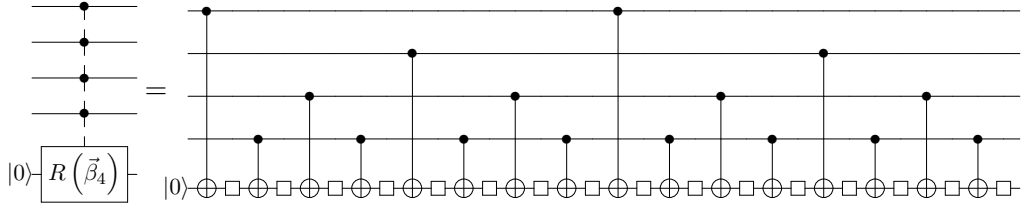
\begin{figure}[!ht]
\[
\begin{gathered}
  \scalebox{0.8}{
  \Qcircuit @C=0.4em @R=1.3em {
  & \ctrl{0} & \qw \\
  & \ctrl{0} \qwxd & \qw \\
  & \ctrl{0} \qwxd & \qw \\
   & \ctrl{0} \qwxd &\qw \\
   |0\rangle \quad & \gate{R\left( \vec{\beta}_{4} \right) } \qwxd & \qw \\
  }}
\end{gathered}
 =
\begin{gathered}
\scalebox{0.8}{
 \Qcircuit @C=0.4em @R=1.6em {
 & \ctrl{4} & \qw & \qw & \qw & \qw & \qw & \qw & \qw & \qw & \qw & \qw & \qw & \qw & \qw & \qw & \qw & \ctrl{4} & \qw & \qw & \qw & \qw & \qw & \qw & \qw & \qw & \qw & \qw & \qw & \qw & \qw & \qw & \qw & \qw \\
 & \qw & \qw & \qw & \qw & \qw & \qw & \qw & \qw & \ctrl{3} & \qw & \qw & \qw & \qw & \qw & \qw & \qw & \qw & \qw & \qw & \qw & \qw & \qw & \qw & \qw & \ctrl{3} & \qw & \qw & \qw & \qw & \qw & \qw & \qw & \qw
 \\
 & \qw & \qw & \qw & \qw & \ctrl{2} & \qw & \qw & \qw  & \qw & \qw & \qw & \qw & \ctrl{2} & \qw & \qw &  \qw & \qw & \qw & \qw & \qw & \ctrl{2} & \qw & \qw &  \qw & \qw & \qw & \qw & \qw & \ctrl{2} & \qw & \qw & \qw & \qw
  \\
 & \qw & \qw & \ctrl{1} & \qw & \qw & \qw & \ctrl{1} & \qw  & \qw & \qw & \ctrl{1} & \qw & \qw & \qw & \ctrl{1} & \qw  & \qw & \qw & \ctrl{1} & \qw & \qw & \qw & \ctrl{1} & \qw  & \qw & \qw & \ctrl{1} & \qw & \qw & \qw & \ctrl{1} & \qw  &\qw
 \\
 |0\rangle \quad & \targ & \gate{} & \targ & \gate{} & \targ & \gate{} & \targ & \gate{} & \targ & \gate{} & \targ & \gate{} & \targ & \gate{} & \targ & \gate{} & \targ & \gate{} & \targ & \gate{} & \targ & \gate{} & \targ & \gate{} & \targ & \gate{} & \targ & \gate{} & \targ & \gate{} & \targ & \gate{} &\qw
  \\
  }}
  \end{gathered}
  \]
  \caption{
  An example of the controlled rotation shown in Fig.~\ref{fig:circTreeArb}
  for level 4 $(\ell = 4)$, where an $\ell$-controlled unitary defined by a vector of rotation angles connected by a dashed vertical stem represents a series of $2^\ell$ CNOT operators interleaved with $2^\ell$ single-qubit rotations.
  The targets of the CNOTs act on the $(\ell +1)$\textsuperscript{th} qubit while the controls are placed
  from top to bottom of the qubit staff as increasing bipartitions.
  The angles defining the $2^\ell$ single-qubit rotations are defined from left-to-right by the
  sequential elements of the angle vector, $\beta_\ell$, that are defined in Eq.~(\ref{eq:thetastobetas}).
   }
  \label{fig:circlevel4eg}
\end{figure}

There are many ways one might consider implementing the above circuit, specifically the $\ell$-controlled rotations.
One way is to build the gates level-by-level in the decomposition of Barenco et. al \cite{Barenco:1995na}.
While a correct decomposition, the number of required CNOT gates for each $n$-controlled unitary scales as $\mathcal{O}(3^{n-1})$ and thus the circuit of $n_Q$ qubits requires a number of CNOT gates scaling as $\mathcal{O}(2^{n_Q+1} 3^{n_Q})$.  Rather than the circuit structure of Fig.~\ref{fig:circNctrlArb}, we propose the circuit of Fig.~\ref{fig:circTreeArb} as an alternative capable of creating the same arbitrary real wavefunctions but with a number of CNOTs scaling only as $\mathcal{O}(2^{n_Q})$ with the number of qubits.
The gate at the right in Fig.~\ref{fig:circTreeArb},
suggesting extrapolation to $n_Q$ qubits with dashed vertical connectivity,
represents a collection of $2^{n_Q-1}$ CNOTs with staggered controls alternating with the same
number of single-qubit rotations.
The structure of these controls must allow all linearly-independent combinations of the $\theta$s to be created through the $\beta$s at every level.  This symmetry can be assured through repeated bisection e.g., level $\ell = 4$  for the 5-qubit system is structured as shown in Fig.~\ref{fig:circlevel4eg},
where the empty boxes represent the $y$-axis rotations with angles defined by the 16-dimensional
vector of angles, $\vec{\beta}_4$.
Systematically, CNOTs of increasing length appear with frequency decreasing in powers of two.
Each level in this procedure contains $2^\ell$ rotations and the same number of CNOTs with the
exception of $\ell = 0$ where there are $0$ CNOTs.
Thus, the total number of these circuit elements required to initialize an arbitrary positive, real wavefunction is
\begin{equation}
  \text{rotations}(n_Q) = \sum_{\ell = 0}^{{n_Q}-1} 2^\ell = 2^{n_Q} -1 \qquad \text{CNOT}(n_Q) = \sum_{\ell = 1}^{n_Q-1} 2^\ell = 2^{n_Q} -2
  \ \ \ .
\end{equation}
Instead of applying high-dimensional controls to apply rotations to various sub spaces of the Hilbert space,
this circuit always implements single-qubit rotations whose angles are linear combinations of the previous angles specifically chosen to balance for an equivalent unitary operator.  Using the fact that
\begin{equation}
  X R(\theta) X = R(-\theta)
\end{equation}
the linear combinations of $\beta$ angles associated with the original $\theta$ angles can be read directly from the circuit structure.  For the 3-qubit example,
\begin{equation}
\begin{aligned}[c]
  \theta_{0,0} &= \beta_{0,0}\\
  \theta_{1,0} &= \sum_{k = 0}^{2^1-1}\beta_{1,k} \\
  \theta_{1,1} &= -\beta_{1,0} + \beta_{1,1} \\
\end{aligned}
\qquad
\begin{aligned}[c]
  \theta_{2,0} &= \sum_{k = 0}^{2^2-1}\beta_{2,k} \\
  \theta_{2,1} &= \beta_{2,0} - \beta_{2,1} - \beta_{2,2} + \beta_{2,3} \\
  \theta_{2,2} &= - \beta_{2,0} - \beta_{2,1} + \beta_{2,2} + \beta_{2,3}  \\
  \theta_{2,3} &= -\beta_{2,0} + \beta_{2,1} - \beta_{2,2} + \beta_{2,3}
\end{aligned}
\label{eq:thetastobetas}
\end{equation}
A block diagonal translation matrix with blocks of size $2^\ell$ each containing all linearly independent vectors of $\{1,-1\}$ can be constructed and inverted to translate the calculations of $\theta_{\ell,k}$ into values for $\beta_{\ell, k}$ to be implemented more efficiently.  An example of this matrix is given in Eq.~\ref{eq:angletranslation} of the next subsection.  It should be noted that this inversion need be calculated only once for any $n_Q$ as it is angle-independent.  The following subsection provides an explicit example of this process for the preparation of a Gaussian wavefunction.

%%%%%%%%%%%%%%%%%%%%%%%%%%%%%%%%%%%%%%%%%%
\FloatBarrier
\subsection{Preparation of a Gaussian}
\FloatBarrier

The Gaussian wavefunction is
of particular importance for bosonic degrees of freedom and the ground states of scalar field theories.
Its preparation on a register of qubits has been previously considered in
Refs.~\cite{2008arXiv0801.0342K,Quantumc53:online,Somma:2016:QSO:3179430.3179434,2018arXiv180302466N} both recursively and dynamically.
Here we largely recover the recursive findings with the circuitry described previously,
though we provide explicit circuits for the method without calculating the angles on qubits,
and take advantage of the symmetry of the Gaussian to reduce the necessary quantum resources with respect to more general implementation strategies~\cite{2002quant.ph..8112G,PhysRevA.71.052330,PhysRevA.83.032302}.
This section, and the discussions of the symmetrized exponential above, could be considered to provide an inexpensive initialization for Somma Inflation~\cite{Somma:2016:QSO:3179430.3179434}, where eigenstates of the harmonic oscillator are prepared efficiently through tuned time evolution from a narrow state with Gaussian-distributed amplitudes.~\footnote{The evolution process is completed with two separate operators:  the free propagator and the field operator.
The free propagator allows the Gaussian wavepacket to expand to encompass a larger number of states.
Propagating next with the field operator allows cancellation of the imaginary components of the wavefunction from the free propagation.
This proposal is similar to an adiabatic evolution between harmonic oscillator Hamiltonians with differing values of $\omega$.
However, Somma Inflation requires the quantum resources of only a single Trotter step.
}

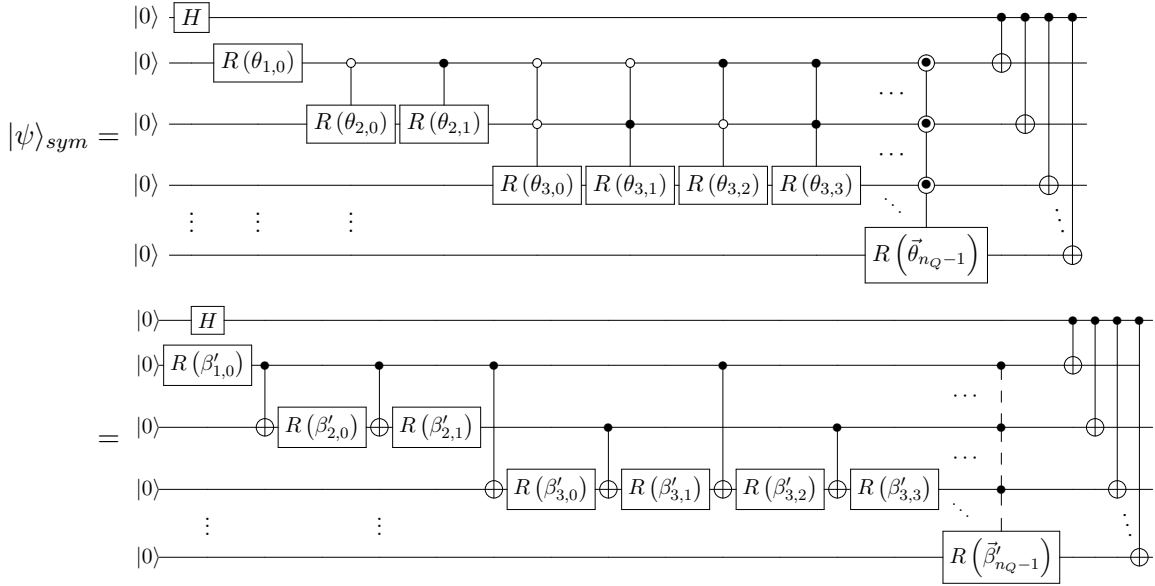
\begin{figure}[!ht]
\begin{align}
|\psi\rangle_{sym} &= \quad
  \begin{gathered}
  \scalebox{0.8}{
    \Qcircuit @C=0.2em @R=.5em {
    |0\rangle \qquad & \gate{H} & \qw & \qw & \qw & \qw & \qw & \qw & \qw & \qw & \ctrl{1} & \ctrl{3} & \ctrl{5} & \ctrl{7} & \qw \\
    |0\rangle \qquad & \qw & \gate{R\left( \theta_{1,0} \right)} & \ctrlo{2} & \ctrl{2} & \ctrlo{2} & \ctrlo{2} & \ctrl{2} & \ctrl{2} & \ctrlco{2} & \targ & \qw & \qw & \qw & \qw \\
    &&&&&&&& \hspace{2.5cm} \cdots
   \\
    |0\rangle \qquad & \qw & \qw & \gate{R \left(\theta_{2,0} \right) } & \gate{R \left(\theta_{2,1} \right) } & \ctrlo{2} & \ctrl{2} & \ctrlo{2} & \ctrl{2} & \ctrlco{2} & \qw & \targ & \qw & \qw & \qw \\
    &&&&&&&& \hspace{2.5cm} \cdots
   \\
    |0\rangle \qquad & \qw & \qw & \qw & \qw & \gate{R\left( \theta_{3,0} \right) } & \gate{R\left( \theta_{3,1} \right) } & \gate{R\left( \theta_{3,2} \right) } & \gate{R\left( \theta_{3,3} \right) } & \ctrlco{2} & \qw & \qw & \targ & \qw & \qw \\
    & \vdots &\vdots&\vdots&&&&& \hspace{2.2cm} \stackrel{\text{\begin{rotate}{0}$\ddots$\end{rotate}}}{} &&& \hspace{0.6cm} \stackrel{\ }{\text{\begin{rotate}{-40}$\ddots$\end{rotate}}} \\
    |0\rangle \qquad & \qw & \qw & \qw & \qw &\qw & \qw & \qw & \qw & \gate{R\left( \vec{\theta}_{n_Q-1}\right) } & \qw & \qw & \qw  & \targ
    }}
  \end{gathered} \nonumber \\
  &= \ \ \begin{gathered}
      \scalebox{0.75}{
   \Qcircuit @C=0.2em @R=.5em {
   |0\rangle \quad & \gate{H} & \qw & \qw & \qw & \qw & \qw & \qw & \qw & \qw & \qw & \qw & \qw & \qw & \qw  &  \ctrl{1} & \ctrl{3} & \ctrl{5} & \ctrl{7} & \qw \\
   |0\rangle \quad & \gate{R\left(\beta'_{1,0}\right)} & \ctrl{2} & \qw & \ctrl{2} &\qw & \ctrl{4} & \qw & \qw & \qw & \ctrl{4} & \qw & \qw & \qw & \ctrl{0}  &  \targ & \qw & \qw & \qw \\
   &&&&&&&&&&&&& \hspace{2.5cm} \cdots
   \\
   |0\rangle \quad & \qw & \targ & \gate{R\left(\beta'_{2,0}\right)} & \targ & \gate{R\left(\beta'_{2,1} \right)}  & \qw & \qw & \ctrl{2} & \qw & \qw & \qw & \ctrl{2} & \qw & \ctrl{0} \qwxdt & \qw & \targ & \qw & \qw & \qw \\
   &&&&&&&&&&&&& \hspace{2.5cm} \cdots
   \\
   |0\rangle \quad & \qw & \qw & \qw &\qw & \qw & \targ & \gate{R\left(\beta'_{3,0}\right)} & \targ & \gate{R\left(\beta'_{3,1}\right)} & \targ & \gate{R\left(\beta'_{3,2}\right)} & \targ & \gate{R\left(\beta'_{3,3}\right)} & \ctrl{0} \qwxdt & \qw & \qw & \targ & \qw & \qw \\
   & \vdots &&&\vdots&&& &&&&& & \hspace{2cm} \stackrel{\text{\begin{rotate}{0}$\ddots$\end{rotate}}}{} &&& \hspace{0.6cm} \stackrel{\ }{\text{\begin{rotate}{-40}$\ddots$\end{rotate}}}
   \\
   |0\rangle \quad & \qw & \qw & \qw & \qw & \qw & \qw & \qw & \qw & \qw & \qw & \qw & \qw & \qw & \gate{R\left( \vec{\beta}'_{n_Q-1} \right) } \qwxdt & \qw & \qw & \qw & \targ & \qw \\
   }}
  \end{gathered} \nonumber
\end{align}
  \caption{
  The upper circuit shows that in Fig.~\ref{fig:circNctrlArb} specialized to the implementation of a real wavefunction that is symmetric in the binary-interpreted Hilbert space.
By leveraging linear combinations of rotation angles, the highly-controlled rotation gates can be replaced by the lower (tree) circuit
described by Fig.~\ref{fig:circTreeArb} and Fig.~\ref{fig:circlevel4eg}.  An explicit example of transforming $\theta \rightarrow \beta'$ angles is given in Eqs.~\eqref{eq:eq15} and~\eqref{eq:eq16} for an $n_Q = 3$ Gaussian wavefunction.
}
  \label{fig:circNctrlArbSYM}
\end{figure}

Beginning with the well-localized~\footnote{Gaussian wavefunctions that are not well localized may, as usual, be implemented with periodic boundary conditions by including image Gaussians at distances defined by the Hilbert space dimension. Once such a target wavefunction is established,
Eq.~\eqref{eq:thetas} and subsequent procedures are unchanged.} Gaussian wavefunction, the state vector is a digital sampling of the continuous Gaussian
\begin{equation}
  |\psi \rangle = \mathcal{N}\sum_{x = 0}^{2^{n_Q}-1} \exp\left[-\frac{\left(x-\mu_{n_Q}\right)^2}{2 \sigma_{n_Q}^2}\right] |x\rangle \qquad \mathcal{N}^{-1} = \sqrt{\sum_{x = 0}^{2^{n_Q}-1} \exp\left[-\frac{\left(x-\mu_{n_Q}\right)^2}{ \sigma_{n_Q}^2}\right]}
  \ \ \ .
\end{equation}
The $n_Q$-dependent factors associated with the mean and standard deviation account for the scaling of the sampling grid as the number of qubits is increased,
\begin{equation}
  \mu_{n_Q} = 2^{n_Q - 1} \mu
   \ \ \ \  ,\ \ \
   \sigma_{n_Q} = 2^{n_Q - 1} \sigma
  \ \ \  ,
\end{equation}
such that adding a qubit increases the sampling resolution rather than distorting the wavefunction.
A value of $\mu = 1$  places the peak of the Gaussian at the site $2^{n_Q-1}$.
To place the peak in the center of the binary Hilbert space, symmetrizing the sample points
as would be done in Ref.~\cite{Klco:2018zqz} for a field-space wavefunction about $\phi = 0$,
the value of $\mu$ is also $n_Q$-dependent $\mu_{\rm center} = 1 - \frac{1}{2^{n_Q}}$.
As an example, the vector of angles calculated from Eq.~\eqref{eq:thetas} that would be used to implement a
3-qubit Gaussian of $\sigma = 0.2$ centered symmetrically in the Hilbert space
using the circuit of Fig.~\ref{fig:circNctrlArb} would be
\begin{equation}
  \vec{\theta} = \left( \theta_{0,0}, \theta_{1,0}, \theta_{1,1}, \theta_{2,0}, \theta_{2,1}, \theta_{2,2}, \theta_{2,3} \right)^T
  = \left(0.785, 1.562, 0.009, 1.562, 1.364, 0.207, 0.009\right)^T  \ .
\end{equation}
The first angle, due to the symmetry between the Hilbert space half spaces is simply $\arctan{1} = \frac{\pi}{4}$.
To translate these angles to the angles implemented in the improved tree circuit of Fig.~\ref{fig:circTreeArb}, one analyses the action of each of the $2^{\ell}$ states of the control inputs at level $\ell$ to produce the block diagonal translation matrix described by Eq.~\eqref{eq:thetastobetas}.  Under inversion, all linearly independent combinations of the $\theta_{\ell, k}$'s at each level $\ell$ are captured in the $\beta$ angles of the tree circuit,
\begin{align}
  \vec{\beta} &= \begin{pmatrix}
     1 & 0 & 0 & 0 & 0 & 0 & 0  \\
 0 & 1 & 1 & 0 & 0 & 0 & 0 \\
 0 & -1 & 1 & 0 & 0 & 0 & 0 \\
 0 & 0 & 0 & 1 & 1 & 1 & 1 \\
 0 & 0 & 0 & 1 & -1 & -1 & 1  \\
 0 & 0 & 0 & -1 & -1 & 1 & 1 \\
 0 & 0 & 0 & -1 & 1 & -1 & 1  \\
  \end{pmatrix}^{-1} \vec{\theta} \nonumber \\&= \left(0.785,  0.776, 0.785, 0.677, 0.0 , 0.099, \
0.785 \right)^T \ \ \ .
\label{eq:angletranslation}
\end{align}
Given that the Gaussian is a symmetric function, it is apparent that the $2^{n_Q}-1$ angles of Eq.~\eqref{eq:thetas} are not providing independent information to the state preparation.  Guided by the symmetrization (or anti-symmetrization) shown in the circuit of Fig.~\ref{fig:symmetrizedExpcircuit}---introducing a qubit at the top of the stave (most-significant in the binary-interpreted Hilbert space) in the $|\pm\rangle$ state to copy the distribution and a string of CNOTs to reflect the second half space---an arbitrary real, symmetric wavefunction can be implemented with fewer quantum resources.
By effectively replacing the final level ($\ell = n_Q-1$) with a symmetrization, this trades $2^{n_Q-1}$ CNOTs for only $n_Q-1$ CNOTs.
As can be intuited, only the first half of the $\vec{\theta}_\ell$ is retained at each level with the control on the level-0 qubit removed.
The first angle thus becomes $R\left( \frac{\pi}{4}\right) = H$ as noted above for any symmetric wavefunction.
In terms of the angles calculated in Eq.~\eqref{eq:thetas}, the circuit with multi-controlled quantum gates becomes
that of  Fig.~\ref{fig:circNctrlArbSYM}.
As shown in this figure, the gate associated with $\theta_{\ell,k}$ is an $(\ell-1)$-controlled rotation still with controls dictated by the binary interpretation of $k$ but now omitting the top qubit.  The remaining relevant angles for the 3-qubit example above are,
\begin{equation}
  \vec{\theta} = \left( \theta_{1,0}, \theta_{2,0}, \theta_{2,1} \right) = \left( 1.5618, 1.5616, 1.364 \right)^T \ \ \ .
  \label{eq:eq15}
\end{equation}
The associated $\beta'$ angles are constructed as linear combinations at each level as above,
\begin{equation}
  \vec{\beta}' =\left( \beta'_{1,0}, \beta'_{2,0}, \beta'_{2,1} \right) =\begin{pmatrix}
    1 & 0 & 0  \\
    0 & 1 & 1  \\
    0 & -1 & 1  \\
  \end{pmatrix}^{-1} \theta = \left( 1.562, 0.099, 1.463 \right)^T
  \ \ \ ,
  \label{eq:eq16}
\end{equation}
and can, similarly to the relationship between the circuits of Fig.~\ref{fig:circNctrlArb} and Fig.~\ref{fig:circNctrlArbSYM}, be implemented by removing the final layer of
the tree circuit of Fig.~\ref{fig:circTreeArb}, relabeling the angles, and adding a qubit at the top of the stave for symmetrization.  This final circuit is shown in the second line of Fig.~\ref{fig:circNctrlArbSYM}.  Thus, the resource costs become,
\begin{equation}
  \text{Rotations}(n_Q) = \sum_{\ell = 0}^{n_Q-2} 2^\ell = 2^{n_Q-1}  \qquad
  {\rm CNOT}(n_Q) = 2^{n_Q -1} + n_Q -3 + \delta_{n_Q,1}\ \ \ ,
\end{equation}
specifically relevant for the preparation of any eigenstate of the harmonic oscillator and
generally relevant for preparing an arbitrary real, symmetric (or antisymmetric) wavefunction centered in the Hilbert space.

%%%%%%%%%%%%%%%%%%%%%%%%%%%%%%%%%%%%%%%%%%%%
\section{Reflections}

There are a number of items to discuss related to the work described in the previous sections.
The first point is related to the impact of field digitization.
We have found that a symmetric exponential wavefunction, which can be initialized with a minimal number of entangling gates,
can be tuned to have a substantial overlap with a Gaussian wavefunction.
While dense sampling of the function in the region of support would reveal a cusp at its symmetry point,
with a limited number of qubits per site and the symmetry point placed mid-way between two states in the Hilbert space,
as considered here, the cusp is not resolved and the sampled evaluations of the
symmetric exponential can be interpolated smoothly.
An additional consideration is that as the near-term devices are noisy, initializing the device with precision better than that of the noise will not improve the quality of the simulation.
Therefore, until the noise in quantum devices drops  below the percent level, a symmetric
exponential wavefunction is likely to provide an adequate initialization state.

As is made clear in examining the 2- and 3-site systems compared with the 1-site system,
for a multi-site lattice scalar field theory, the number of states in the Hilbert space rapidly becomes too large to completely sample,
as is well known.
The dimensionality of the measurement-error system(s) nominally scales
with the square of the Hilbert-space dimensionality,
becoming unmanageable even before the measurements of quantities of interest are considered.
New algorithms will be required to isolate important quantities from the field theory computation to inform the quality of state preparation that do not scale with the Hilbert-space dimensionality, likely exploiting auxiliary qubits.

In establishing the symmetric exponential as a potentially-useful state to prepare, we made use of the fact that duplication and reflection are straightforward at the circuit level, requiring only a Hadamard gate and $n_Q-1$ CNOT gates, respectively.
For wavefunctions that have vanishing derivative at the edges and have a constant logarithmic derivative within the half Hilbert space,
such as an exponential function or a constant,  such functions can be initialized without entangling gates.
Therefore, a step wavefunction, an exponential, and a uniform wavefunction can be initialized without CNOT gates.

More complicated periodic functions can also be initialized relatively simply.
For instance, one period of a $\cos\phi$ or $\sin\phi$ function can be initialized using the general circuit described previously.
Additional qubits can be included, and when acted on with a Hadamard gate lead to a wavefunction of multiple periods of the
$\cos\phi$ or $\sin\phi$.
This can also be accomplished by first initializing a quarter-period, then anti-symmetrizing and symmetrizing using Hadamard and Pauli $\hat{Z}$ gates.

In the Introduction, the small-$n_Q$ approximate wavefunction preparation in this work was placed into the context of state preparation in the asymptotic regime of many qubits as a possible first step for Somma Inflation---the controlled evolution capable of expanding an initial Gaussian wavefunction to one with support over a larger portion of the Hilbert space.
The quantum resources required to implement this inflation are equivalent to that of a single first-order Trotter step of time evolution, and will thus benefit from any future algorithmic improvements of the already-efficient time evolution operator.
We find that the symmetrized exponential wavefunction is less costly than implementing the period of inflation (required two-qubit gates grow linearly rather than quadratically).
Thus, in noisy hardware where an exponential approximation to the Gaussian wavefunction may introduce sub-leading theoretical systematics, the symmetrized exponential may be used without further post processing.
For situations in which higher precision is desired in the wavefunction structure, either an adiabatic approach or the perturbative scheme of improvement proposed below may be implemented on small qubit registers.
If inflating to larger qubit registers, the errors in the initial state will be propagated to the final Gaussian state.  Further exploration is needed to optimize the properties of the initial narrow Gaussian wavefunction digitization and extent such that the inflated Gaussian is created to saturate the Nyquist Shannon sampling theorem (see e.g., Refs.~\cite{Macridin:2018oli,Klco:2018zqz}) and thus efficiently utilize the available quantum resources.

\begin{figure}
  \centering
  \includegraphics[width=0.45\textwidth]{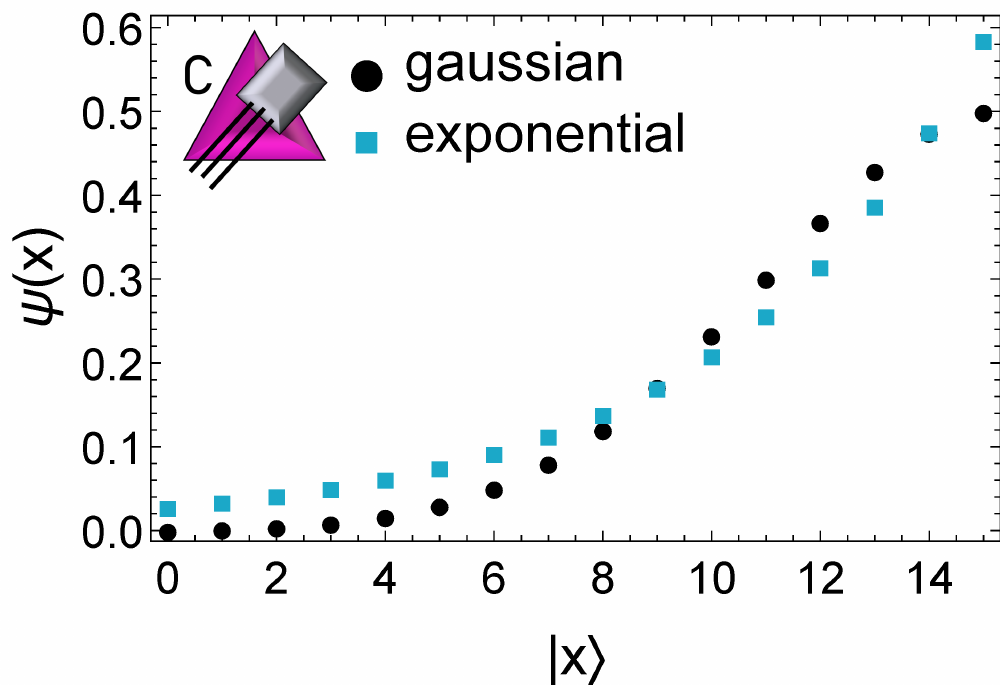}
  \includegraphics[width=0.45\textwidth]{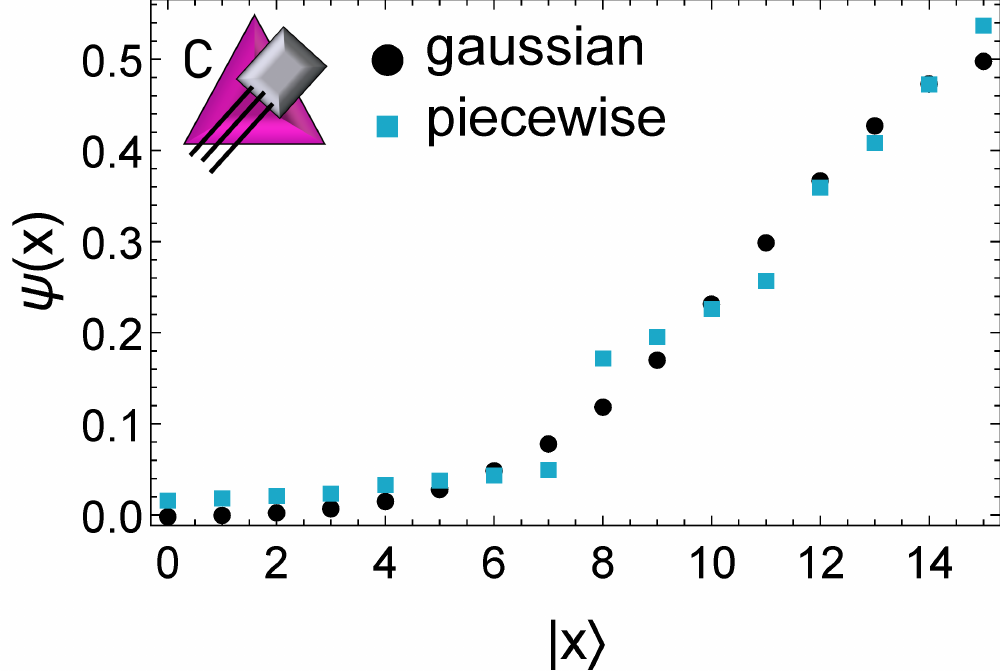}
  \caption{
  (Color online)
  The left panel shows an exponential wavefunction (blue, square points),
  initialized on a $n_Q=4$ quantum register with three rotation angles calculated as in Eq.~\eqref{eq:expangles}.
  This exponential wavefunction has an optimal overlap with
  the Gaussian wavefunction (black, circular points).
  The right panel shows a wavefunction (blue, square points) obtained by fitting the three rotation angles to
  maximize the overlap with the Gaussian wavefunction.
  }
  \label{fig:pert}
\end{figure}
Given the results that we have obtained, it is interesting to consider the possibility that a perturbative expansion
in the application of entangling gates may be possible, that would, for example,
order-by-order transform a symmetric exponential wavefunction, or some other variant thereof,
into a Gaussian wavefunction.
Such an expansion could be helpful in working with noisy quantum devices, as the noise would provide an upper limit to the
required precision of the perturbative approximation.
As a starting point, the single qubit rotations that define the exponential function can be modified to optimize the overlap with
a Gaussian wavefunction, giving rise to a piece-wise wavefunction.
While providing a better approximation to a Gaussian wavefunction than the exponential function,
as shown in Fig.~\ref{fig:pert} for the case of $n_Q=4$,
this introduces higher-frequency components in field conjugate momentum space.
We have not yet explored the utility of the piecewise wavefunction, nor the impact of the high-frequency conjugate momentum states,  nor developed the
quantum circuit perturbation theory, but intend to do so in subsequent work.

An important question to address is that of the utility of an initial state that is a tensor product of symmetric exponential wavefunctions at each point of a lattice scalar field theory.
Recalling the seminal work of Jordan, Lee and Preskill~\cite{Jordan:2011ci}, who have detailed how to
perform calculations in scalar field theory using a quantum computer,
initializing a quantum computer into the ground state of an interacting scalar field theory is
beyond the capabilities of a quantum computer due to its complexity.
They proposed to compute, for instance, scattering S-matrix elements by initializing each spatial site into the ground state
of a harmonic oscillator, a tensor-product state corresponding to a non-interacting theory without inter-site interaction.
Wavepackets of the non-interacting theory are prepared, and then the entire system is adiabatically evolved from the non-interacting theory to the interacting theory while retaining the asymptotic separation of the wavepackets.
The system is then evolved forward in time.
A similar process can be employed when the sites are prepared in a symmetric exponential wavefunction.
As such states are the lowest-lying eigenstate of an attractive Dirac-delta function, $V^{(\delta)}(\phi) = - g \delta(\phi)$, where $g$ is a coupling constant, the scalar field theory initialized into a tensor-product state of these wavefunctions
can be adiabatically evolved from
$V^{(\delta)}(\phi) $ to $V(\phi)={1\over 2} m^2\phi^2 -\frac{1}{2} \phi \nabla_a^2 \phi+ {1\over 4!}\lambda \phi^4$.
Therefore, the process outlined in Ref.~\cite{Jordan:2011ci} to accomplish scattering in scalar field theory remains unchanged when symmetric exponential wavefunctions are initialized, but the interactions that are adiabatically evolved are different.  Of course, it is likely to be the case that the {\it devil is in the details} of actual quantum simulations, which remain to be performed.

%%%%%%%%%%%%%%%%%%%%%%%%%%%%%%%%
\section{Summary and Conclusions}

State preparation on quantum devices is a key ingredient in the quantum simulation of systems
of importance in biology, chemistry, physics and other areas.
If the properties and dynamics of low-lying states of a quantum many-body system are the focus,
a quantum device used to perform the simulation needs to be initialized to have significant overlap with these states.
Without this, particularly in the near-term, isolating the physics of interest will be extremely challenging,
and in most cases, not possible, through subsequent evolution and measurement of the system.

As the number of entangling gates involved in preparing a symmetric exponential wavefunction grows linearly with the number of qubits, they may provide good approximations  to  Gaussian  wavefunctions with support over a  small  or  modest number of states due to the inherent digitization of the field and resulting truncation in field-conjugate-momentum space.
Improvement of this approximation can be made perturbatively or adiabatically, reaching the digitized Gaussian wavefunction using resources scaling exponentially with the region of support of the wavefunction.
Inflating this Gaussian wavefunction may be implemented efficiently through simple unitary transformations~\cite{Somma:2016:QSO:3179430.3179434}.
Our studies of these wavefunctions, which include ideal quantum simulations, noisy quantum simulations and
simulations on IBM~Q~Experience quantum devices, with measurement-error correction and CNOT-gate error mitigation,
show that they may provide good starting points for subsequent time-evolution,
with overlaps $\sim 0.95$ or greater for the small systems we have explored.
Clearly, further studies and simulations are required to determine the extent of their utility.
Our results show that even with the small systems that are presently available, the systematic errors and noise in initializing a Gaussian wavefunction are significantly more than in initializing a symmetric exponential wavefunction.
In some of our initializations of a Gaussian wavefunction with 4 qubits per site,
the CNOT errors were sufficiently large such that they could not be reduced through a standard extrapolation procedure.
Following from the success in preparing the symmetrized exponential, there may be utility in seeking other approximate wavefunctions that also
require fewer entangling gates than a Gaussian.

In order to accommodate fluctuations in gate operations and measurement errors during production running on the quantum devices,
we employed a workflow comprised of a sequence of circuits that were run immediately prior to the symmetrized exponential wavefunction
circuit.  As detailed in the text,
the results obtained from these other circuits, such as  an Hadamard-gate in vacuum,
were used to implement a set of data acceptance criteria (cuts),  analogous to those used in the analysis of experimental data,
that had to be satisfied to retain the results of the symmetrized exponential wavefunction circuit.
Establishing these cuts
was found to be a valuable component in analyzing the results of circuits executed on the quantum devices.
These types of integrated workflows are likely to be useful throughout the NISQ era.

We used IBM's {\tt Poughkeepsie} to produce the results presented in this work from a quantum device,
and used its noise parameters in noisy simulations.
This was determined largely by availability, throughput and the gate- and qubit-fidelities that were available during this production.
The interconnect fabric on {\tt Poughkeepsie} prevents the preparation of symmetrized exponentials beyond $n_Q=4$ qubits per spatial site with $(n_Q-1)$ CNOTs~\footnote{If the number of CNOTs is allowed to double, then the symmetrization can be reconfigured for nearest-neighbor connectivity in a 1D string of qubits and thus the interconnect fabric on {\tt Poughkeepsie} can support the preparation of an 18-qubit symmetrized exponential (see Appendix~\ref{app:Connectivity} for details).}, and also prevents
Trotterized time-evolution for $n_Q=3$ qubits per spatial site and larger.
It would appear that improvement in qubit- and gate-fidelities for sufficiently-connected qubit clusters are currently required to prepare and time evolve
even a small lattice scalar field theory on such quantum devices.

The three stages of preparing the symmetric exponential wavefunction are informative about the
qubit- and gate-fidelities of the quantum device, and may serve as a calibration benchmark for future simulations.
The first stage involves the recursive generation of an exponential distribution distributed across half of the states in the Hilbert space.
As this only involves y-axis rotations of all but one of the qubits,
the fidelity of these rotations can be ascertained (after measurement-error correction),
and these angles can be iteratively tuned to optimize the initialized state with regard to the target exponential function.
The second stage of preparation is a single Hadamard gate applied to the last qubit, duplicating the exponential wavefunction in the first half of the Hilbert space into the second half.
The fidelity of this operation can be determined and the gate operation tuned, both ``in vacuum'' and ``in-medium'' to optimize the symmetry of the wavefunction in the quantum register.
The last stage of preparation corresponds to CNOT gates connecting each qubit to the last.  As the CNOT fidelity is significantly worse than that of the single qubit gates, it is this stage where substantial noise and systematic twisting of the control qubit are observed. The CNOT errors are mitigated with the standard techniques of extrapolation, but this increases the uncertainty in the calculation, and is the limiting stage of the state preparation.   The twisting of the control qubit introduces an additional asymmetry into the wavefunction that is visually apparent.
We found that introducing cuts on the quality of the wavefunction at each stage of this process with time-correlated calibration circuits provided
valuable improvement of the state preparation.

While our present work has been cast in the context of scalar field theory, it has broader application.
One such example is in the state preparation of a weakly-bound non-relativistic system such as the deuteron.
At leading order in the pionless effective field theory (EFT)~\cite{vanKolck:1998bw,Kaplan:1998we,Kaplan:1998tg,bedaque2002}, which recovers effective range theory
for nucleon-nucleon scattering, the deuteron wavefunction is dominated by the s-wave exponential wavefunction,
with contributions from the tensor force inducing s-d mixing suppressed in the power-counting.
While recent quantum calculations employing the pionless EFT~\cite{PhysRevLett.120.210501,Lu:2018pjk,Shehab:2019gfn} have been performed in Fock-space,
similar calculations could be performed in position space, in which the exponential function
is the exact wavefunction at long-distances.
The perturbative expansion that defines the pionless EFT could be implemented as
a perturbative expansion in quantum circuits, which will be explored in subsequent work.
Our 2-site and 3-site calculations demonstrate that we are now able to begin to simulate 2-dim. and 3-dim. systems in real space.

In this work, we find benefits in the use of tuned
symmetric exponential wavefunctions---requiring only $(n_Q-1)$ entangling gates---as approximations to Gaussian wavefunctions---requiring $\mathcal{O}\left(2^{n_Q}\right)$ entangling gates for small $n_Q$---for the initialization of scalar field theories in future quantum simulations.
This is one example of the existence of wavefunctions closely
resembling target wavefunctions on digitized Hilbert spaces, but requiring reduced entanglement.
Identification of such wavefunction pairs allows the precision of state preparation to be advantageously matched to the precision of the quantum hardware.
While our results have been implemented with IBM's~Q~Experience quantum devices,  the conceptual developments are device independent.
It would be interesting to
explore these developments on other quantum architectures, such as trapped ions or cold atoms.

%%%%%%%%%%%%%%%%%%%%%%%%%%%%%%%%%%%%%%%%%%%
%%%%%%%%%%%%%%%%%%%%%%%%%%%%%%%%%%%%%%%%%%%
\appendix

%%%%%%%%%%%%%%%%%%%%%%%%%%%%%%%%%%%%%%%%%%%
\FloatBarrier
\section{Connectivity}
\label{app:Connectivity}
\FloatBarrier

\begin{figure}[!ht]
\centering
\includegraphics[width=0.35\textwidth]{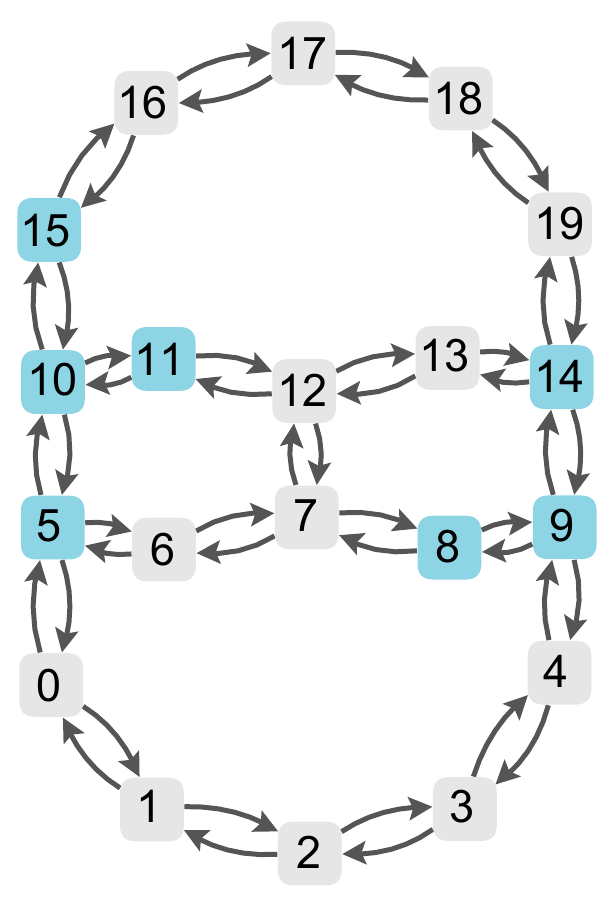}
\hspace{0.8cm}
\includegraphics[width=0.35\textwidth]{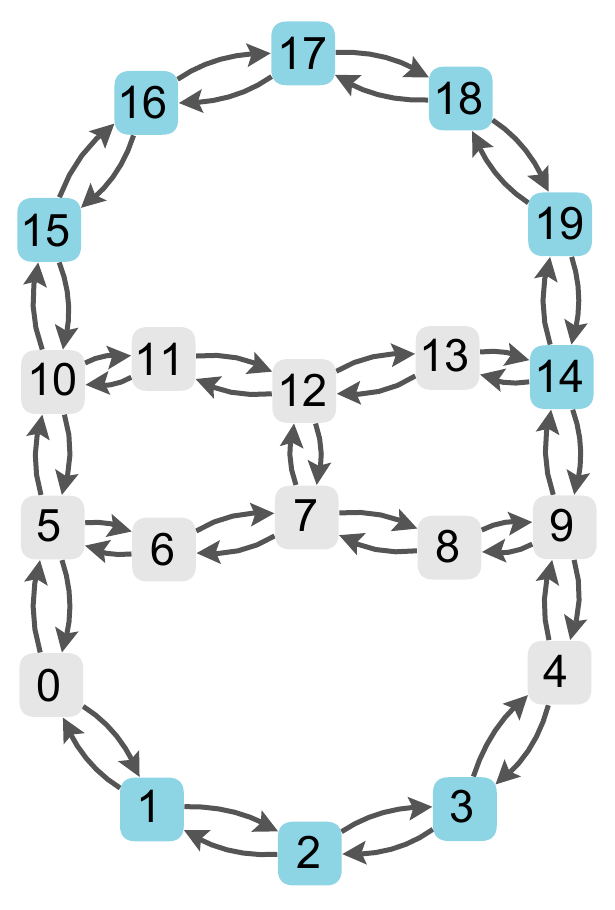}
\caption{
(Color online)
Connectivity map for IBM Q device {\tt Poughkeepsie}.
The partitions used in this work are colored in blue, while remaining spectator qubits on the chip are colored in light gray.
Directed connections indicate the availability of CNOT gates from control to target.
The left panel shows the partitions used for single-site quantum simulation on {\tt Poughkeepsie}
and 2-site noise simulations (see Table~\ref{tab:qubitmaps}).
The right panel shows the partitions used for 2-site and 3-site quantum simulation on {\tt Poughkeepsie}
(see Table~\ref{tab:qubitmaps}).
}
\label{fig:PouConnectivity}
\end{figure}
The available connectivity of a quantum device determines the quantum states that can be accessed.
High degrees of inter-qubit connectivity, while allowing more flexibility of circuit implementation,
bring significant  challenges in maintaining the locality and fidelity of single- and multi-qubit gates.
The device used in this work, IBM {\tt Poughkeepsie}, has been designed as interconnected rings
with qubits maximally connected to three other qubits.
Preparing a Gaussian wavefunction, or other arbitrary real positive wavefunctions,
requires all-to-all connectivity (see Sec.~\ref{sec:GaussianCircuits}).
The largest number of qubits that can thus be used to digitize a Gaussian on {\tt Poughkeepsie} (without increasing the CNOT count via SWAP operators) is two.
However, as demonstrated in the main text of this paper, the preparation of a symmetrized exponential function can be designed to have high overlap with the Gaussian wavefunction while requiring less entanglement.
More significantly, preparation of the exponential no longer requires all-to-all connectivity,
but rather a single central qubit's connectivity to all others (see the circuit shown in  Fig.~\ref{fig:symmetrizedExpcircuit}).
The maximum instance of this spoke-hub type connectivity can be found at a junction between {\tt Poughkeepsie}'s interconnected rings.
For example, the qubits used in this work to prepare the symmetrized exponential on four qubits were $\left[5, 11, 15, \mathbf{10}\right]$ where qubit 10, with its hub-connectivity, was used as the most-significant qubit in the binary interpretation of the Hilbert space through which the exponential is copied and symmetrized.
These qubits and their further connectivity across the device are depicted in Fig.~\ref{fig:PouConnectivity}.
For the initialization of a scalar field, this capability of 4-qubit digitization (with exact digitized wavefunctions) has been found to be sufficient when considering the attainable precision in the low-energy eigenspace with respect to the expected level of hardware noise~\cite{Klco:2018zqz}.
Time evolution of a single site, however, requires all-to-all connectivity between the qubits
digitizing the site (e.g., all 2-qubit $ZZ$ operators to implement the mass term).
Together, these requirements and capabilities emphasize the importance of a co-design process
between algorithmic advances and hardware design.

\begin{figure}[!ht]
  \[
  \begin{gathered}
  \Qcircuit @C=1.1em @R=.6em {
  & \gate{H} & \ctrl{1} & \ctrl{2} & \ctrl{3} & \ctrl{5}& \qw \\
  & \qw & \targ & \qw  & \qw & \qw & \qw\\
  & \qw & \qw & \targ & \qw & \qw & \qw \\
  & \qw & \qw & \qw & \targ & \qw & \qw \\
  & && \hspace{1.6cm} \stackrel{\ }{\text{\begin{rotate}{-14}$\ddots$\end{rotate}}} \\
  & \qw & \qw & \qw & \qw & \targ & \qw
  }
  \end{gathered}
  =
  \begin{gathered}
  \Qcircuit @C=1.1em @R=.6em {
    & \qw & \qw & \qw & \qw & \qw & \targ & \qw & \qw & \qw & \qw & \gate{H} &  \qw  \\
    & \gate{H} & \qw & \qw & \qw& \targ & \ctrl{-1} & \targ & \qw & \qw & \qw & \gate{H} &  \qw \\
    & \gate{H} & \qw & \qw & \targ & \ctrl{-1} & \qw & \ctrl{-1}& \targ & \qw & \qw & \gate{H} &  \qw \\
    & \gate{H} & \qw & \targtail & \ctrl{-1} & \qw & \qw & \qw & \ctrl{-1} & \targtail & \qw & \gate{H} & \qw \\
    && \hspace{0.8cm} \stackrel{\ }{\text{\begin{rotate}{85}$\ddots$\end{rotate}}} &&&&&& \hspace{1.8cm} \stackrel{\ }{\text{\begin{rotate}{-10}$\ddots$\end{rotate}}}\\
    & \gate{H} & \ctrlstick & \qw & \qw & \qw & \qw & \qw & \qw & \qw & \ctrlstick & \gate{H} & \qw \\
  }
  \end{gathered}
  \]
  \caption{
  Circuit equivalance that allows a transition from the spoke-hub geometry to  linear nearest-neighbor geometry
  for implementation of  wavefunction symmetrization.
  }
  \label{fig:CNOTcircuitIdentity}
\end{figure}
It is possible to modify the connectivity required for the symmetrization of Fig.~\ref{fig:symmetrizedExpcircuit} to be more conducive to the {\tt Poughkeepsie} connectivity.  Using the circuit equality of Fig.\ref{fig:CNOTcircuitIdentity}, the spoke-hub geometry can be exchanged for a linear, nearest-neighbor connectivity with open boundary conditions.
Upon this change, it is no longer the maximum single-qubit connectivity that limits the size of the symmetrized exponential that can be prepared on the device, but rather the longest qubit string connected minimally by uni-directional CNOTs that can be weaved throughout the chip.
On {\tt Poughkeepsie}, with its interconnected-ring geometry, this indicates that it is possible (by available connectivity) to prepare a symmetrized exponential digitized onto 18 qubits---a size greatly exceeding necessary digitization for a single site of the scalar field.
Implementing this modified CNOT structure requires $(2n_Q-3)$ nearest-neighbor CNOTs to implement the symmetrization rather than the $(n_Q-1)$ CNOTs required for the spoke-hub configuration and is thus twice as susceptible to the dominant quantum noise arising with the implementation of 2-qubit entangling gates.

%%%%%%%%%%%%%%%%%%%%%%%%%%%%%%%%%%%%%%%%%%%
\section{Measurement-Error Correction}
\label{app:MeasurementCorrection}

\begin{figure}[!ht]
  \includegraphics[width = 0.44 \textwidth]{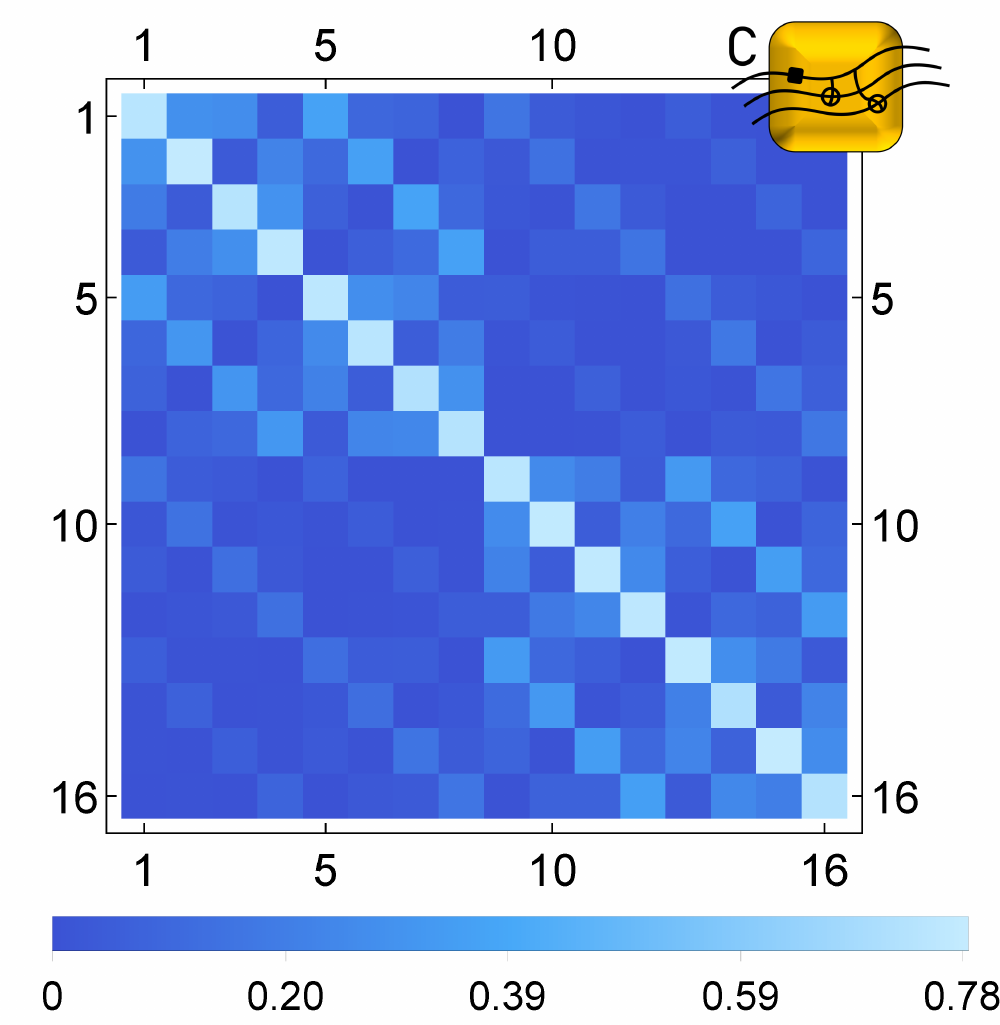}
  \includegraphics[width = 0.44 \textwidth]{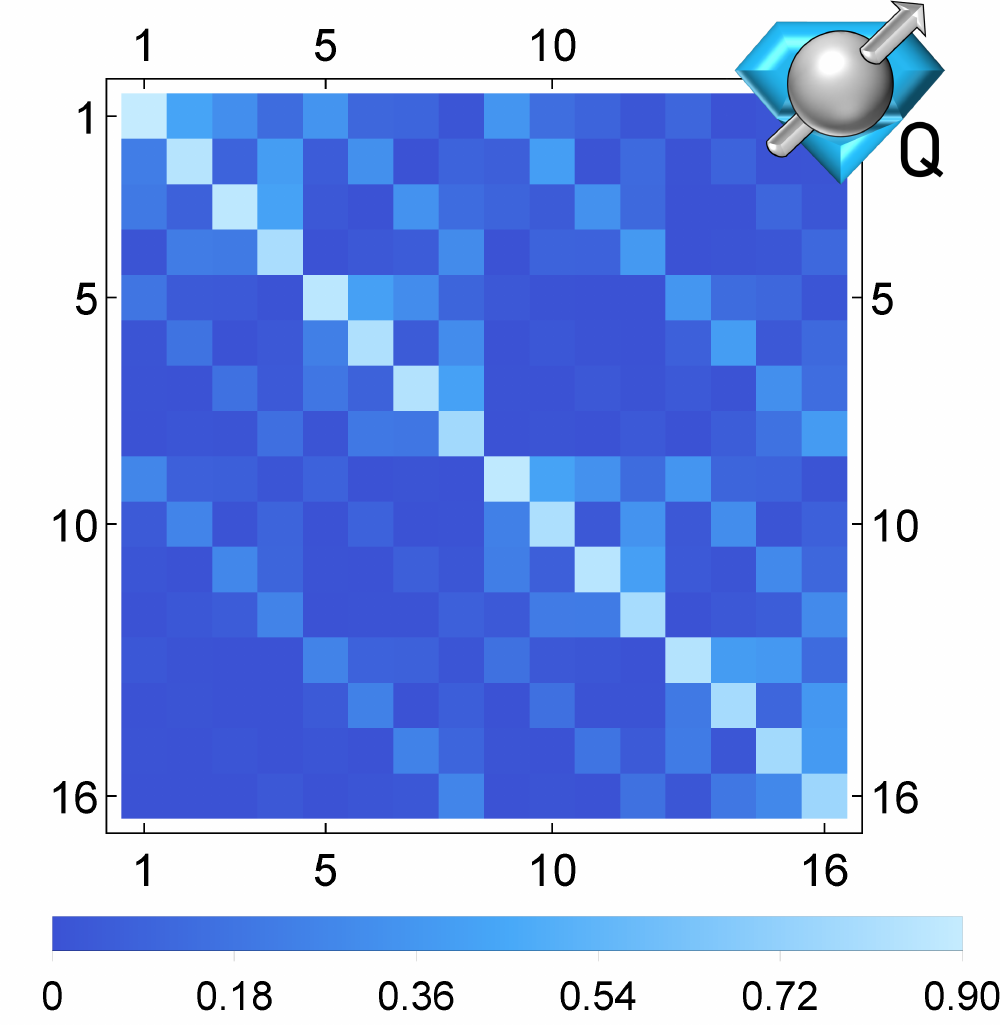}\\
  \includegraphics[width = 0.5\textwidth]{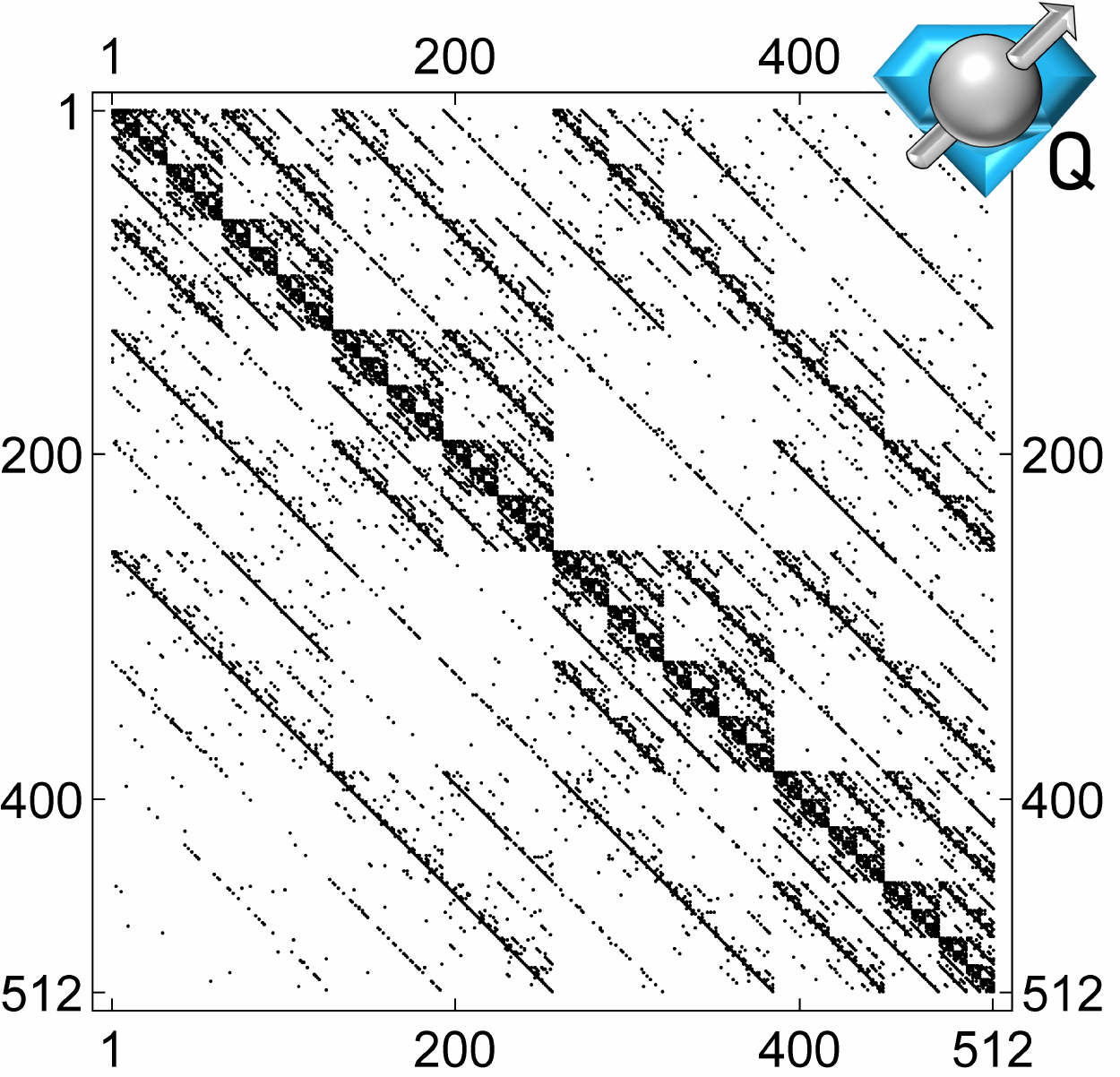}
  \caption{Measurement-error  calibration matrices extracted from the simulator (upper-left)
  with the extracted {\tt Poughkeepsie} noise model and from the {\tt Poughkeepsie} quantum hardware (upper-right) for 4-qubit systems.
  Lower panel shows the measurement-error calibration matrix for the 9-qubit system utilized for 3-site state
  preparation with a monochrome color scheme to support clear visualization of the 512-dimensional space.
  }
  \label{fig:PouCalibrationMatrices}
\end{figure}

Some errors of the hardware implementation can be attributed to the preparation of
$|{\bf 0}\rangle^{\otimes n_Q}$  state,  and the ability to faithfully apply projective measurements to the resulting state.
Such errors can be seen by simply preparing $|{\bf 0}\rangle^{\otimes n_Q}$
and immediately measuring in the computational basis.
Ideally, all measurements are recorded to be in the $|0\rangle$ basis state.
However, due to the errors expressed above, this is an insufficient assumption for reliable computation on current devices.
For example, performing this calibration measurement on four qubits of {\tt Poughkeepsie}
used to digitize the symmetrized exponential onto 16 states results in only $\sim 90\%$ of the
measurements assigned to the $|0\rangle$ state,
as can be seen on the diagonal of Fig.~\ref{fig:PouCalibrationMatrices}.
States deviating by a single qubit flip ($|1\rangle, |2\rangle, |4\rangle,$ and $|8\rangle$) tend to account for the next largest contribution,
at a few percent each.
While it is not possible to fully isolate the role of state preparation errors from the role of assignment errors in the measurement, we choose to interpret the above observations as due to measurement error and use this calibration to correct future measurements on the device.
If the measurement errors are stochastic, symmetric between the basis states (i.e., the probability of measuring $|1\rangle$ when the quantum state is in the $|0\rangle$ state is equal to the probability of measuring a $|0\rangle$ when the quantum state is in the $|1\rangle$ state) and did not depend on multi-qubit correlations, the above single-circuit calibration measurement would be sufficient to reconstruct the necessary measurement calibration matrix.
However, it was observed in this work that such assumptions do not lead to sufficient measurement-error correction.
Instead, the procedure promoted by IBM is used,
in which $2^{N_Q}$ circuits are prepared in sequentially-increasing basis states through application of
single-qubit Pauli $\hat{X}$ gates and then measured in the computational basis.
This informs the complete structure of the noisy measurement-error matrix
\begin{equation}
  M = \begin{pmatrix}
    \begin{pmatrix}
      \\
      \\
      \vec{m}_{|0\rangle}
      \\
      \\
      \\
    \end{pmatrix}
    &
    \begin{pmatrix}
      \\
      \\
      \vec{m}_{|1\rangle}
      \\
      \\
      \\
    \end{pmatrix}
    & \cdots
    &
    \begin{pmatrix}
      \\
      \\
      \vec{m}_{|2^n\rangle}
      \\
      \\
      \\
    \end{pmatrix}
  \end{pmatrix}
\end{equation}
where the vector $\vec{m}_{|x\rangle}$ is the vector of measurement probabilities observed from the device after preparation of the $|x\rangle$ computational basis state.
In order to correct future measurements after this calibration of every basis state, vectors of measured probabilities may simply be acted upon by $M^{-1}$.
However, the resulting vector of measurements, $M^{-1} \vec{m}_\psi$, may contain negative probabilities (when a basis state does not appear frequently in the sampling) and is not constrained to unit probability.
To maintain physical properties of the measurement-error corrected vector of probabilities,
$\vec{m}_{\rm mc}$, one can instead implement the inversion through a minimization procedure for the following condition,
\begin{equation}
  \vec{m}_{\psi} - M \vec{m}_{\rm mc} = \vec{0}
\end{equation}
subject to the constraints that all elements of $\vec{m}_{\rm mc}$ are positive and sum to unity. The entirety of this procedure is described in the IBM Q Experience documentation and implemented in the {\tt ignis} tools for noise characterization and error correction.
While this procedure for 16 or 512 circuits to characterize $M$ for $n_Q = 4$ or $9$ was reasonable,
the exponentially-growing number of circuits needed to explore the state preparation and measurement
errors of each basis state may become unrealistic.
At some threshold, a characterization of the measurement-error correction in this way cannot be implemented on a time scale commensurate with the temporal fluctuations of the device.
Reducing the resource cost for measurement calibration using potentially architecture-specific knowledge of device properties will be essential for working with larger numbers of qubits going forward.

\begin{figure}[!ht]
  \includegraphics[width = 0.8\textwidth]{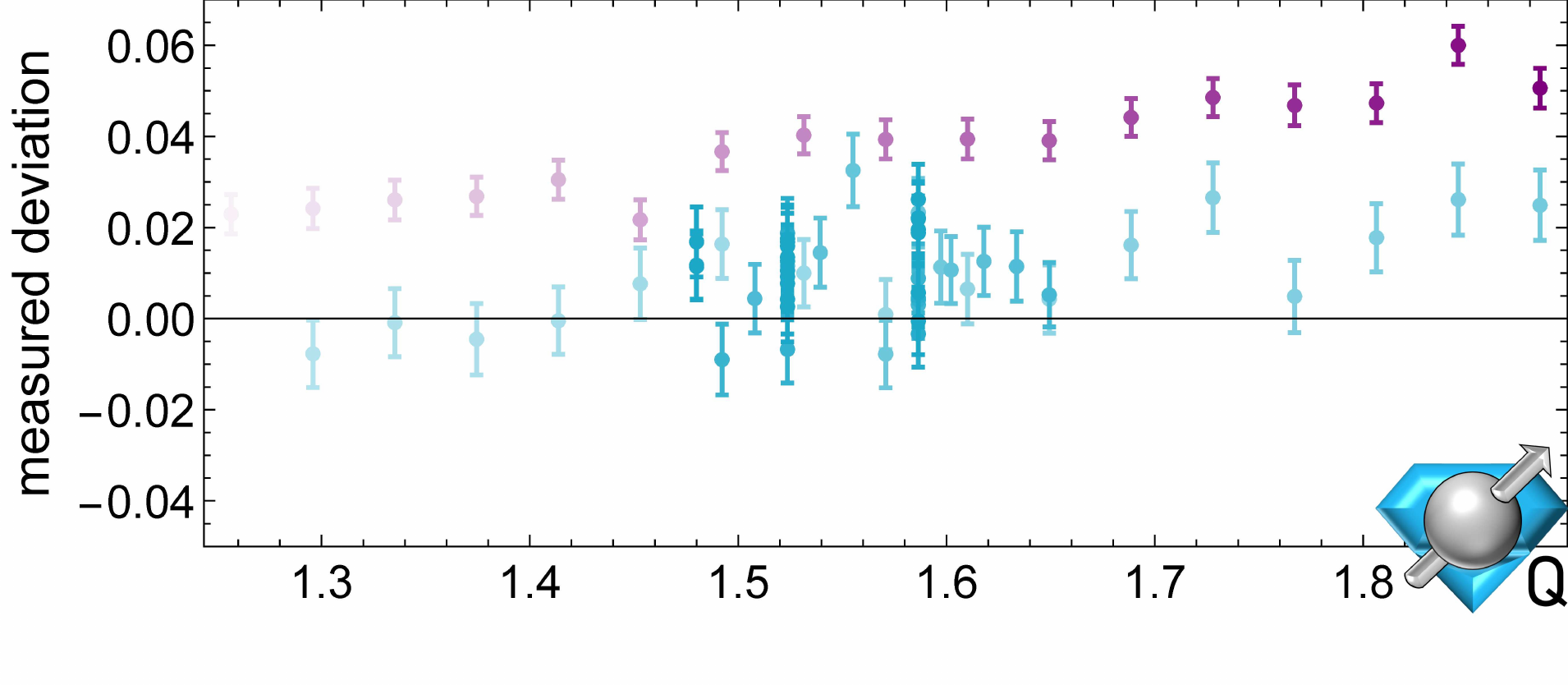} \\
  \includegraphics[width = 0.8\textwidth]{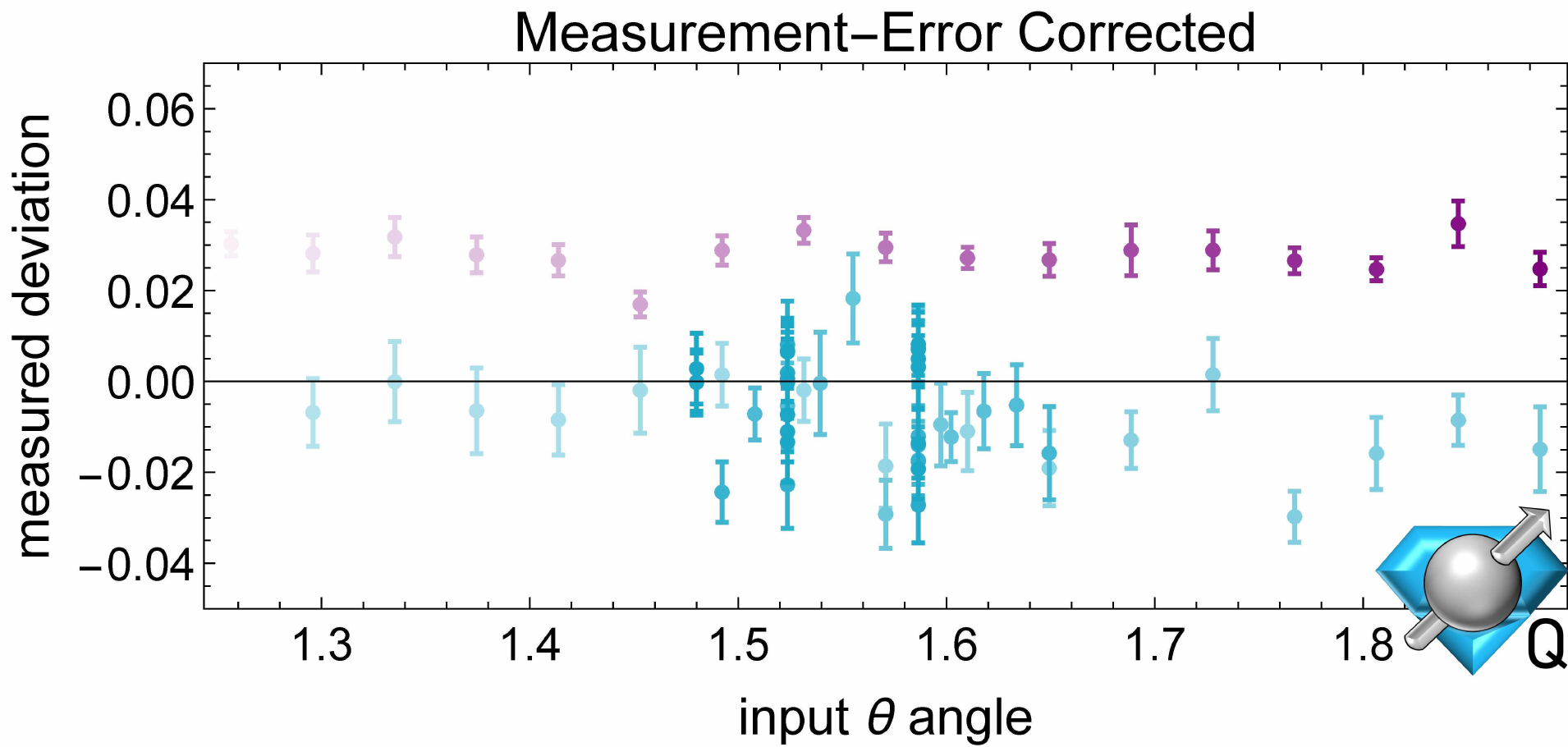}
  \caption{
  Deviations between the input angle
  and the effective angle applied to the first argument of $U_3(\theta, 0, \pi)$ as a function of input angle.
  The upper panel shows the raw results from {\tt Poughkeepsie}, while the lower panel shows the measurement-error
  corrected results.
  The purple points have the $U_3$ implemented in vacuum,
  while the blue points have the $U_3$ implemented in the medium of the half-space exponential on the three initial qubits.
  The saturation of points and error bars indicate their order of implementation with early-to-late times shown as light-to-dark color.
  }
  \label{fig:measurementcorrection}
\end{figure}
The importance of measurement-error correction in this work can be emphasized in the data of Fig.~\ref{fig:measurementcorrection}.
The data in this figure was acquired over a range of input angles, $\theta$, of the $U_3(\theta,0,\pi)$ gate,
both in vacuum(purple) and in medium(blue), defined by the 3-qubit half-space exponential.
The vertical axis of this figure shows the deviation between the measured angle (as informed by the relative probabilities in the dominant two states---$|0\rangle, |8\rangle$ in vacuum and $|7\rangle, |15\rangle$ in medium) and the input angle.
In the top panel of raw results,
it would be tempting to conclude that there exists a $\theta$-dependence to the angle deviation
and thus suggest a $\theta$-dependence to any procedure designed to improve the symmetry of the
wavefunction when copied into the second half of the Hilbert space.
However, upon constrained-inversion measurement-error correction with 16 calibration circuits measuring intended
computational basis states, the $\theta$-dependence of the angle deviation is effectively removed and
angle deviations become roughly $\theta$- and time-independent.
The darkness of each data point in Fig.~\ref{fig:measurementcorrection} indicates the order of its implementation in time.
From light to dark, the results in vacuum were produced over a period of 65 minutes.
Also from light to dark, the results for the rotation in medium were produced over a period of 720 minutes.
While the on-line calibration techniques employed in the main text provide some robustness to temporal fluctuations
of the device, we find that the stability of {\tt Poughkeepsie} over these time scales was
essential to the quality of attainable probability distributions.

%%%%%%%%%%%%%%%%%%%%%%%%%%%%%%%%%%%%%%%%%%%
\section{Resource Icons}
\label{app:ResourceIcons}

\begin{figure}[!ht]
  \includegraphics[width = 0.95\textwidth]{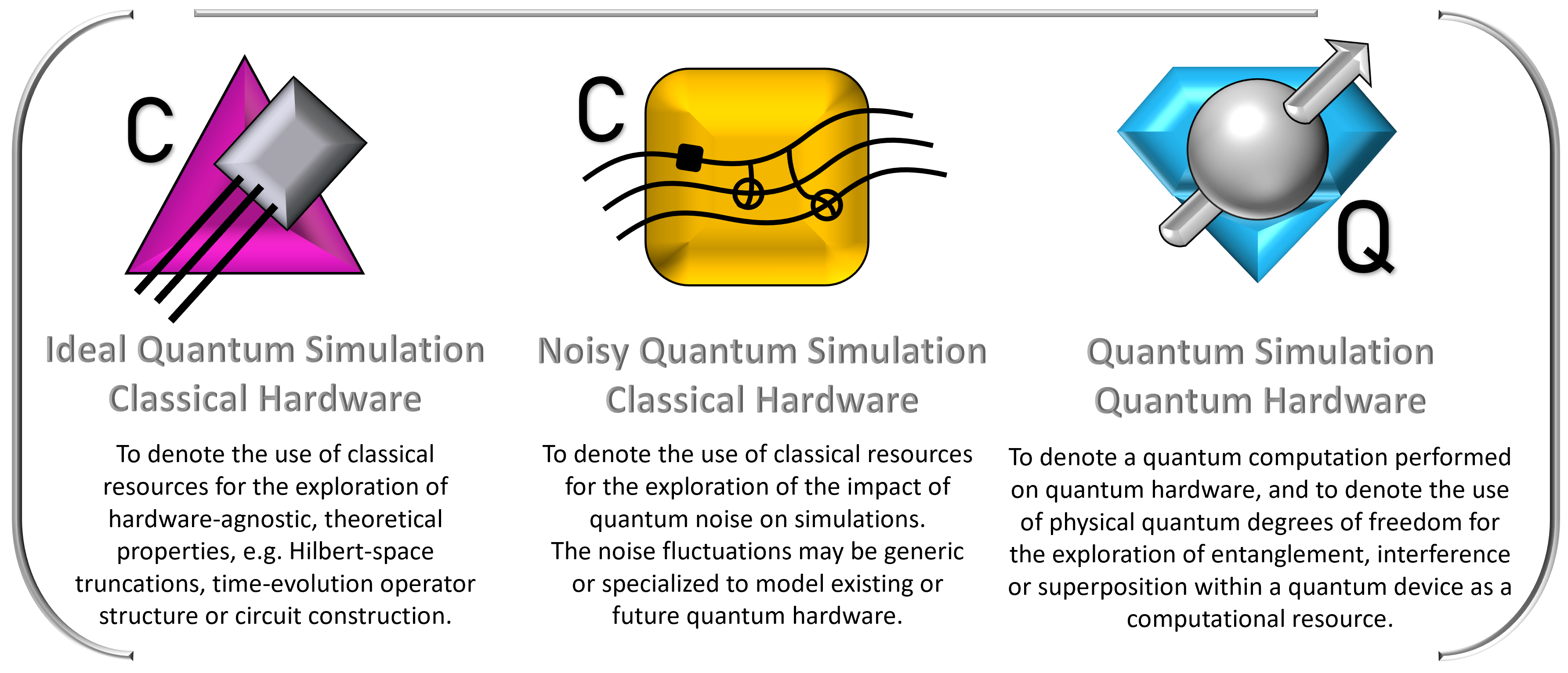}
  \caption{ (Color Online)
  Three icons delineating the nature of  classical and  quantum simulations.
  The left (purple triangle) icon indicates that classical computing resources were used to generate the results of a calculation.
  This includes the use of ideal (noiseless) quantum simulators, or standard manipulation packages
  such as {\tt Mathematica} or {\tt MatLab}.
  The center (yellow square) icon indicates that classical computing resources were used to produce
  results from a noisy quantum simulator.
  The right (blue diamond) icon indicates that results were obtained using a quantum computing device.
  }
  \label{fig:icons}
\end{figure}

With the increasing number of simulations of systems on quantum devices,
on noisy quantum simulators and on ideal simulators,  we anticipate that readily discernible icons will be helpful
in conveying the nature of present and future simulations.
We have created three such icons, as shown in Fig.~\ref{fig:icons},
 and have used them  in presenting our work as a trial of their utility.
The types of calculations and simulations they are each intended to identify are detailed in Fig.~\ref{fig:icons} and its caption.
These icons can be obtained from the authors, or from the website:
\url{https://sites.google.com/uw.edu/UWFundamentalQISQC/icons}.

%%%%%%%%%%%%%%%%%%%%%%%%%%%%%%%%%%%%%%%%%%%
\section{Data Tables}
In this appendix,  we provide the numerical values of simulation results and theoretical calculations
appearing in the figures of this paper.
The 3-site results associated with the lower panel of Fig.~\ref{fig:2site}
and the lower panel of Fig.~\ref{fig:measurementcorrection}
are provided in supplementary files.

\begin{table}[!ht]
  \centering
  \begin{tabular}{c|c|c}
  \multicolumn{1}{c}{}& \multicolumn{1}{c}{$n_Q = 3$} & \\
    \hline
    \hline
    $|x\rangle $ & $\psi^2(x)_{\rm th}$ & $\psi^2(x)$ \\
    \hline
    \hline
0&0.001&0.001(1)\\
1&0.008&0.011(3)\\
2&0.059&0.068(8)\\
3&0.432&0.429(16)\\
4&0.432&0.419(16)\\
5&0.059&0.066(8)\\
6&0.008&0.006(2)\\
7&0.001&0.000(0)\\
    \hline
    \hline
  \end{tabular}
  \hspace{1cm}
  \begin{tabular}{c|c|c}
  \multicolumn{1}{c}{}& \multicolumn{1}{c}{$n_Q = 4$} & \\
    \hline
    \hline
    $|x\rangle $ & $\psi^2(x)_{\rm th}$ & $\psi^2(x)$ \\
    \hline
    \hline
0&0.0003&0.001(1)\\
1&0.0008&0.001(1)\\
2&0.0021&0.001(1)\\
3&0.0058&0.005(2)\\
4&0.0157&0.013(4)\\
5&0.0428&0.036(6)\\
6&0.1163&0.133(11)\\
7&0.3162&0.313(15)\\
\end{tabular}
\begin{tabular}{c|c|c}
\multicolumn{3}{c}{}\\
\multicolumn{3}{c}{}\\
8&0.3162&0.318(15)\\
9&0.1163&0.118(10)\\
10&0.0428&0.034(6)\\
11&0.0157&0.018(4)\\
12&0.0058&0.007(3)\\
13&0.0021&0.002(1)\\
14&0.0008&0.000(0)\\
15&0.0003&0.000(0)\\
    \hline
    \hline
  \end{tabular} \\ \vspace{0.4cm}
  \begin{tabular}{c|c|c}
  \multicolumn{1}{c}{}& \multicolumn{1}{c}{$n_Q = 5$} & \\
    \hline
    \hline
    $|x\rangle $ & $\psi^2(x)_{\rm th}$ & $\psi^2(x)$ \\
    \hline
    \hline
0&0.0001&0.000(0)\\
1&0.0002&0.000(0)\\
2&0.0003&0.000(0)\\
3&0.0005&0.000(0)\\
4&0.0008&0.000(0)\\
5&0.0013&0.002(1)\\
6&0.0022&0.003(2)\\
7&0.0036&0.005(2)\\
\end{tabular}
\begin{tabular}{c|c|c}
\multicolumn{3}{c}{}\\
\multicolumn{3}{c}{}\\
8&0.0059&0.012(3)\\
9&0.0098&0.010(3)\\
10&0.0162&0.015(4)\\
11&0.0266&0.032(6)\\
12&0.0439&0.038(6)\\
13&0.0724&0.076(8)\\
14&0.1194&0.110(10)\\
15&0.1968&0.201(13)\\
\end{tabular}
\begin{tabular}{c|c|c}
\multicolumn{3}{c}{}\\
\multicolumn{3}{c}{}\\
16&0.1968&0.193(12)\\
17&0.1194&0.112(10)\\
18&0.0724&0.066(8)\\
19&0.0439&0.049(7)\\
20&0.0266&0.029(5)\\
21&0.0162&0.018(4)\\
22&0.0098&0.012(3)\\
23&0.0059&0.010(3)\\
\end{tabular}
\begin{tabular}{c|c|c}
\multicolumn{3}{c}{}\\
\multicolumn{3}{c}{}\\
24&0.0036&0.003(2)\\
25&0.0022&0.001(1)\\
26&0.0013&0.001(1)\\
27&0.0008&0.001(1)\\
28&0.0005&0.000(0)\\
29&0.0003&0.001(1)\\
30&0.0002&0.000(0)\\
31&0.0001&0.000(0)\\
    \hline
    \hline
  \end{tabular}
  \caption{
  Numerical values of the probabilities appearing in the top row of Fig.~\ref{fig:ExpSim}.
  The second column is the probability calculated from the exact exponential wavefunction,
  while the third is the probability determined from $10^3$ samples from an ideal implementation
  of the circuit in Fig.~\ref{fig:symmetrizedExpcircuit}.
  }
  \label{tab:cleansampledEXP}
\end{table}
\begin{table}[!ht]
  \centering
  \begin{tabular}{c|c|c|c|c}
    \multicolumn{2}{c}{}& \multicolumn{2}{c}{$n_Q = 3$} & \\
    \hline
    \hline
    $|x\rangle $ & $\psi^2(x)_{\rm th}$ & $\sigma_\theta = 10^{-1}$ & $ \sigma_\theta = 0.3 \times 10^{-1} $ & $\sigma_\theta = 10^{-2}$ \\
    \hline
    \hline
    0&0.001&0.139(11)&0.044(6)&0.012(3)\\
1&0.008&0.151(11)&0.056(7)&0.018(4)\\
2&0.059&0.113(10)&0.141(11)&0.069(8)\\
3&0.432&0.113(10)&0.256(14)&0.405(16)\\
4&0.432&0.106(10)&0.256(14)&0.412(16)\\
5&0.059&0.116(10)&0.133(11)&0.065(8)\\
6&0.008&0.143(11)&0.062(8)&0.014(4)\\
7&0.001&0.119(10)&0.052(7)&0.005(2)\\
    \hline
    \hline
  \end{tabular}
\caption{
Numerical values of the probabilities appearing in the bottom left panel of Fig.~\ref{fig:ExpSim}.
The second column is the probability calculated from the exact exponential wavefunction,
while the third through fifth is the probability determined from $10^3$ samples from an implementation
of the circuit in Fig.~\ref{fig:symmetrizedExpcircuit} with Gaussian gate error
characterized by standard deviation $\sigma_\theta$.
}
\label{tab:gausssampledEXPnq3}
\end{table}
\begin{table}
  \centering
  \begin{tabular}{c|c|c|c|c}
    \multicolumn{2}{c}{}& \multicolumn{2}{c}{$n_Q = 4$} & \\
    \hline
    \hline
    $|x\rangle $ & $\psi^2(x)_{\rm th}$ & $\sigma_\theta = 10^{-1}$ & $ \sigma_\theta = 0.3 \times 10^{-1} $ & $\sigma_\theta = 10^{-2}$ \\
    \hline
    \hline
0&0.0003&0.061(8)&0.021(5)&0.005(2)\\
1&0.0008&0.060(8)&0.017(4)&0.003(2)\\
2&0.0021&0.066(8)&0.032(6)&0.005(2)\\
3&0.0058&0.060(8)&0.028(5)&0.008(3)\\
4&0.0157&0.059(7)&0.053(7)&0.021(5)\\
5&0.0428&0.063(8)&0.067(8)&0.051(7)\\
6&0.1163&0.064(8)&0.100(9)&0.117(10)\\
7&0.3162&0.065(8)&0.160(12)&0.272(14)\\
\end{tabular}
\begin{tabular}{c|c|c|c|c}
\multicolumn{5}{c}{}\\
\multicolumn{5}{c}{}\\
8&0.3162&0.066(8)&0.153(11)&0.283(14)\\
9&0.1163&0.052(7)&0.102(10)&0.134(11)\\
10&0.0428&0.058(7)&0.077(8)&0.052(7)\\
11&0.0157&0.067(8)&0.075(8)&0.022(5)\\
12&0.0058&0.060(8)&0.045(7)&0.014(4)\\
13&0.0021&0.061(8)&0.029(5)&0.006(2)\\
14&0.0008&0.066(8)&0.023(5)&0.003(2)\\
15&0.0003&0.072(8)&0.018(4)&0.004(2)\\
    \hline
    \hline
  \end{tabular}
\caption{
Numerical values of the probabilities appearing in the bottom middle panel of Fig.~\ref{fig:ExpSim}.
The second column is the probability calculated from the exact exponential wavefunction,
while the third through fifth is the probability determined from $10^3$ samples from
an implementation of the circuit in Fig.~\ref{fig:symmetrizedExpcircuit} with Gaussian gate error
characterized by standard deviation $\sigma_\theta$.
}
\label{tab:gausssampledEXPnq4}
\end{table}
\begin{table}
  \centering
  \begin{tabular}{c|c|c|c|c}
    \multicolumn{2}{c}{}& \multicolumn{2}{c}{$n_Q = 5$} & \\
    \hline
    \hline
    $|x\rangle $ & $\psi^2(x)_{\rm th}$ & $\sigma_\theta = 10^{-1}$ & $ \sigma_\theta = 0.3 \times 10^{-1} $ & $\sigma_\theta = 10^{-2}$ \\
    \hline
    \hline
0&0.0001&0.032(6)&0.012(3)&0.004(2)\\
1&0.0002&0.038(6)&0.012(3)&0.000(0)\\
2&0.0003&0.030(5)&0.010(3)&0.003(2)\\
3&0.0005&0.035(6)&0.006(2)&0.000(0)\\
4&0.0008&0.032(6)&0.019(4)&0.002(1)\\
5&0.0013&0.021(5)&0.008(3)&0.002(1)\\
6&0.0022&0.034(6)&0.020(4)&0.008(3)\\
7&0.0036&0.033(6)&0.018(4)&0.004(2)\\
8&0.0059&0.028(5)&0.033(6)&0.012(3)\\
9&0.0098&0.031(5)&0.033(6)&0.011(3)\\
10&0.0162&0.026(5)&0.035(6)&0.013(4)\\
11&0.0266&0.032(6)&0.033(6)&0.031(5)\\
12&0.0439&0.035(6)&0.058(7)&0.039(6)\\
13&0.0724&0.025(5)&0.054(7)&0.057(7)\\
14&0.1194&0.023(5)&0.077(8)&0.111(10)\\
15&0.1968&0.033(6)&0.066(8)&0.189(12)\\
\end{tabular}
\begin{tabular}{c|c|c|c|c}
\multicolumn{5}{c}{}\\
\multicolumn{5}{c}{}\\
16&0.1968&0.031(5)&0.065(8)&0.167(12)\\
17&0.1194&0.027(5)&0.086(9)&0.128(11)\\
18&0.0724&0.028(5)&0.056(7)&0.068(8)\\
19&0.0439&0.031(5)&0.061(8)&0.048(7)\\
20&0.0266&0.032(6)&0.028(5)&0.030(5)\\
21&0.0162&0.030(5)&0.028(5)&0.024(5)\\
22&0.0098&0.026(5)&0.029(5)&0.012(3)\\
23&0.0059&0.031(5)&0.032(6)&0.016(4)\\
24&0.0036&0.045(7)&0.020(4)&0.005(2)\\
25&0.0022&0.034(6)&0.020(4)&0.004(2)\\
26&0.0013&0.028(5)&0.018(4)&0.005(2)\\
27&0.0008&0.033(6)&0.018(4)&0.002(1)\\
28&0.0005&0.031(5)&0.014(4)&0.000(0)\\
29&0.0003&0.037(6)&0.011(3)&0.000(0)\\
30&0.0002&0.034(6)&0.011(3)&0.002(1)\\
31&0.0001&0.034(6)&0.009(3)&0.003(2)\\
    \hline
    \hline
  \end{tabular}
\caption{
Numerical values of the probabilities appearing in the bottom right panel of Fig.~\ref{fig:ExpSim}.
The second column is the probability calculated from the exact exponential wavefunction,
while the third through fifth is the probability determined from $10^3$ samples from an implementation
of the circuit in Fig.~\ref{fig:symmetrizedExpcircuit} with Gaussian gate error characterized by standard deviation $\sigma_\theta$.
}
\label{tab:gausssampledEXPnq5}
\end{table}

\begin{table}
  \centering
  \setlength\tabcolsep{0.6em}
  \begin{tabular}{c|c|c|c|c|c|c|c|c}
  \hline
  \hline
  $\theta$ & $|0\rangle_{\rm vac}$ & $|8\rangle_{\rm vac}$ & $|7\rangle_{\rm med}$ & $|15\rangle_{\rm med}$ & $|0\rangle_{\rm vac}^{\rm mc}$ & $|8\rangle_{\rm vac}^{\rm mc}$ & $|7\rangle_{\rm med}^{\rm mc}$ & $|15\rangle_{\rm med}^{\rm mc}$ \\
  \hline
  \hline
1.26&0.632(5)&0.303(5)&0.277(5)&0.155(4)&0.678(5)&0.316(5)&0.337(7)&0.187(3)\\
1.30&0.618(5)&0.320(5)&0.276(5)&0.164(4)&0.662(5)&0.335(6)&0.337(6)&0.198(4)\\
1.34&0.602(5)&0.336(5)&0.266(5)&0.166(4)&0.645(5)&0.354(5)&0.324(6)&0.203(6)\\
1.37&0.585(6)&0.353(5)&0.252(5)&0.173(4)&0.623(4)&0.375(5)&0.301(6)&0.212(5)\\
1.41&0.566(6)&0.365(5)&0.251(5)&0.184(4)&0.602(5)&0.388(5)&0.302(6)&0.228(5)\\
1.45&0.543(6)&0.392(5)&0.233(5)&0.178(4)&0.574(4)&0.423(5)&0.277(6)&0.220(6)\\
1.49&0.539(6)&0.398(5)&0.237(5)&0.189(4)&0.568(5)&0.429(4)&0.281(6)&0.236(6)\\
1.53&0.524(6)&0.413(6)&0.227(5)&0.202(4)&0.552(6)&0.446(6)&0.268(5)&0.252(6)\\
1.57&0.506(6)&0.432(6)&0.213(5)&0.213(5)&0.530(5)&0.468(5)&0.249(6)&0.266(6)\\
1.61&0.488(6)&0.451(6)&0.213(5)&0.224(5)&0.507(5)&0.492(4)&0.247(6)&0.282(6)\\
1.65&0.471(6)&0.471(6)&0.200(4)&0.230(5)&0.486(6)&0.513(6)&0.231(6)&0.291(6)\\
1.69&0.451(6)&0.479(6)&0.202(4)&0.240(5)&0.466(6)&0.526(5)&0.229(5)&0.302(7)\\
1.73&0.441(6)&0.497(6)&0.194(4)&0.239(5)&0.451(4)&0.547(4)&0.221(5)&0.303(7)\\
1.77&0.405(5)&0.498(6)&0.174(4)&0.253(5)&0.412(6)&0.551(7)&0.193(5)&0.322(5)\\
1.81&0.394(5)&0.523(6)&0.176(4)&0.263(5)&0.400(6)&0.578(5)&0.197(5)&0.335(6)\\
1.85&0.397(5)&0.542(6)&0.168(4)&0.263(5)&0.400(6)&0.599(6)&0.185(5)&0.338(6)\\
1.88&0.368(5)&0.565(6)&0.160(4)&0.273(5)&0.366(7)&0.628(6)&0.173(7)&0.351(7)\\
\hline
\hline
\end{tabular}
\caption{
Numerical values of the probabilities appearing in Fig.~\ref{fig:HadamardTuning} of the main text.
The first column is the symmetrization angle input as the $U_3(\theta, 0, \pi)$ implementation of the Hadamard.
The next 4 columns are the probabilities in the two most populous states after implementation of this operator in vacuum
and in medium.
The final four columns are the same quantities after constrained measurement-error
correction (see Appendix~\ref{app:MeasurementCorrection}).}
\label{tab:HadamardTuningdatatable}
\end{table}
\begin{table}
  \centering
  \begin{tabular}{c|c|c|c|c}
  \multicolumn{5}{c}{raw} \\
  \hline
  \hline
  $ |x\rangle$ & $r = 1$ & $r = 3$ & $r = 5$ & $r = 7$ \\
  \hline
  \hline
  0&0.019(2)&0.023(2)&0.018(1)&0.022(2)\\
1&0.019(2)&0.020(2)&0.017(1)&0.019(2)\\
2&0.023(2)&0.027(2)&0.030(2)&0.034(2)\\
3&0.028(2)&0.030(2)&0.024(2)&0.024(2)\\
4&0.048(2)&0.049(2)&0.057(3)&0.069(3)\\
5&0.068(3)&0.064(3)&0.054(3)&0.044(2)\\
6&0.140(4)&0.138(4)&0.153(4)&0.144(4)\\
7&0.201(4)&0.174(4)&0.140(4)&0.108(3)\\
8&0.221(5)&0.196(4)&0.199(4)&0.199(4)\\
9&0.098(3)&0.107(3)&0.119(4)&0.116(4)\\
10&0.065(3)&0.072(3)&0.083(3)&0.080(3)\\
11&0.027(2)&0.044(2)&0.048(2)&0.063(3)\\
12&0.015(1)&0.021(2)&0.018(1)&0.028(2)\\
13&0.010(1)&0.014(1)&0.014(1)&0.016(1)\\
14&0.009(1)&0.012(1)&0.015(1)&0.015(1)\\
15&0.008(1)&0.011(1)&0.012(1)&0.018(1)\\
  \hline
  \hline
  \end{tabular}
  \begin{tabular}{c|c|c|c|c}
  \multicolumn{5}{c}{measurement-error corrected} \\
    \hline
  \hline
  $ |x\rangle$ & $r = 1$ & $r = 3$ & $r = 5$ & $r = 7$ \\
  \hline
  \hline
  0&0.001(1)&0.006(2)&0.001(1)&0.003(2)\\
1&0.006(1)&0.006(2)&0.003(1)&0.007(2)\\
2&0.008(2)&0.012(2)&0.016(2)&0.020(3)\\
3&0.015(2)&0.019(2)&0.014(3)&0.015(2)\\
4&0.040(3)&0.041(2)&0.050(3)&0.064(3)\\
5&0.065(4)&0.064(3)&0.052(2)&0.041(2)\\
6&0.146(5)&0.144(5)&0.165(6)&0.158(4)\\
7&0.255(4)&0.220(6)&0.176(4)&0.134(4)\\
8&0.237(5)&0.204(6)&0.208(5)&0.210(4)\\
9&0.112(3)&0.122(5)&0.136(3)&0.132(4)\\
10&0.066(3)&0.073(2)&0.085(3)&0.080(3)\\
11&0.028(2)&0.051(2)&0.054(2)&0.072(3)\\
12&0.012(1)&0.018(2)&0.013(2)&0.025(2)\\
13&0.008(1)&0.013(2)&0.013(2)&0.015(2)\\
14&0.002(1)&0.007(3)&0.010(2)&0.009(2)\\
15&0.000(0)&0.002(2)&0.006(2)&0.015(2)\\
  \hline
  \hline
  \end{tabular}
  \begin{tabular}{c|c}
  \multicolumn{2}{c}{extrapolated} \\
    \hline
  \hline
  $ |x\rangle$ & prob. \\
  \hline
  \hline
0&-0.001(2)\\
1&0.006(2)\\
2&0.007(3)\\
3&0.014(4)\\
4&0.039(5)\\
5&0.066(6)\\
6&0.147(8)\\
7&0.273(7)\\
8&0.253(9)\\
9&0.106(6)\\
10&0.063(5)\\
11&0.016(4)\\
12&0.008(2)\\
13&0.005(2)\\
14&-0.000(2)\\
15&-0.001(1)\\
  \hline
  \hline
  \end{tabular}
  \caption{
  Numerical values for results presented in Fig.~\ref{fig:CNOTextrapolation} of the main text.
  }
  \label{tab:CNOTexp1sitedatatable}
\end{table}

\begin{table}
  \centering
  \begin{tabular}{c|c|c}
  \multicolumn{3}{c}{$\theta = 1.52$} \\
  \hline
  \hline
  $r$ & $A^{(2)}_{CE}$ & $A^{(2)}_{SE}$ \\
  \hline
1&0.065(9)&0.067(6)\\
3&0.061(7)&0.044(5)\\
5&0.056(8)&0.057(5)\\
  \hline
  \hline
  \multicolumn{3}{c}{}\\
  \end{tabular}
  \hspace{1cm}
  \begin{tabular}{c|c|c}
  \multicolumn{3}{c}{$\theta = 1.52$} \\
  \hline
  \hline
  $r$ & $A^{(2)}_{CE}$ & $A^{(2)}_{SE}$ \\
  \hline
1&0.016(4)&0.042(6)\\
3&0.031(8)&0.032(7)\\
5&0.031(8)&0.056(5)\\
7&0.035(7)&0.092(5)\\
  \hline
  \hline
  \end{tabular}
  \hspace{1cm}
  \begin{tabular}{c|c|c}
  \multicolumn{3}{c}{$\theta = 1.59$} \\
  \hline
  \hline
  $r$ & $A^{(2)}_{CE}$ & $A^{(2)}_{SE}$ \\
  \hline
  \hline
1&0.026(7)&0.032(5)\\
3&0.030(8)&0.028(5)\\
5&0.027(6)&0.056(7)\\
7&0.031(7)&0.079(6)\\
  \hline
  \hline
  \end{tabular}
  \caption{Numerical values of the asymmetries appearing in Fig.~\ref{fig:CNOTAsymmetry} of the main text.}
  \label{tab:1siteAsymmetrydatatable}
\end{table}

\begin{table}
  \centering
  \begin{tabular}{c|c}
  \hline
  \hline
  $\theta$ & $A^{(2)}_{CE}$ \\
  \hline
  \hline
1.26&0.172(8)\\
1.30&0.151(9)\\
1.34&0.135(7)\\
1.37&0.102(8)\\
1.41&0.095(8)\\
1.45&0.060(8)\\
1.49&0.056(8)\\
1.53&0.029(7)\\
1.57&0.032(6)\\
1.61&0.039(8)\\
1.65&0.074(9)\\
1.69&0.088(7)\\
1.73&0.103(8)\\
1.77&0.146(8)\\
1.81&0.164(9)\\
1.85&0.175(8)\\
1.88&0.202(8)\\
\hline
\hline
  \end{tabular}
  \hspace{1cm}
  \begin{tabular}{c|c|c}
    \hline
    \hline
    $\theta$ & $ A^{(2)}_{CE}$ & $ A^{(2)}_{SE}$ \\
    \hline
    \hline
1.57&0.036(7)&0.067(18)\\
1.59&0.020(6)&0.063(21)\\
1.56&0.033(8)&0.048(17)\\
1.60&0.032(8)&0.059(17)\\
1.54&0.030(7)&0.067(20)\\
1.62&0.038(8)&0.067(19)\\
1.52&0.029(7)&0.053(19)\\
1.63&0.060(8)&0.070(21)\\
1.51&0.034(6)&0.052(16)\\
1.65&0.074(8)&0.053(18)\\
1.49&0.045(6)&0.050(18)\\
    \hline
    \hline
  \end{tabular}
  \hspace{1cm}
  \begin{tabular}{c|c}
  \hline
  \hline
  $|x\rangle$ & probability \\
  \hline
  \hline
0&0.001(2)\\
1&0.004(2)\\
2&0.007(3)\\
3&0.012(3)\\
4&0.039(4)\\
5&0.065(8)\\
6&0.153(12)\\
7&0.250(16)\\
8&0.230(19)\\
9&0.121(12)\\
10&0.061(7)\\
11&0.033(3)\\
12&0.011(3)\\
13&0.005(2)\\
14&0.005(2)\\
15&0.001(1)\\
  \hline
  \hline
  \end{tabular}
  \caption{
  Numerical values of the asymmetries and probabilities presented in Fig.~\ref{fig:HadamardWindowExp} in the main text.
  The vertical order in the center table correlates with increasing time throughout the production.
  }
  \label{tab:1siteWindowdatatable}
\end{table}

\begin{table}
  \centering
  \begin{tabular}{c|c|c|c}
  \hline
  \hline
  $\theta$ & $r$ & $A_{CE}^{(2)}$ & $A_{SE}^{(2)}$ \\
  \hline
  \hline
  1.41&1&0.092(10)&0.123(10)\\
1.41&3&0.104(11)&0.106(8)\\
1.41&5&0.081(11)&0.115(6)\\
1.49&1&0.053(10)&0.075(8)\\
1.49&3&0.048(9)&0.065(6)\\
1.49&5&0.039(9)&0.108(6)\\
1.53&1&0.045(7)&0.060(6)\\
1.53&3&0.058(8)&0.066(5)\\
\end{tabular}
\begin{tabular}{c|c|c|c}
\multicolumn{4}{c}{}\\
\multicolumn{4}{c}{}\\
1.53&5&0.040(8)&0.087(4)\\
1.57&1&0.036(9)&0.045(8)\\
1.57&3&0.029(6)&0.052(6)\\
1.57&5&0.035(8)&0.095(7)\\
1.57&1&0.026(5)&0.038(6)\\
1.57&3&0.038(7)&0.053(4)\\
1.57&5&0.035(7)&0.094(4)\\
1.61&1&0.028(7)&0.029(4)\\
\end{tabular}
\begin{tabular}{c|c|c|c}
\multicolumn{4}{c}{}\\
\multicolumn{4}{c}{}\\
1.61&3&0.025(5)&0.053(4)\\
1.61&5&0.034(7)&0.095(4)\\
1.65&1&0.037(9)&0.047(8)\\
1.65&3&0.046(10)&0.043(6)\\
1.65&5&0.041(8)&0.092(7)\\
1.73&1&0.065(11)&0.064(9)\\
1.73&3&0.059(10)&0.071(8)\\
1.73&5&0.045(8)&0.105(8)\\
  \hline
  \hline
  \end{tabular}
  \caption{
  Numerical values of the asymmetries appearing in the top left panel of Fig.~\ref{fig:2site} of the main text.
  }
  \label{tab:asymmetry2sitedatatable}
\end{table}
\begin{table}
  \centering
  \begin{tabular}{cccccccc}
0.0003(4)&0.0001(2)&0.0017(15)&0.0015(15)&0.0002(2)&0.0002(3)&0.0004(5)&0.0001(1)\\
0.0004(5)&0.0001(3)&0.0015(14)&0.0124(29)&0.0052(30)&0.0036(16)&0.0011(9)&0.0002(2)\\
0.0004(5)&0.0038(20)&0.0103(30)&0.0418(63)&0.0416(54)&0.0093(28)&0.0023(16)&0.0010(9)\\
0.0036(27)&0.0072(39)&0.0296(52)&0.1686(105)&0.1571(90)&0.0355(44)&0.0167(35)&0.0067(24)\\
0.0030(27)&0.0030(24)&0.0261(49)&0.1376(81)&0.1380(72)&0.0317(44)&0.0116(28)&0.0032(20)\\
0.0003(4)&0.0005(6)&0.0051(24)&0.0297(42)&0.0368(45)&0.0069(24)&0.0004(7)&0.0000(1)\\
0.0000(1)&0.0002(3)&0.0017(12)&0.0008(12)&0.0058(26)&0.0003(4)&0.0006(6)&0.0003(3)\\
0.0000(1)&0.0001(2)&0.0004(7)&0.0028(23)&0.0001(5)&0.0020(13)&0.0007(7)&0.0001(2)\\
  \end{tabular}
  \caption{
  Numerical values of the probabilities appearing in the top right panel of Fig.~\ref{fig:2site} of the main text.
  To translate this back to the linearized Hilbert space, this data should be read left-to-right and top-to-bottom.
  }
  \label{tab:lego2sitedatatable}
\end{table}
\begin{table}
  \centering
  \begin{tabular}{c|c|c|c}
  \hline
  \hline
  $|x\rangle$ & Gaussian & Exponential & Piecewise \\
  \hline
  \hline
  0&0.0012&0.0260&0.0158\\
1&0.0025&0.0320&0.0180\\
2&0.0050&0.0394&0.0208\\
3&0.0096&0.0485&0.0237\\
4&0.0177&0.0596&0.0331\\
5&0.0309&0.0733&0.0376\\
6&0.0513&0.0902&0.0435\\
7&0.0809&0.1110&0.0494\\
\end{tabular}
\begin{tabular}{c|c|c|c}
\multicolumn{4}{c}{}\\
\multicolumn{4}{c}{}\\
8&0.1213&0.1366&0.1719\\
9&0.1729&0.1681&0.1954\\
10&0.2343&0.2068&0.2260\\
11&0.3018&0.2544&0.2569\\
12&0.3695&0.3130&0.3591\\
13&0.4302&0.3851&0.4083\\
14&0.4760&0.4738&0.4722\\
15&0.5007&0.5829&0.5368\\
\hline
\hline
  \end{tabular}
  \caption{
  Numerical values of the wavefunctions appearing in Fig.~\ref{fig:pert} of the main text.
  }
  \label{tab:PWdatatable}
\end{table}
\begin{table}
  \centering
  \scalebox{0.65}{
  \begin{tabular}{cccccccccccccccc}
0.776(5)&0.037(2)&0.036(2)&0.002(0)&0.069(3)&0.003(1)&0.003(1)&0.000(0)&0.024(2)&0.001(0)&0.001(0)&0.000(0)&0.002(0)&0.000(0)&0.000(0)&0.000(0)\\
0.038(2)&0.783(5)&0.001(0)&0.031(2)&0.004(1)&0.069(3)&0.000(0)&0.002(1)&0.001(0)&0.023(2)&0.000(0)&0.000(0)&0.000(0)&0.002(0)&0.000(0)&0.000(0)\\
0.028(2)&0.001(0)&0.776(5)&0.039(2)&0.002(0)&0.000(0)&0.072(3)&0.004(1)&0.001(0)&0.000(0)&0.025(2)&0.001(0)&0.000(0)&0.000(0)&0.003(1)&0.000(0)\\
0.001(0)&0.029(2)&0.037(2)&0.780(5)&0.000(0)&0.002(0)&0.004(1)&0.069(3)&0.000(0)&0.002(0)&0.002(0)&0.024(2)&0.000(0)&0.000(0)&0.000(0)&0.003(1)\\
0.065(3)&0.004(1)&0.003(1)&0.000(0)&0.778(5)&0.036(2)&0.032(2)&0.001(0)&0.002(0)&0.000(0)&0.000(0)&0.000(0)&0.023(2)&0.001(0)&0.001(0)&0.000(0)\\
0.003(1)&0.059(3)&0.000(0)&0.003(1)&0.035(2)&0.777(5)&0.002(0)&0.029(2)&0.000(0)&0.001(0)&0.000(0)&0.000(0)&0.001(0)&0.026(2)&0.000(0)&0.001(0)\\
0.002(1)&0.000(0)&0.058(3)&0.004(1)&0.031(2)&0.002(0)&0.769(5)&0.038(2)&0.000(0)&0.000(0)&0.002(0)&0.000(0)&0.001(0)&0.000(0)&0.025(2)&0.002(0)\\
0.000(0)&0.003(1)&0.003(1)&0.061(3)&0.001(0)&0.032(2)&0.033(2)&0.775(5)&0.000(0)&0.000(0)&0.000(0)&0.001(0)&0.000(0)&0.001(0)&0.001(0)&0.026(2)\\
0.023(2)&0.001(0)&0.001(0)&0.000(0)&0.002(1)&0.000(0)&0.000(0)&0.000(0)&0.777(5)&0.034(2)&0.029(2)&0.001(0)&0.063(3)&0.004(1)&0.002(1)&0.000(0)\\
0.001(0)&0.023(2)&0.000(0)&0.001(0)&0.000(0)&0.001(0)&0.000(0)&0.000(0)&0.035(2)&0.782(5)&0.002(0)&0.030(2)&0.004(1)&0.069(3)&0.000(0)&0.003(1)\\
0.001(0)&0.000(0)&0.021(2)&0.001(0)&0.000(0)&0.000(0)&0.002(0)&0.000(0)&0.031(2)&0.001(0)&0.781(5)&0.034(2)&0.002(0)&0.000(0)&0.066(3)&0.004(1)\\
0.000(0)&0.000(0)&0.001(0)&0.023(2)&0.000(0)&0.000(0)&0.000(0)&0.002(0)&0.002(0)&0.027(2)&0.033(2)&0.778(5)&0.000(0)&0.003(1)&0.002(1)&0.065(3)\\
0.002(0)&0.000(0)&0.000(0)&0.000(0)&0.021(2)&0.001(0)&0.002(0)&0.000(0)&0.065(3)&0.003(1)&0.002(0)&0.000(0)&0.782(5)&0.036(2)&0.028(2)&0.001(0)\\
0.000(0)&0.002(1)&0.000(0)&0.000(0)&0.001(0)&0.020(2)&0.000(0)&0.001(0)&0.004(1)&0.063(3)&0.000(0)&0.001(0)&0.030(2)&0.765(5)&0.001(0)&0.031(2)\\
0.000(0)&0.000(0)&0.002(0)&0.000(0)&0.001(0)&0.000(0)&0.024(2)&0.001(0)&0.003(1)&0.000(0)&0.066(3)&0.004(1)&0.031(2)&0.002(1)&0.785(5)&0.036(2)\\
0.000(0)&0.000(0)&0.000(0)&0.003(1)&0.000(0)&0.001(0)&0.001(0)&0.025(2)&0.000(0)&0.002(1)&0.002(1)&0.068(3)&0.001(0)&0.033(2)&0.030(2)&0.774(5)\\
  \end{tabular}
  }
\caption{
Numerical values of the measurement-error calibration matrix elements appearing in the top left panel
of Fig.~\ref{fig:PouCalibrationMatrices}.
}
\label{tab:MCc16datatable}
\end{table}
\begin{table}
  \centering
  \scalebox{0.65}{
  \begin{tabular}{cccccccccccccccc}
0.900(3)&0.081(3)&0.053(3)&0.006(1)&0.056(3)&0.004(1)&0.004(1)&0.000(0)&0.058(3)&0.007(1)&0.003(1)&0.001(0)&0.004(1)&0.000(0)&0.000(0)&0.000(0)\\
0.023(2)&0.842(4)&0.003(1)&0.070(3)&0.002(0)&0.054(3)&0.000(0)&0.003(1)&0.002(1)&0.071(3)&0.000(0)&0.005(1)&0.000(0)&0.003(1)&0.000(0)&0.000(0)\\
0.020(2)&0.003(1)&0.868(4)&0.077(3)&0.001(0)&0.000(0)&0.056(3)&0.006(1)&0.003(1)&0.001(0)&0.055(3)&0.005(1)&0.000(0)&0.000(0)&0.004(1)&0.001(0)\\
0.000(0)&0.022(2)&0.021(2)&0.794(5)&0.000(0)&0.001(0)&0.002(0)&0.049(2)&0.000(0)&0.003(1)&0.003(1)&0.061(3)&0.000(0)&0.000(0)&0.001(0)&0.005(1)\\
0.016(1)&0.001(0)&0.001(0)&0.000(0)&0.862(4)&0.075(3)&0.050(2)&0.004(1)&0.001(0)&0.000(0)&0.000(0)&0.000(0)&0.059(3)&0.006(1)&0.004(1)&0.000(0)\\
0.000(0)&0.012(1)&0.000(0)&0.001(0)&0.023(2)&0.808(4)&0.001(0)&0.050(2)&0.000(0)&0.001(0)&0.000(0)&0.000(0)&0.003(1)&0.071(3)&0.001(0)&0.005(1)\\
0.000(0)&0.000(0)&0.011(1)&0.001(0)&0.017(1)&0.003(1)&0.832(4)&0.075(3)&0.000(0)&0.000(0)&0.001(0)&0.000(0)&0.001(0)&0.000(0)&0.054(3)&0.006(1)\\
0.000(0)&0.000(0)&0.000(0)&0.009(1)&0.000(0)&0.018(1)&0.016(1)&0.767(5)&0.000(0)&0.000(0)&0.000(0)&0.001(0)&0.000(0)&0.002(0)&0.011(1)&0.065(3)\\
0.038(2)&0.003(1)&0.002(1)&0.000(0)&0.003(1)&0.000(0)&0.000(0)&0.000(0)&0.874(4)&0.078(3)&0.055(3)&0.006(1)&0.057(3)&0.004(1)&0.003(1)&0.000(0)\\
0.001(0)&0.035(2)&0.000(0)&0.003(1)&0.000(0)&0.003(1)&0.000(0)&0.000(0)&0.026(2)&0.804(4)&0.001(0)&0.056(3)&0.001(0)&0.051(2)&0.000(0)&0.003(1)\\
0.001(0)&0.000(0)&0.039(2)&0.004(1)&0.000(0)&0.000(0)&0.002(1)&0.001(0)&0.023(2)&0.002(1)&0.847(4)&0.073(3)&0.001(0)&0.000(0)&0.046(2)&0.004(1)\\
0.000(0)&0.001(0)&0.002(0)&0.034(2)&0.000(0)&0.000(0)&0.000(0)&0.003(1)&0.001(0)&0.021(2)&0.021(2)&0.780(5)&0.000(0)&0.001(0)&0.002(0)&0.046(2)\\
0.001(0)&0.000(0)&0.000(0)&0.000(0)&0.035(2)&0.003(1)&0.003(1)&0.000(0)&0.010(1)&0.001(0)&0.001(0)&0.000(0)&0.832(4)&0.069(3)&0.060(3)&0.006(1)\\
0.000(0)&0.000(0)&0.000(0)&0.000(0)&0.001(0)&0.032(2)&0.000(0)&0.002(1)&0.000(0)&0.010(1)&0.000(0)&0.000(0)&0.020(2)&0.776(5)&0.004(1)&0.059(3)\\
0.000(0)&0.000(0)&0.000(0)&0.000(0)&0.001(0)&0.000(0)&0.033(2)&0.004(1)&0.000(0)&0.000(0)&0.013(1)&0.001(0)&0.021(2)&0.001(0)&0.768(5)&0.065(3)\\
0.000(0)&0.000(0)&0.000(0)&0.001(0)&0.000(0)&0.001(0)&0.001(0)&0.036(2)&0.000(0)&0.000(0)&0.000(0)&0.010(1)&0.001(0)&0.017(1)&0.043(2)&0.734(5)\\
  \end{tabular}
  }
\caption{
Numerical values of the measurement-error calibration matrix elements appearing in the top right
panel of Fig.~\ref{fig:PouCalibrationMatrices}.
}
\label{tab:MCq16datatable}
\end{table}
\begin{table}
  \centering
  \begin{tabular}{c|c}
   \multicolumn{2}{c}{Vacuum} \\
  \hline
  \hline
  $\theta_{\rm in}$ & $\theta_{\rm in} - \theta_{\rm meas}$ \\
  \hline
  \hline
1.30&0.024(4)\\
1.34&0.026(4)\\
1.37&0.027(4)\\
1.41&0.030(4)\\
1.45&0.022(4)\\
1.49&0.037(4)\\
1.53&0.040(4)\\
1.57&0.039(4)\\
1.61&0.039(4)\\
1.65&0.039(4)\\
1.69&0.044(4)\\
1.73&0.049(4)\\
1.77&0.047(4)\\
1.81&0.047(4)\\
1.85&0.060(4)\\
1.88&0.051(4)\\
\hline
\hline
  \end{tabular}
  \hspace{1cm}
  \begin{tabular}{c|c}
  \multicolumn{2}{c}{In-Medium} \\
  \hline
  \hline
  $\theta_{\rm in}$ & $\theta_{\rm in} - \theta_{\rm meas}$ \\
  \hline
  \hline
1.30&-0.008(7)\\
1.34&-0.001(7)\\
1.37&-0.005(8)\\
1.41&-0.000(7)\\
1.45&0.008(8)\\
1.49&0.016(8)\\
1.53&0.010(7)\\
1.57&0.001(8)\\
1.61&0.007(8)\\
1.65&0.004(7)\\
1.69&0.016(7)\\
1.73&0.027(8)\\
1.77&0.005(8)\\
1.81&0.018(7)\\
1.85&0.026(8)\\
1.88&0.025(8)\\
1.60&0.011(8)\\
1.57&-0.008(7)\\
\end{tabular}
\begin{tabular}{c|c}
\multicolumn{2}{c}{}\\
\multicolumn{2}{c}{}\\
1.59&0.023(8)\\
1.56&0.033(8)\\
1.60&0.011(7)\\
1.54&0.014(8)\\
1.62&0.013(8)\\
1.52&0.012(8)\\
1.63&0.011(8)\\
1.51&0.004(8)\\
1.59&0.005(8)\\
1.65&0.005(7)\\
1.59&0.003(7)\\
1.49&-0.009(8)\\
1.59&0.009(8)\\
1.52&-0.007(7)\\
1.59&0.004(7)\\
1.52&0.019(8)\\
1.59&-0.003(7)\\
1.52&0.013(7)\\
\end{tabular}
\begin{tabular}{c|c}
\multicolumn{2}{c}{}\\
\multicolumn{2}{c}{}\\
1.59&0.026(8)\\
1.52&0.017(8)\\
1.59&0.006(8)\\
1.52&0.009(8)\\
1.59&0.022(8)\\
1.52&0.004(8)\\
1.59&0.020(7)\\
1.52&0.008(8)\\
1.59&0.019(7)\\
1.52&0.016(7)\\
1.59&-0.001(7)\\
1.52&0.013(7)\\
1.59&0.022(8)\\
1.52&0.018(7)\\
1.52&0.003(8)\\
1.52&0.011(8)\\
1.48&0.017(8)\\
1.48&0.012(8)\\
1.48&0.011(7)\\
\hline
\hline
  \end{tabular}
  \caption{
  Numerical values of the raw angle deviation results presented in the top panel of Fig.~\ref{fig:measurementcorrection}.
  The ordering in these tables reflects the time-ordering of the measurements on the quantum device as shown
  by the saturation of  points and error bars in Fig.~\ref{fig:measurementcorrection}.
  }
  \label{tab:MCinputanglesDatatable}
\end{table}

\begin{table}
  \centering
  \begin{tabular}{c|c}
   \multicolumn{2}{c}{Vacuum} \\
  \hline
  \hline
  $\theta_{\rm in}$ & $\theta_{\rm in} - \theta_{\rm meas}$ \\
  \hline
  \hline
1.3&0.028(4)\\
1.34&0.032(4)\\
1.37&0.028(4)\\
1.41&0.027(3)\\
1.45&0.017(3)\\
1.49&0.029(3)\\
1.53&0.033(3)\\
1.57&0.029(3)\\
1.61&0.027(2)\\
1.65&0.027(4)\\
1.69&0.029(6)\\
1.73&0.029(4)\\
1.77&0.027(3)\\
1.81&0.025(3)\\
1.85&0.035(5)\\
1.88&0.025(4)\\
\hline
\hline
\end{tabular}
\hspace{1cm}
  \begin{tabular}{c|c}
  \multicolumn{2}{c}{In-Medium} \\
  \hline
  \hline
  $\theta_{\rm in}$ & $\theta_{\rm in} - \theta_{\rm meas}$ \\
  \hline
  \hline
1.26&-0.334(10)\\
1.3&-0.007(7)\\
1.34&-0.000(9)\\
1.37&-0.006(9)\\
1.41&-0.008(8)\\
1.45&-0.002(9)\\
1.49&0.001(7)\\
1.53&-0.002(7)\\
1.57&-0.019(9)\\
1.61&-0.011(9)\\
1.65&-0.019(8)\\
1.69&-0.013(6)\\
1.73&0.002(8)\\
1.77&-0.030(6)\\
1.81&-0.016(8)\\
1.85&-0.009(6)\\
1.88&-0.015(9)\\
1.6&-0.009(9)\\
\end{tabular}
\begin{tabular}{c|c}
\multicolumn{2}{c}{}\\
\multicolumn{2}{c}{}\\
1.57&-0.029(7)\\
1.59&0.008(8)\\
1.56&0.018(10)\\
1.6&-0.012(5)\\
1.54&-0.000(11)\\
1.62&-0.007(8)\\
1.52&-0.005(9)\\
1.63&-0.005(9)\\
1.51&-0.007(6)\\
1.59&-0.017(9)\\
1.65&-0.016(10)\\
1.59&-0.014(9)\\
1.49&-0.024(7)\\
1.59&-0.012(7)\\
1.52&-0.023(10)\\
1.59&-0.018(8)\\
1.52&0.008(10)\\
1.59&-0.027(8)\\
1.52&0.002(9)\\
\end{tabular}
\begin{tabular}{c|c}
\multicolumn{2}{c}{}\\
\multicolumn{2}{c}{}\\
1.59&0.007(6)\\
1.52&0.006(6)\\
1.59&-0.014(8)\\
1.52&-0.007(8)\\
1.59&0.008(9)\\
1.52&-0.013(9)\\
1.59&0.005(5)\\
1.52&-0.000(6)\\
1.59&0.007(8)\\
1.52&0.007(7)\\
1.59&-0.019(7)\\
1.52&0.007(7)\\
1.59&0.003(9)\\
1.52&0.007(6)\\
1.52&-0.011(7)\\
1.52&0.001(9)\\
1.48&0.003(8)\\
1.48&-0.000(6)\\
1.48&-0.000(7)\\
  \hline
  \hline
  \end{tabular}
  \caption{
  Numerical values of the measurement-error corrected angle deviation results presented in the bottom panel of
  Fig.~\ref{fig:measurementcorrection}.
  The ordering in these tables reflects the time-ordering of the measurements on the quantum device as shown
  by the saturation of  points and error-bars in Fig.~\ref{fig:measurementcorrection}.
  }
  \label{tab:MCinputanglesDatatable}
\end{table}
\begin{acknowledgements}
We would like to thank Aidan Murran, John Preskill, Alessandro Roggero, Jesse Stryker, and Nathan Wiebe for stimulating discussions.
We would also like to thank our Oak Ridge collaborators,
Eugene Dumitrescu,
Pavel Lougovski, Alex McCaskey, Titus Morris, Raphael Pooser
and George Siopsis for previous discussions related to lattice scalar field theory.
We would like to thank Donny Greenberg and the ORNL Q Hub for facilitating the use of IBM's quantum devices.
Finally, we would like to thank David Hertzog for providing us with quiet space in CENPA to develop the quantum circuitry presented in this work.
This work is supported by U.S. Department of Energy grant No. DE-FG02-00ER41132 and by the U.S. Department
of Energy, Office of Science, Office of Advanced Scientific Computing Research (ASCR) quantum algorithm teams
program, under field work proposal number ERKJ333.
NK was supported in part by a Microsoft PhD Fellowship.
We gratefully acknowledge use of the IBM Q experience for generating the quantum results shown in this work.
Classical calculations were done using~\emph{Wolfram Mathematica 11.1}
and the quantum circuits appearing in this paper were typeset using the latex package~\emph{Qcircuit},
originally developed by Bryan Eastin and Steven Flammia.
\end{acknowledgements}

\bibliography{gspbib}
\end{document}